\documentclass[twoside,12pt]{article}
\usepackage{epsfig}
\usepackage{amsmath,amssymb}
\usepackage{enumerate}

\newcommand{\be}{\begin{equation}}
\newcommand{\ee}{\end{equation}}

\usepackage{lineno}
\usepackage{hyperref}

\oddsidemargin- 1.cm
\begin{document}

\title{ \vspace{0.3cm} Photon-photon physics at the LHC and laser beam experiments, present and future}
\author{ 
L.\ Schoeffel$^1$, C.\ Baldenegro$^{2}$, H. Hamdaoui$^3$, S.\ Hassani$^1$, C.\ Royon$^2$, M. Saimpert$^4$ \\
\\
{\small $^1$H. CEA Saclay, Irfu/DPhP, Gif-sur-Yvette, France} \\
{\small $^2$ University  of Kansas, Lawrence, Kansas, U.S.} \\
{\small $^3$ Facult\'e des sciences, Universit\'e Mohammed V, Rabat, Morocco} \\
{\small $^4$ CERN, Geneva, Switzerland.}
}
\maketitle
\begin{abstract} 

\noindent
Under certain running conditions, the CERN Large Hadron Collider (LHC) can be considered as a photon-photon collider. Indeed, in proton-proton, 
proton-ion, ion-ion collisions, when incoming particles pass very close to each other in very peripheral collisions, the incoming 
protons or ions remain almost intact and continue their path along the beam axis. Then, only the electromagnetic (EM) fields of these ultra-relativistic 
charged particles (protons or ions) interact to leave a signature in the central detectors of the LHC experiments. 
The interest is that the photon-photon interactions happen at unprecedented energies 
(a few TeV per nucleon pairs)
where the quantum electrodynamics (QED) theory 
can be tested in extreme conditions and unforeseen laws of nature could be discovered.  
In this report, we propose a focus on a particular reaction, called light-by-light scattering in which two incoming photons interact, 
producing another pair of photons. We describe how 
experimental results have been obtained at the LHC.
In addition, we discuss prospects for on-shell photon-photon interactions in dedicated laser beam facilities. Potential signatures
of new physics might manifest as resonant deviations in the refractive index, induced by anomalous light-by-light scattering effects.
Importantly, we explain how this process can be used to probe the physics beyond the standard model such as theories that include large extra dimensions. 
Finally, some perspectives and ideas are given for future data taking or experiments.

\end{abstract}
\newpage
\tableofcontents
\newpage


\section{Introduction}

Collisions of ions at ultra-relativistic energies, like lead($Pb^{82+}$)-lead($Pb^{82+}$) collisions at the LHC, are usually studied in cases where the ions interact hadronically, producing large multiplicities and a volume in which  participants to the reactions can form a domain of hot, dense quark-gluon matter  \cite{Citron:2018lsq,Ollitrault:2008zz,moi}. However, in addition to central collisions, 
it is  of interest to look also at very peripheral (or ultra-peripheral)
collisions (UPC), which are characterized by the condition that the impact parameter $b$ of the reaction is
larger than the sum of the ions radii. Then, the two ions do not interact directly via the nuclear
interaction, which is mostly short range, but mainly through their electromagnetic (EM) fields.
Indeed, relativistic ions  generate strong EM fields (with magnitudes of $E \sim 10^{25}$ V/m for lead-lead collisions at the LHC with a center of mass energy per nucleon pair of $5$ TeV), where the electric and magnetic fields of each ion (measured at a point of transverse distance of order $b$ from the ion center of mass) are perpendicular and almost transverse to the direction of propagation. These EM fields may be treated as a flux of nearly-real photons and can give rise to photonuclear (photon versus ion) or photon-photon interactions.   By definition, these interactions are expected to occur mainly in UPC collisions. 
The order of magnitude given above ($E \sim 10^{25}$ V/m for ultra-relativistic lead) is easy to justify.

\begin{figure}[!h]   
\centering
\includegraphics[width=0.7\textwidth]{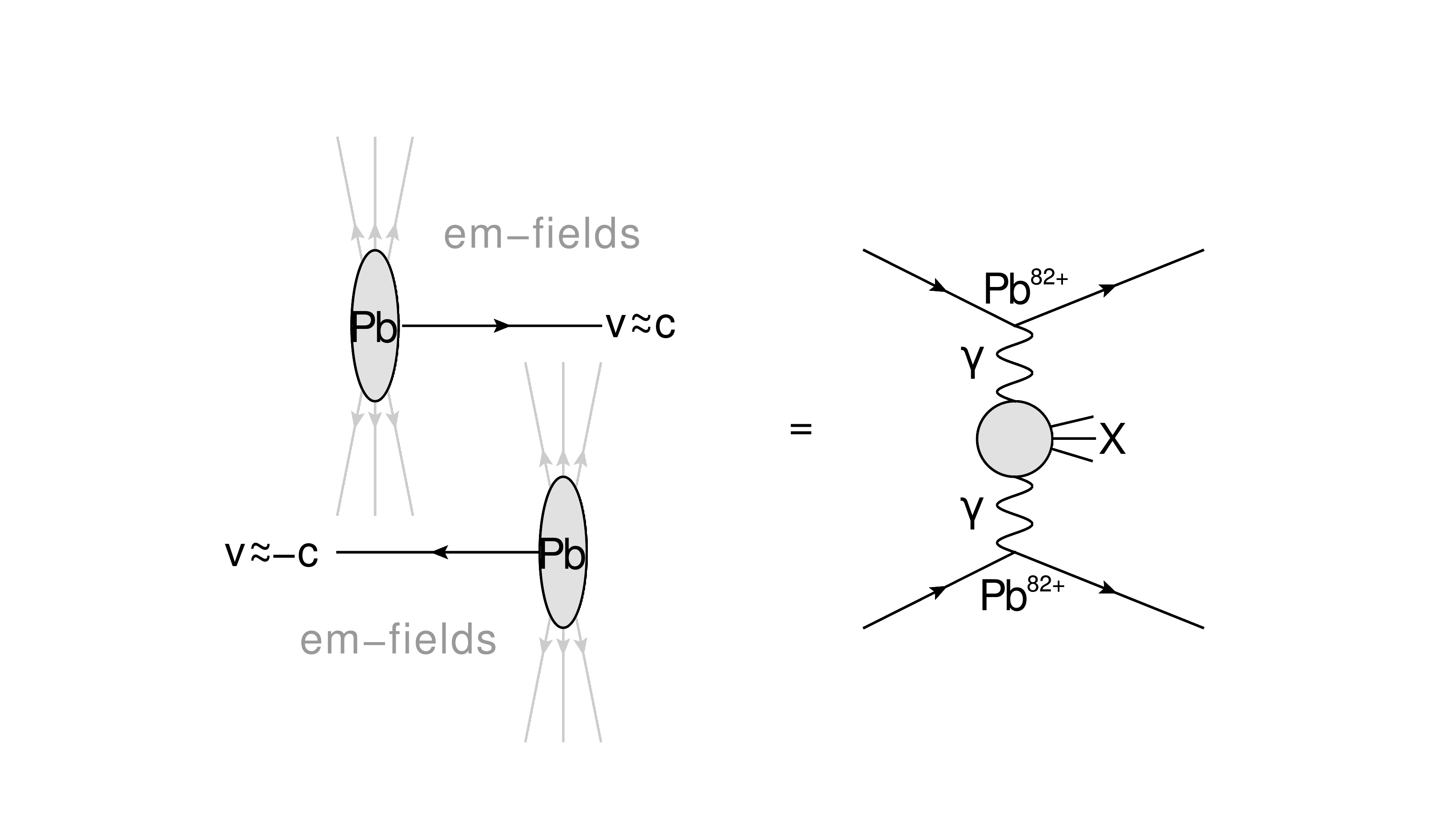}
\caption{\label{fig0}
{\small
Illustration of an ultra peripheral collision of two lead ions. The EM interaction between the ions is described (equivalently) as an exchange of photons that  couple to form a given generic final state X. }
}
\end{figure}

We start from the definition of the electric field measured at a transverse distance $b$ from the ion center of mass: $E \sim Ze \gamma/b^2$ where
$\gamma$ is the Lorentz factor, $Z$ the charge number and $e$ the elementary electric charge. Then, we convert this expression as: $E \sim Ze \gamma/b^2 \sim  (Z=82) \gamma \alpha_{em} E_{cr}$ where 
$\alpha_{em}=e^2/(4\pi) \simeq 1/137$ and 
$E_{cr}$ the critical (Schwinger) electric field. What we call the critical electric field $E_{cr}$ is defined as: $e E_{cr} \frac{1}{m_e} = m_e$, which gives an order of magnitude of $E_{cr} \sim 10^{18}$ V/m.
Thus, we find: $E \sim  (Z=82) \gamma \alpha_{em} E_{cr} \sim 10^{25}$ V/m $>> E_{cr}$. We understand that the physics relevant to describe these processes is
the quantum electrodynamics (QED) at high intensity but with an interaction time still very short $\delta t \sim b/\gamma$, meaning that  perturbation theory can (in principle) still be used. 
An important physics parameter that is characteristic of the high intensity regime of QED is the ratio: ${\cal Y} = E/E_{cr}$. When this ratio is approaching unity or above, the high intensity regime is reached. A broader discussion on this property is done in section (\ref{TT}) where the ratio ${\cal Y}$
is examined and calculated in a given laser beam experiment.
This is an essential purpose of the measurements presented in this document to confirm or infirm this statement.
In the following, we focus the presentation of the physics case of photon-photon interactions in UPC
\cite{Baur:2001jj}, where the impact parameter of the collision is large (larger than the sum of the sizes of the colliding ions in the transverse plane). Thus,  the strong interactions are suppressed 
while the EM fields of the colliding ions interact
(see Fig. \ref{fig0}).

One key point in such EM interactions, which means the interactions of the EM fields from the colliding ions at a point (area) in the transverse plane of the collision, or photon-photon interactions is that the photons are nearly real. Thus the actions of all charges in the ions are coherent and the photon flux for each ion scales as $Z^2$ (where $Z$ is the number of protons of the ions): $Z=82$ in case of lead($Pb^{82+}$)-lead($Pb^{82+}$) collisions and $Z=79$ in case of gold($Au^{79+}$)-gold($Au^{79+}$) collisions. Therefore, photon-photon effective luminosities are then enhanced by a factor $Z^4=4.5 \ 10^7$ for lead-lead collisions, $Z^4=3.9 \ 10^7$ for gold-gold collisions,  compared to proton-proton collisions.
However, as the instantaneous luminosity in lead-lead is about $7$ orders of magnitude below the one in proton-proton, the global factor between lead-lead and proton-proton effective luminosities in $\gamma\gamma$ is not that large.
In practice, with the present parameters of the LHC machine, we find that: ${\cal L}_{PbPb}(\gamma\gamma)/{\cal L}_{pp}(\gamma\gamma) \sim 10$.

What matters also is what we call the pile-up environment which is in general quite different for ion-ion and proton-proton collisions. We expose all this in details in the next sections.
First results have been obtained in $Au Au \rightarrow Au Au (\gamma \gamma) \rightarrow Au Au e^+ e^-$ at the Relativistic Heavy Ion Collider (RHIC) for a center of mass energy per nucleon pair of $200$ GeV  \cite{Adams:2004rz,Afanasiev:2009hy}.
Then, results have been obtained at the LHC   first in proton-proton  collisions 
at center of mass energies of $7$ TeV, $8$ TeV and $13$ TeV with several final states produced: $pp \rightarrow pp (\gamma \gamma) \rightarrow pp \mu^+ \mu^-$,
$pp \rightarrow pp (\gamma \gamma) \rightarrow pp e^+ e^-$ \cite{Chatrchyan:2011ci,Chatrchyan:2012tv,Aad:2015bwa,Aaboud:2017oiq} or
$pp \rightarrow pp (\gamma \gamma) \rightarrow pp W^+ W^-$ \cite{Chatrchyan:2013akv,Khachatryan:2016mud,Aaboud:2016dkv}
and then in lead-lead collisions at a center of mass energy per nucleon pair of $5.02$ TeV  ($2.76$ TeV)
with $Pb Pb \rightarrow Pb Pb (\gamma \gamma) \rightarrow Pb Pb \mu^+ \mu^-$ \cite{atlaspbpb}
($Pb Pb \rightarrow Pb Pb (\gamma \gamma) \rightarrow Pb Pb e^+ e^-$ \cite{alice}).

Let us mention that colliding energies coincide, making a very coherent set of data for ion-ion and proton-proton collisions. 
For example, a center of mass energy of $5$ TeV per nucleon pair for lead-lead scattering correspond to a proton-proton center of mass energy
of $5*208/82 \sim 13$ TeV. Similarly, for $2.7$ TeV per nucleon pair for lead-lead scattering, we get $2.7*208/82 \sim 7$ TeV in proton-proton.
Also, as the interaction time is of the form $\delta t \sim b/\gamma$, this means that the maximum energy for an interacting photon is inversely 
proportional to the impact parameter ($\omega_{max} \sim \gamma/b$) and thus, for a given nucleon-nucleon energy, is maximum for proton-proton, then for proton-ion 
(larger for proton-argon than for proton-lead)
and finally the photon energy is minimal for 
ion-ion (larger for argon-argon than for lead-lead). 
Following these statements, 
we show in this review how to reach an optimal use these properties in order to cover a large range of possibilities in 
searches beyond the standard predictions.
Finally, 
photon-photon interactions (in lead-lead collisions) at the LHC have provided
 the observation of the so-called light-by-light scattering:
$ Pb Pb \rightarrow Pb Pb (\gamma \gamma) \rightarrow Pb Pb \gamma \gamma$
\cite{dEnterria:2013zqi,lbl1,lbl2,lbl3,Sirunyan:2018fhl}.
We can mention already  that for this reaction,
the two photons that collide  produce another set of two photons, which corresponds to an elastic collision of two photons and thus
 justifies the label of the reaction: light-by-light scattering.

In this report, we describe all these aspects of photon-photon physics at the LHC, mainly in the ultra-peripheral collision configuration. We intend to present extensively the experimental results in proton-proton, ion-ion or proton-ion collisions, with an emphasis of rare phenomenon as the recent observation of the elastic scattering of two photons in lead-lead collisions. For the latter case, with some motivations explained in the report, we extend the discussion from lead beam experiments feasible at the LHC to laser beam experiments feasible in the laboratory. For each case, we intent to describe  the level of understanding, experimental and theoretical, that we have achieved. We hope this work could  be useful  to help interpreting correctly any deviation which may be observed by current experiments and plan ahead the future of the UPC physics program at LHC or HL-LHC
(high luminosity LHC).

\section{Formalism of photon-photon interactions (in UPC)}

\subsection{Position of the problem}

First, let us pose the physics problem of photon-photon interactions in the case of proton-proton collisions. This is the simplest case and will have an immediate extension for proton-ion or ion-ion collisions. The theory is well known, since this is directly related to interactions in QED and a precise determination of the electromagnetic form factors of the proton. Our purpose is to pose some notations, explain how we can build the generators on existing calculations and emphasize the potential theory uncertainties.
As mentioned in the introduction, calculations of the cross-section for photon-photon interactions in proton-proton collisions are based on the 
equivalent photon approximation (EPA) \cite{fermi,ww,t1,t2,Budnev:1973tz}. This relies on the property that the EM field of a charged particle, here a proton, moving at high velocity becomes more and more transverse with respect to the direction of propagation. As a consequence, an observer in the laboratory frame cannot distinguish between the EM field of a relativistic proton and the transverse component of the EM field associated with equivalent photons. Therefore, using the EPA, we can write the cross-section for a reaction where the two photons fuse to give a generic final state $X$:
\begin{equation}
\sigma(p p \rightarrow p p (\gamma \gamma) \rightarrow p p X) =\int \int f(\omega_1)f(\omega_2) \sigma_{\gamma \gamma \rightarrow X}(\omega_1,\omega_2) 
{d\omega_1}{d\omega_2},
\label{start}
\end{equation}
where $f(.)$ are the equivalent photon flux functions and
 $\omega_{1,2}$ represent the energies of the photons, integrated over.
 These functions can be obtained easily from the Poynting vector associated to the EM field produced by the ultra-relativistic proton.
They read:
$$
f(\omega)=
\frac{e^2}{\pi \omega}
\int \frac{d^2 \vec{k}_\perp}{(2 \pi)^2} 
  \left(\frac{F(k_\perp^2+\frac{\omega^2}{\gamma^2})}{k_\perp^2+\frac{\omega^2}{\gamma^2}}\right)^2
|\vec{k}_\perp|^2,
$$
where $F(.)$ represents the proton form factor, including electric and magnetic components.
Here $\gamma$ is the Lorentz contraction factor, 
 $\omega$ and $\vec{k}_\perp$ represent the energy and transverse momentum of photons respectively.
For each photon, the maximum energy is  the energy of the incident proton $\sqrt{s}/2$. However, there is also the 
constraint that the highest available energy for one photon is of the order of the inverse Lorentz contracted radius of the proton, 
$\gamma/r_p$,
where $r_p$ represents the proton radius and $\gamma$ the Lorentz factor equal to $\sqrt{s}/(2 m_p)$ here.
Let us note that the two photon center-of-mass energy squared is $W_{\gamma \gamma}^2 = 4\omega_1\omega_2$, and the rapidity of the two photons system is defined as
$y_{\gamma \gamma} = 0.5 \ln[{\omega_1}/{\omega_2}]$.
In Eq. (\ref{start}), the photon distributions $f(.)$ are already integrated over the virtuality  ($Q_{1,2}^2$)
of the photons. As this dependence is of the order of $1/Q_{1,2}^2$, this justifies the approximation that both
photons are nearly-real.

Let us mention that there is another way of deriving  Eq. \ref{start} without using explicitly the EPA. We can write brute force the cross section 
for the process (for example producing two leptons in the final state): $p+p \rightarrow pp (\gamma(k_1) \gamma(k_2)) \rightarrow  p+p+ L^-(k_3)L^+(k_4)$. This gives:
$$
\sigma \sim
\int d^2 {\bf b} \int d\Gamma_{(3,4)}   
\left| 
\int d\Gamma_{(1,2)} A_1^\mu (k_1,\vec{b}) T_{\mu \nu}(k_1,k_2,k_3,k_4) A_2^\nu (k_2)
\right|^2.
$$
where the momenta of particles are labeled $(k_1,k_2,k_3,k_4)$ and the phase space integrands are written as $d\Gamma_{(1,2)}$ and
$d\Gamma_{(3,4)}$.
To  simplify the notations, we have included the normalization with flux factor and the conservation law $(2\pi)^4 \delta^4(k_1+k_2-k_3-k_4)$ into the transition
amplitude $T_{\mu \nu}(k_1,k_2,k_3,k_4) $,
computable from first principles, which describes the interaction of the two incoming photons with the produced outgoing particles. 
The functions $A_1^\mu (k_1,\vec{b})$ and $A_2^\nu (k_2)$ are the EM potentials produced by 
the colliding protons, whose velocities are $u_1=\gamma(1,v,0,0)$ and $u_2=\gamma(1,-v,0,0)$, with the condition that the impact parameter of the reaction is $b$.
The amplitude $T_{\mu \nu}(k_1,k_2,k_3,k_4) $ is calculable from QED rules. Then, standard calculations give back an expression of the form
of Eq. \ref{start} under the approximation that the photons are nearly-real. This is why this formula seems to be quite robust and is a good starting point for discussion.

We can remark that for practical issues, the situation may be more complex. Indeed, each proton can either survive and, then, is scattered at  a small angle, as considered above. This is the case of elastic emission. 
Elastic two-photon processes yield very clean event topologies at the LHC: two very forward protons measured away from the interaction point
and a few centrally produced particles (forming the final state $X$).
But, it is also possible that one or both protons  dissociate into a hadronic state. This is the case of inelastic emission. 
Indeed, in case of collisions of ions the inelastic
excitation will contribute also to the photon spectrum. This has been studied for the excitation of
the most important excited state, called the giant dipole resonance (GDR). For the collisions of protons, it
was found that the photon emission from individual quarks gives a contribution that is larger than the elastic one. 
We come back on these effects in the next sections when describing the measurements.

\subsection{Impact parameter dependences}

In posing the problem, an important remark is that basic Eq. (\ref{start}) is incomplete and 
requires significant corrections due to proton absorptive effects. 
Indeed, the proton has a finite size already encoded in the form factor $F(.)$ (in the above expression) but that will also carry
other important deviations to Eq. (\ref{start}). 
These effects are mainly related to proton-proton strong-interaction exchanges that accompany the photon-photon interaction and that lead to the production of additional hadrons in the final state. 
These effects are of the order of $20$ \% in the measurements performed at the LHC and thus cannot be neglected (see next sections). 
This includes both the strong proton-proton absorptive correction and the photon–proton coherence condition (the impact parameter of the reaction needs to be larger than the size of the incoming protons).
It can be shown that these corrections can be accounted for making the replacement
\cite{Dyndal:2014yea,Dyndal:2015hrp,Harland-Lang:2018iur}:
\begin{equation}
f(\omega_1)f(\omega_2) \rightarrow \int \int n({\bf b}_{1},\omega_1) n({\bf b}_{2},\omega_2) 
d^2 {\bf b}_{1} d^2 {\bf b}_{2},
\label{rep}
\end{equation}
where $n({\bf b},\omega)$ are generalized photon flux, which still depend on the energy of the photon but also of the 
distance ${\bf b}$ to the proton trajectory (measured in the plane transverse to this trajectory).
By symmetry, this is clear that $n({\bf b},\omega)$ depends only on the modulus of ${\bf b}$.
Based on the same logic as for the EPA, it reads \cite{Dyndal:2014yea,Dyndal:2015hrp,Harland-Lang:2018iur}:
\begin{equation}
n({b},\omega) = \frac{e^2/(4\pi)}{\pi^2 \omega}
\left |
\int d{k}_\perp {k}_\perp^2
 \frac{F(k_\perp^2+\frac{\omega^2}{\gamma^2})}{k_\perp^2+\frac{\omega^2}{\gamma^2}}
J_1(b k_\perp)
\right |^2 
\label{photond1}
\end{equation}

\begin{figure}[hbtp]
\centering
  \includegraphics[scale=0.5]{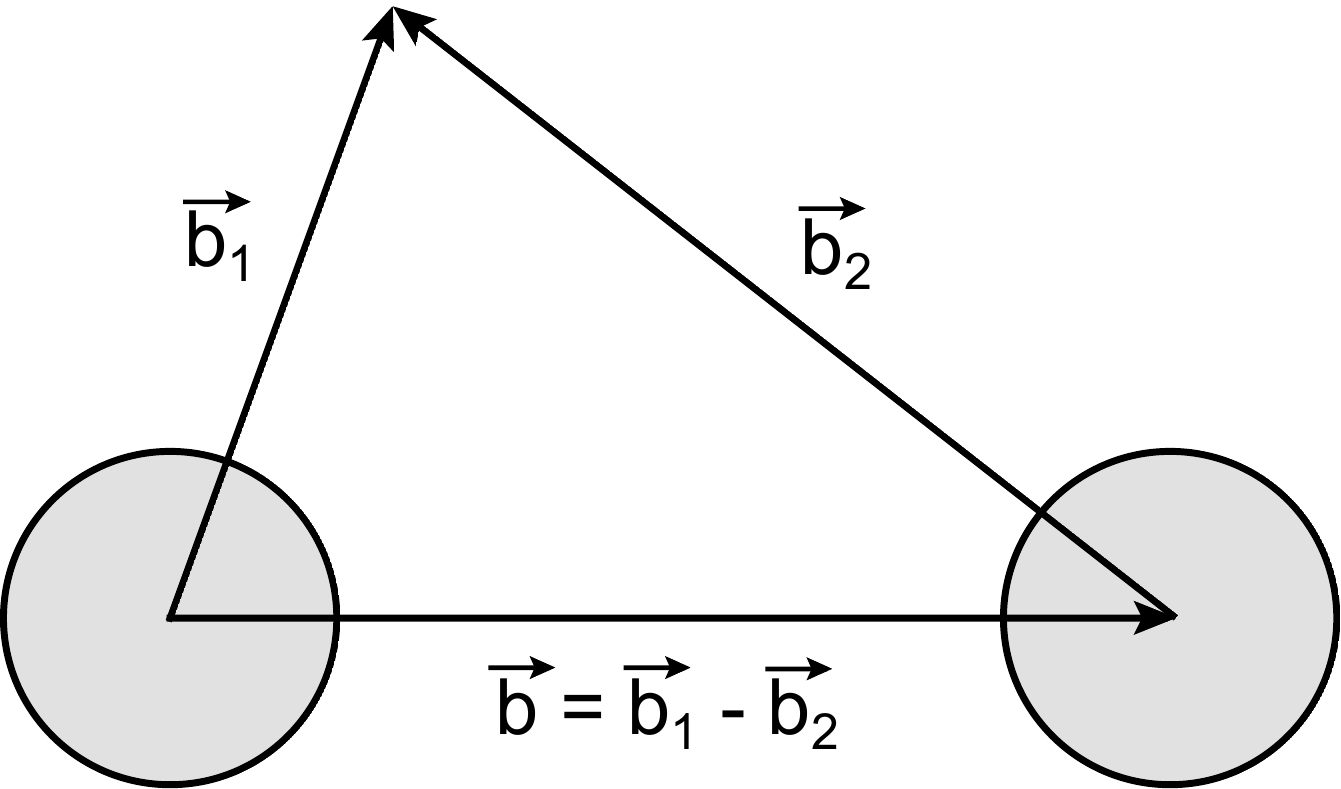}
   \caption[]{
   {\small
   Schematic view of the two protons and the transverse distances ${\bf b}_{1}$ and  ${\bf b}_{2}$.
The difference ${\bf b}={\bf b}_{1}-{\bf b}_{2}$  is also pictured. This is clear from this view that
the geometrical non-overlapping  condition of the two protons corresponds to $|{\bf b}_{1}-{\bf b}_{2}| > 2r_p$.}  } 
  \label{pict}
\end{figure}

 In Eq. (\ref{rep}), there are integrations on the parameters ${\bf b}_{1}$ and  ${\bf b}_{2}$ 
 pictured in Fig. \ref{pict}. This illustrates  that 
 the bounds of integrations on ${\bf b}_{1}$ and  ${\bf b}_{2}$  are linked and that integrations 
 on these parameters
cannot be performed independently.
Indeed, there are important geometrical constraints to  encode:
the two photons  need to interact at the same point outside the two protons, of radii $r_p$ (here the radius projected on the
transverse plane w.r.t. proton trajectory), while the proton-halos do not overlap.
This implies minimally that $b_1 > r_p$, $b_2 > r_p$ and $|{\bf b}_{1}-{\bf b}_{2}| > 2r_p$ 
(see Fig. \ref{pict}).
The last condition clearly breaks the factorization in the variables ${\bf b}_{1}$ and ${\bf b}_{2}$ of
the integral  in Eq. (\ref{rep}).

In the following of the reasoning let us discuss the case of ion-ion (that we label generically as $A-A$) collisions.
It seems very logical that all the arguments mentioned above also hold for ion-ion,
after changing $r_p$ by the ion radius (labeled $r_A$) and 
with two changes in Eq. (\ref{photond1}): (i) the charge factor is not $e^2$ but $Z^2e^2$ and (ii) the form factors are
related to the ion charge and magnetic distributions. 
The point (i) is not so obvious for $A-A$ collisions. Indeed, it assumes 
the coherent action of all the charges (whose number is $Z$) in $A$.  For 
this condition to be correct, the wavelength of the photon needs to be larger than the size of $A$, such that
 it does not resolve the individual nucleons but sees the 
coherent action of them. 
This means that
the coherence condition limits the virtuality $Q^2=-q^2$ of the photon to 
very low values:
$Q^2 \lesssim 1/R_A^2$,
where $R_A$ is the charge radius of $A$. 
For protons, $R_p \simeq 0.7$ fm and for lead, $R_{Pb} \simeq 7$ fm.
We have already mentioned above that the (main) dependence in $1/Q^2$ of functions
$n(.)$, which itself comes from the rapid decrease of the 
nuclear electromagnetic form factor for high $Q^2$ values, justifies this assumption
and the interacting photons can be considered as nearly-real.
In this report, we restrict the discussion to coherent photon exchange only.

At this level, as a consequence, we can set more accurately the bounds on the
energies of interacting photons and also on their transverse momenta.
The four-momentum  of one emitted photon (from the ion or the proton moving at velocity $v$) reads:
$q^\mu=(\omega,\vec q_\perp,q_3=\omega/v)$. With $
Q^2=\frac{\omega^2}{\gamma^2}+q_\perp^2$, we obtain the bounds:
$\omega<\omega_{max} \approx \frac{\gamma}{R_A}$
and 
$q_\perp \lesssim \frac{1}{R_A}$.
At the LHC, this gives $\omega_{max} \simeq 1.8$ TeV ($1.1$ TeV) in proton-proton collisions at a center of mass energy of $13$ TeV ($8$ TeV) and
 $\omega_{max} \simeq 70$ GeV in lead-lead collisions at a center of mass energy per nucleon pair of $5$ TeV.
In particular, this means that all processes are not accessible to lead-lead collisions at $\sqrt{s_{NN}}=5$ TeV. The reaction
$Pb Pb \rightarrow Pb Pb (\gamma \gamma) \rightarrow Pb Pb W^+ W^-$ is forbidden and thus the exclusive production of
 pairs of $W$ bosons from photon-photon interaction is only possible in proton-proton collisions at the LHC, for example at a center of mass energy of $8$ TeV.

\subsection{Getting the general expression of the cross section}

Here, we have all the elements to pose the cross section for the photon-photon reaction in the case of ion-ion collisions: 
$AA \rightarrow AA (\gamma \gamma)  \rightarrow AA X$. This gives:
\be
 \sigma  = 
\int_{b_1, b_2>r_A}
\theta \left(|{\bf b}_1-{\bf b}_2|-2R_A \right)  
n({\bf b}_{1},\omega_1) n({\bf b}_{2},\omega_2) 
\sigma(\gamma\gamma\rightarrow X) 
d^2 {\bf b}_{1} d^2 {\bf b}_{2}
d \omega_1 d\omega_2.
\label{rep_ion}
\ee
From this expression, we deduce the cross section scales as $Z^4$, as stated in the introduction. 
This expression also holds for proton-proton collisions with $A\equiv p$ and $Z=1$.
As a detail, the condition
$\theta \left(|{\bf b}_1-{\bf b}_2|-2R_A \right)$ can be written differently using what is called an opacity function, but this does not make any practical difference and the interpretation is still that it represents a condition for non overlap. Then, this is not difficult to finish the calculations once we
know the functions $n(.)$ for ions, that derive from the EM form factors of the relativistic ions. They are measured using the elastic scattering of electrons from ions. In general $n(.)$ reads:
$$
n ( b, \omega) = \frac{Z^2 \alpha_{em}}{\pi^2} 
\frac{1}{b^2 \omega} 
\left( \int u^2 J_1 \left( u \right) F \left( \sqrt{\frac{ \left(\frac{b \omega}{\gamma} \right)^2 +u^2}{b^2}} \right) 
\frac{1}{ \left( \frac{b \omega}{\gamma} \right)^2 + u^2} du \right)^2,
$$
where $J_1(.)$ is the Bessel function of the first kind. 
It is clear that
there exists simple form of the form factor  that provides simple calculations. For example,
introducing monopole form factor: $ F(q^2) = \frac{\Lambda^2}{\Lambda^2 + q^2}$
with $\Lambda = \sqrt{\frac{6}{<r^2>}}$, the expression of $n(.)$ becomes:
$$
n (b, \omega) = \frac{Z^2 \alpha_{em}}{\pi^2} \frac{1}{\omega} 
 \left(  \frac{\omega}{\gamma} 
 K_1 \left(  \frac{b \omega}{\gamma} \right) - 
 \sqrt{\frac{\omega^2}{\gamma^2} + \Lambda^2} \;
 K_1 \left(  b \sqrt{\frac{\omega^2}{\gamma^2}+\Lambda^2} \right) \right)^2 \; ,
$$
where $K_1(.)$ is the modified Bessel function of the second kind. The interest of this expression is that it can be easily encoded in 
computer computations. In a second step, this is not complicated to consider a more realistic form for the form factor $F(.)$ using the charge distribution $\rho(r)$
in the ion. For example, a quite general function for $\rho(r)$ can be defined as:
$\rho(r) = \frac{\rho_0}{1+\exp((r-r_0)/a)}$.
Where $r_0$ is the characteristic size of the ion and a characteristic length scale.
Then, the form factor:
$F(q^2)=\int_0^\infty \frac{4\pi}{q} \rho(r) \sin(qr) dr$ can be directly integrated  and encoded in the photon flux above.

Of course, all the above discussion is useful if 
the elementary cross section $\sigma(\gamma\gamma\rightarrow X)$, where $X$ is a pair of leptons or photons, is computable in QED.
This is a reasonable question as the natural expansion parameter $Z \alpha_{em}$ is not a small parameter for ions.
However, what matters also is the fact that the Lorentz factor is very large at LHC energies and  then $1/\ln(\gamma_L^2)$ becomes a small parameter
 \cite{Harland-Lang:2018iur}. Finally, it has been shown that calculations can be done in the framework of perturbation theory  \cite{Harland-Lang:2018iur}  and
it follows that Eq. (\ref{rep_ion}) can be used to compute the observed process at the LHC
in ion-ion collisions, as well as in proton-proton or proton-ion collisions.
These photon equivalent distributions including impact parameter dependences are included in several Monte-Carlo (MC) simulations
used at the LHC. For ion-ion an proton-proton collisions,  all UPC physics processes can be simulated using
{\sc Superchic} \cite{Harland-Lang:2018iur},
{\sc Starlight} \cite{Klein:2016yzr,Auerbach:1983hld} and
{\sc Fpmc} \cite{Boonekamp:2011ky,Baldenegro:2019whq}, while 
{\sc Lpair} \cite{lpair} is restricted to the proton-proton case.
For all these MC simulations, several possibilities for the form factors of the ions or protons can be used.

\begin{figure}[]
\centering
\includegraphics[width=.4\textwidth]{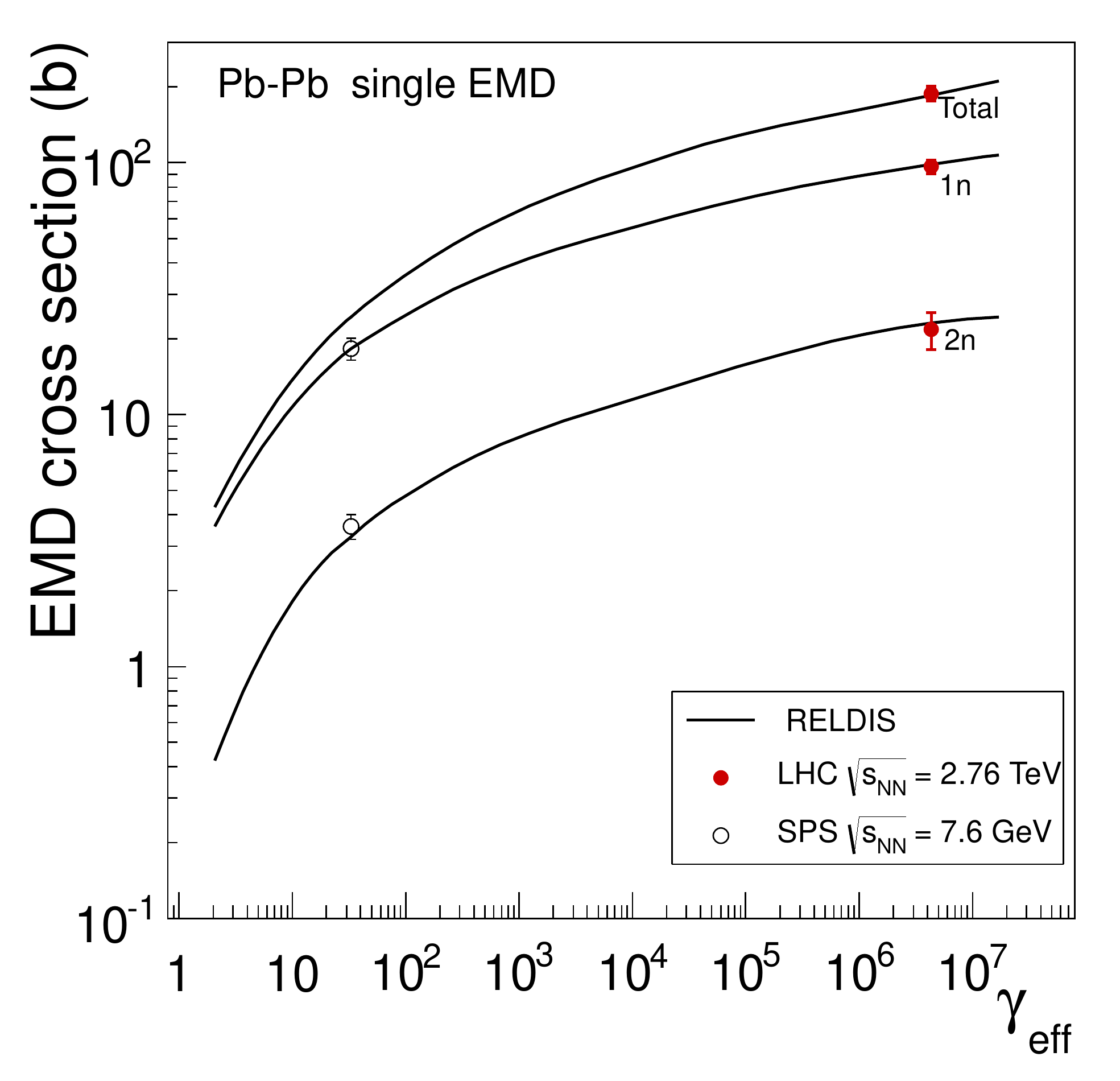}
\caption{\label{fig:ALICE_EDM_ZDC} Total single EMD cross sections and partial EMD cross sections for emission of one and two neutrons as a function of the effective Lorentz factor $\gamma_\mathrm{eff}$. The RELDIS predictions are represented by the solid lines.}
\end{figure}

Let us note also that the nuclear break-up probability can easily be incorporated into Eq. (\ref{rep_ion}) through a multiplicative factor in the integrand
for one or more neutrons emitted by the colliding lead ion.
This probability can be largely understood using the study of Giant Dipole Resonance (GDR), for example in the ALICE experiment.
{The first measurement of neutron emission in electromagnetic dissociation of lead ions at the
LHC was presented by the ALICE experiment in $\sqrt{s}_\mathrm{NN} = 2.76$ TeV. The measurement is performed using the neutron zero degree calorimeters (ZDC) of the ALICE experiment. The ZDC is used to detect neutral particles close to beam rapidity. At low neutron multiplicities, one is able to reconstruct a direct signature of electromagnetic dissociation (EMD) in UPC heavy-ion collisions. Cross section measurements of EMD in heavy-ion collisions provide inputs for Monte-Carlo generators that include these effects. The measured cross sections of single and mutual EMD of lead ions with neutron emission are $\sigma_\mathrm{single-EMD} =$ 187.4 $\pm$ 0.2 $^{+13.2}_{-11.2}$ b.}
{The experimental results can then be compared to theoretical predictions of the Relativistic ELectromagnetic DISsociation (RELDIS) model. RELDIS is able to describe electromagnetic interactions between ultra-relativistic ions including single and double virtual photon absorption, excitation of giant resonances, intra-nuclear cascades of produced hadrons and statistical decay of excited residual ions. The RELDIS model predictions are found to be in very good agreement with ALICE results, as shown in Fig.~\ref{fig:ALICE_EDM_ZDC} for the EMD cross section as function of the effective Lorentz boost factor.}

\section{On the validity of this formalism at the LHC}

The purpose of the present section is to prove experimentally that this formalism is reasonable in the LHC era. This means that
it allows a correct description of the data using the formalism encoded in different MC mentioned above
\cite{Harland-Lang:2018iur,Klein:2016yzr,Auerbach:1983hld,Boonekamp:2011ky,Baldenegro:2019whq,lpair},
depending on the precise process
under investigation.
We start the discussion by the heaviest final states, the exclusive production of pairs of $W$ bosons, then
pairs of leptons. At the end of this part, this will bring us to the physics case we want to emphasize in this report,
namely the light-by-light scattering.

\subsection{Exclusive production of pairs of $W$ bosons at the LHC}

\subsubsection{Experimental observation}

The exclusive production of pairs of $W$ bosons,
$pp \rightarrow pp (\gamma \gamma) \rightarrow pp W^+ W^-$ have been  observed at the LHC
 \cite{Chatrchyan:2013akv,Khachatryan:2016mud,Aaboud:2016dkv}.
 Below we describe the measurements and results by the ATLAS and CMS experiments  \cite{Chatrchyan:2013akv,Khachatryan:2016mud,Aaboud:2016dkv},
 performed using $20.2$ fb$^{-1}$ of proton-proton collisions at
$8$ TeV recorded during 2012. 
Due due a  kinematic requirement that follows from the formalism described above, this process can only been observed in proton-proton collisions
and is forbidden in lead-lead.
 The initial idea for the analysis is to
 consider the opposite-charge different-flavor 
 $\mu^{\pm}e^{\mp}$ final states.
This garanties that the exclusive production of $W^+W^-$  has a clean signature, since no extra activity
is produced in addition to the two (identified) leptons. Also, it is a key process to probe the$ WW \gamma\gamma$ quartic gauge
couplings (next subsection). 
Without tagging the initial state protons combined with having neutrinos in the final states, it is not
possible to separate the elastic from the dissociative events, where a proton is outgoing in an excited state of low mass (labeled as
$p^*$).
As already discussed in another physics case in a previous section, we 
identify single dissociative events (SD): $pp \rightarrow pp (\gamma \gamma) \rightarrow p^*p W^+ W^-$
and double dissociative events (DD): $pp \rightarrow pp (\gamma \gamma) \rightarrow p^*p^* W^+ W^-$.
Therefore, the elastic, SD and DD 
processes are together considered as signal. We cannot discriminate one with respect to the other.
In a second step,
the dissociative contribution is estimated from data, using
an exclusive  sample of pairs of muons with
an invariant mass above $160$ GeV.
The data-driven scale
factor $f_\gamma$ is given by the ratio of the number of data after subtracting off the number
of background events to the number of elastic events
predicted by a purely elastic process simulation. So the total expected signal for analysis is the prediction for the elastic $W^+W^-$  production
scaled by $f_\gamma$:
$$
f_\gamma = \frac{Elastic+SD+DD}{Elastic(MC)} = \frac{N_{observed}-N_{background}}{N_{elastic}}.
$$

The events selected for analysis satisfy any of the following triggers: (i) a single-muon trigger with
$p_{\rm T}^\mu>24$ GeV, (ii) a single-electron trigger with
$p_{\rm T}^e>24$ GeV or (iii) an electron-muon trigger with
$p_{\rm T}^e>12$ GeV and $p_{\rm T}^\mu>8$ GeV.
The pileup conditions in 2012 are quite high. Therefore, a new track-based
strategy for selecting exclusive events has been developed for this analysis. The first analysis requirement is to select two
leptons with
$p_{\rm T} \equiv p_{\rm T}^{e\mu}>30$ GeV and invariant masses above $20$ GeV.
Then, the event vertex is  reconstructed as the average of
the longitudinal impact parameters ($z_0$) of the leptons.
Also, 
in order to suppress most of the inclusive background,
the two leptons are required
to be within $1$ mm in $z$ (along the beam line) of each other. At the next step, the longitudinal distance between the lepton vertex
and other extra tracks ($\Delta z_0$) is computed. Finally, the exclusivity selection is defined as the condition that only events with zero
extra track within a $1$ mm longitudinal distance from the lepton vertex (also written as $\Delta z_0^{iso}=1$ mm) .

Finally, 23 candidates are observed in data,
while $9.3$ signal (elastic+SD+DD types of events -see above-) and $8.3$ background events are expected.
The observed significance over the background only
hypothesis is $3 \sigma$, constituting an evidence for exclusive
production of pairs of $W$ bosons in proton-proton collisions.

\begin{figure}[!]
\centering
\hspace*{-1cm}
\includegraphics[scale=0.6]{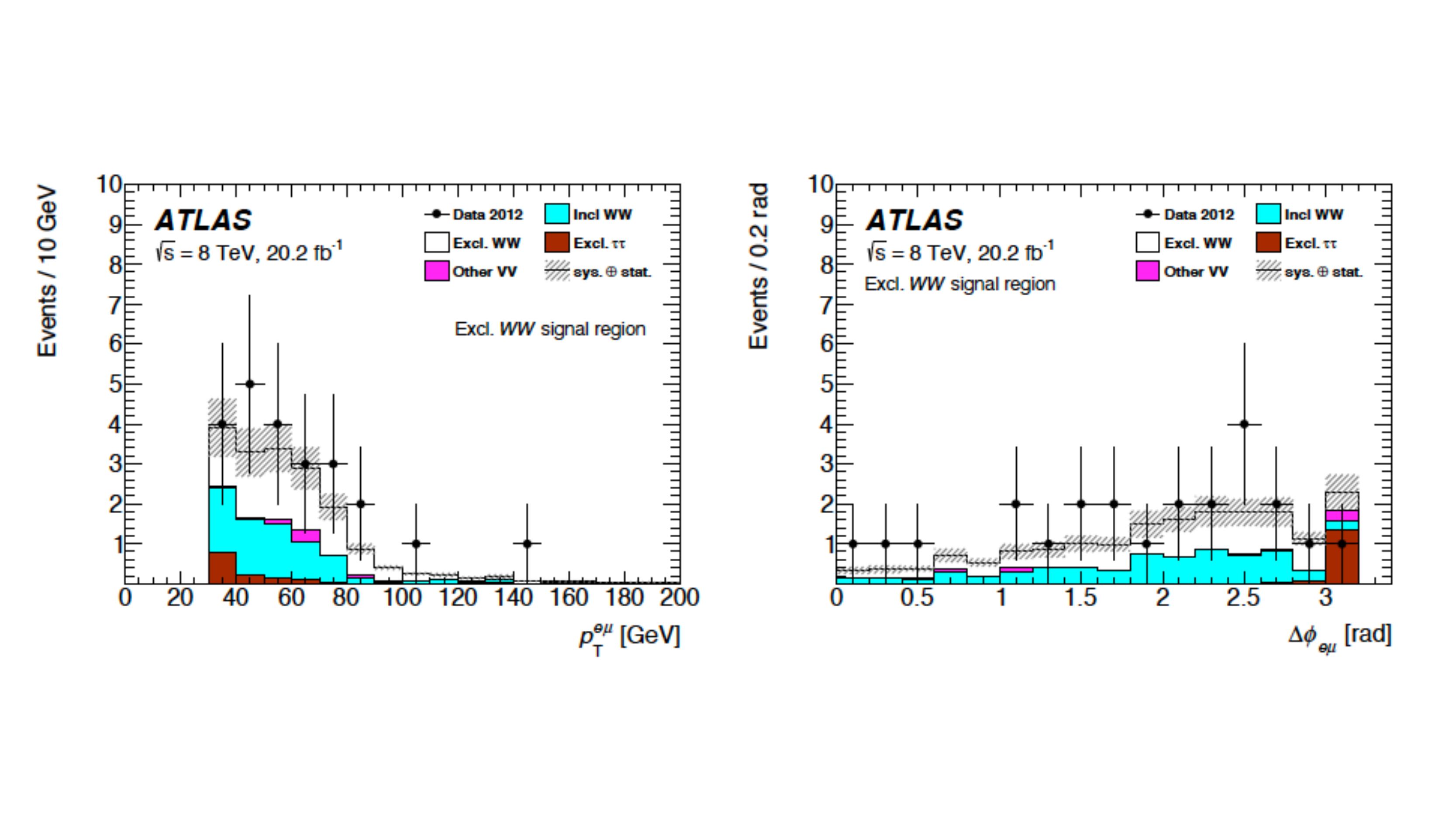}
\vspace*{-2.7cm}
  \caption{ 
  {\small
  Kinematic distributions in the exclusive $W^+W^-$ signal region 
  comparing the simulations to data from ATLAS \cite{Aaboud:2016dkv}.
  The exclusive $W^+W^-$ signal is pictured in withe. }}
  \label{ww}
\end{figure}

\begin{figure}[!]
\centering
\hspace*{-1cm}
\includegraphics[scale=0.55]{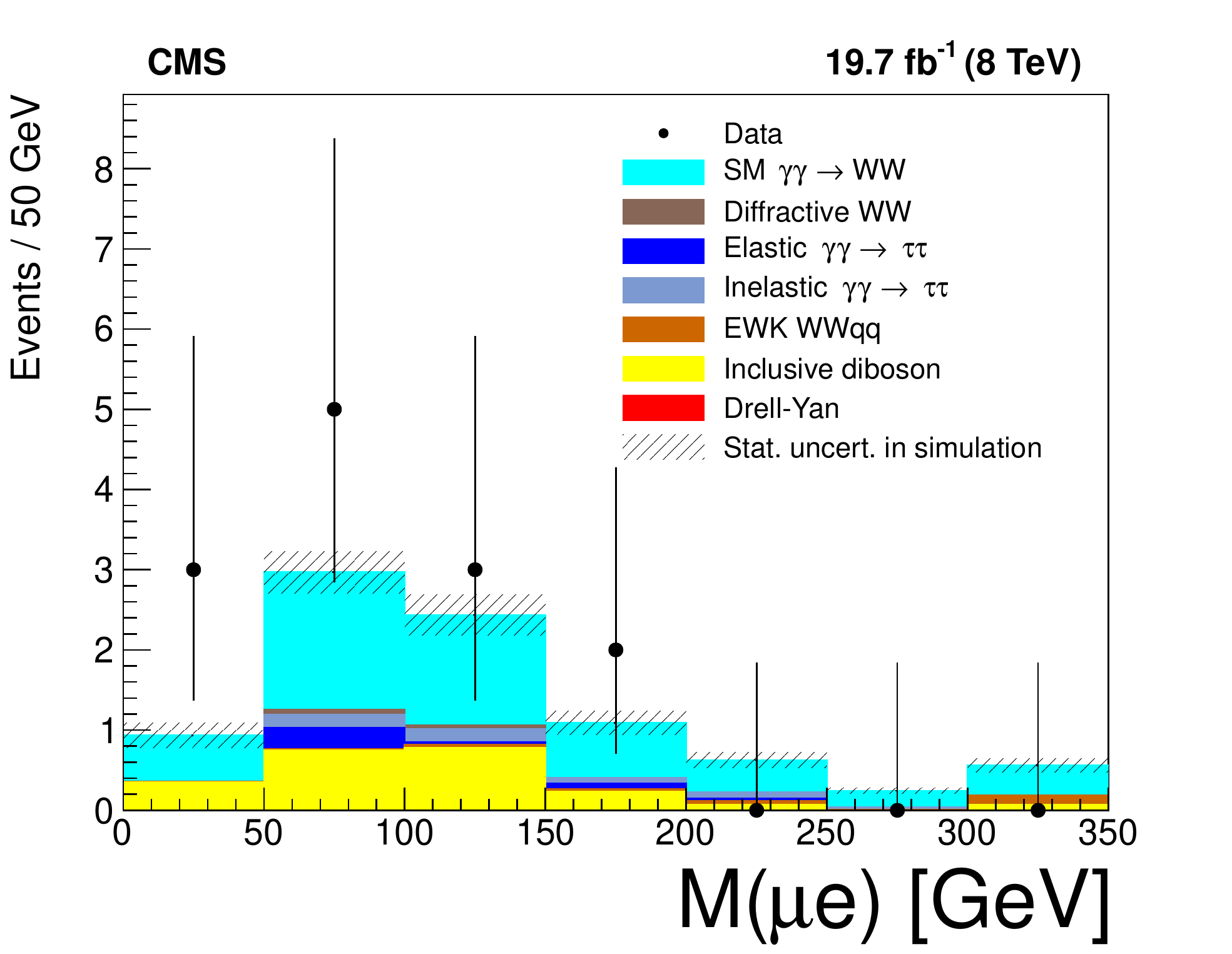}
\vspace*{0cm}
  \caption{ 
  {\small
  Kinematic distributions in the exclusive $W^+W^-$ signal region 
  comparing the simulations to data from CMS \cite{Chatrchyan:2013akv}. The exclusive $W^+W^-$ signal is pictured in light blue.}}
  \label{wwcms}
\end{figure}

Fig. \ref{ww} shows the distributions in the signal region 
($pp \rightarrow pp (\gamma \gamma) \rightarrow pp W^+ W^-$+SD+DD) of the transverse momentum of the electron-muon pair $p_{\rm T}^{e\mu}$ and 
its azimuthal separation $\Delta \phi^{e\mu}$. 
The simulation describes correctly the data.
From this measurement, the cross section for $\gamma \gamma \rightarrow W^+W^-\rightarrow \mu^{\pm}e^{\mp} + X$ 
(for proton-proton collisions at $8$ TeV) can be determined in the signal region
(for the fiducial phase space measured)
and then extrapolated to the full $W^+W^- \rightarrow \mu^{\pm}e^{\mp} + X$ phase space.
Then,
the cross-section times branching ratio is found to be:
$$
 \sigma^{pp}(\gamma \gamma \rightarrow W^+W^-\rightarrow \mu^{\pm}e^{\mp} + X)(8 \ {\rm TeV}) = 6.9 \pm 2.2 \pm 1.4 \ \ {\rm fb}.
$$
For the same energy in the center of mass, the CMS experiment is obtaining a very comparable result with a cross section of
$10.8 \pm 5.1 \pm 4.1$ fb
\cite{Chatrchyan:2013akv,Khachatryan:2016mud}. Kinematical distributions in the signal region 
($pp \rightarrow pp (\gamma \gamma) \rightarrow pp W^+ W^-$+SD+DD) of the invariant mass of of the electron-muon pair $M(e\mu)$ and 
its acoplanarity $1-\vert \Delta \phi(e\mu)\vert/\pi$ are presented in Fig. \ref{wwcms}.
We observe that the descriptions of the data (ATLAS and CMS) by the simulations, exclusive signal plus backgrounds, is quite good within the
experimental and theoretical uncertainty.

\subsubsection{Anomalous quartic couplings}

In the measurement described above, the contribution of the elastic part of the signal region comes from the 
reaction $\gamma \gamma \rightarrow W^+W^-$ which is possible in the SM through its Lagrangian density:
\begin{equation}
L(\gamma\gamma WW) = -e^{2} \left( W^{+}_{\mu}W^{-\mu}A_{\nu}A^{\nu} - W^{+}_{\mu}W^{-}_{\nu}A^{\mu}A^{\nu} \right).
\end{equation}
Where $A^{\mu}$ is the photon field and $W^{\mu}$ is the $W$ boson field. 
This is basically what we have pictured as the exclusive signals
(Fig. \ref{ww} and \ref{wwcms}) once all experimental requirements and efficiencies are taken into account.
At this point,
this is possible to push the discussion a bit forward in exploring potential non standard physics that could be hidden in
the measurement. Indeed,
this is  well established that these measurements are good candidates to search for anomalous gauge couplings (aQGC), which would reveal a sign of new physics. From the conclusion of the experimental observation, we know in advance that we will not conclude to any sign of deviation. However, the exercise is interesting to derive upper limits on the aQGC coupling parameters.

The most general Lagrangian densities leading to aQGC
$\gamma\gamma WW$ (and $\gamma\gamma ZZ$ )  can be written as:
\begin{equation}
\begin{aligned}
L^{0}_{6} &= \frac{e^{2}}{8}\frac{a^{W}_{0}}{\Lambda^{2}}F_{\mu\nu}F^{\mu\nu}W^{+\alpha}W^{-}_{\alpha}-\frac{e^{2}}{16\cos^{2}\Theta_{W}}\frac{a^{Z}_{0}}{\Lambda^{2}}F_{\mu\nu}F^{\mu\nu}Z^{\alpha}Z_{\alpha} \\
L^{C}_{6} &= \frac{-e^{2}}{16}\frac{a^{W}_{C}}{\Lambda^{2}}F_{\mu\alpha}F^{\mu\beta}(W^{+\alpha}W^{-}_{\beta}+W^{-\alpha}W^{+}_{\beta})-\frac{e^{2}}{16\cos^{2}\Theta_{W}}\frac{a^{Z}_{C}}{\Lambda^{2}}F_{\mu\alpha}F^{\mu\beta}Z^{\alpha}Z_{\beta}.
\end{aligned}
\end{equation}
Where $Z^{\alpha}$ stands for the $Z$ boson field. $\Lambda$ is a new scale which is introduced such that the Lagrangian density has the correct dimension and can be interpreted as a scale of new physics. Finally, $a_0$, $a_C$ represents the  new coupling constants.
Also,
the $WW$ and $ZZ$ photon-photon cross sections rise quickly at high energies when 
any of the anomalous parameters are taken to be non-zero. Therefore, the cross section rise has to be 
regulated by a form factor which vanishes in the high energy limit. A possibility is then to
modify the coupling parameters by form factors 
that have the desired behavior, i.e. they modify the coupling at small 
energies only slightly but suppress it  when the center-of-mass energy $W_{\gamma\gamma}$ 
increases. Here, we consider a functional form as
$ a\rightarrow \frac{a}{(1+W^2_{\gamma\gamma}/\Lambda^2)^n}$, with $n=$ and $\Lambda$ given below in the figures.

\begin{figure}[!]
\centering
\includegraphics[scale=0.5]{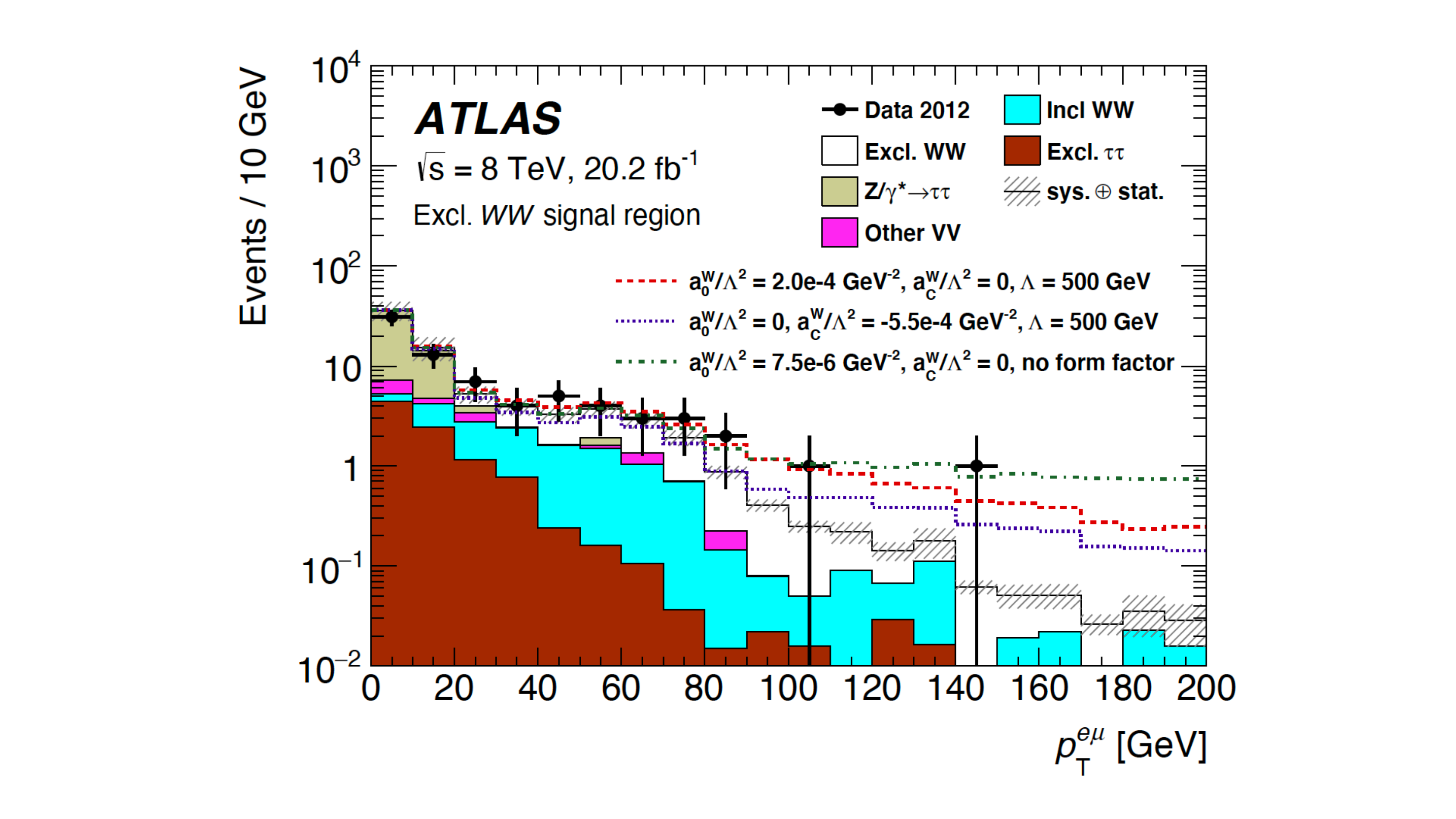}
\vspace*{-0.7cm}
  \caption{ 
  {\small The $p_{\rm T}^{e\mu}$ distribution for data \cite{Aaboud:2016dkv} compared to the standard model prediction for events satisfying all the exclusive
  $W^+W^-$ selection requirements apart from the one on $p_{\rm T}^{e\mu}$ itself.
  Also shown are the various predictions for aQGC coupling parameter values.
   }}
  \label{aqgc1}
\end{figure}

\begin{figure}[!]
\centering
\includegraphics[scale=0.5]{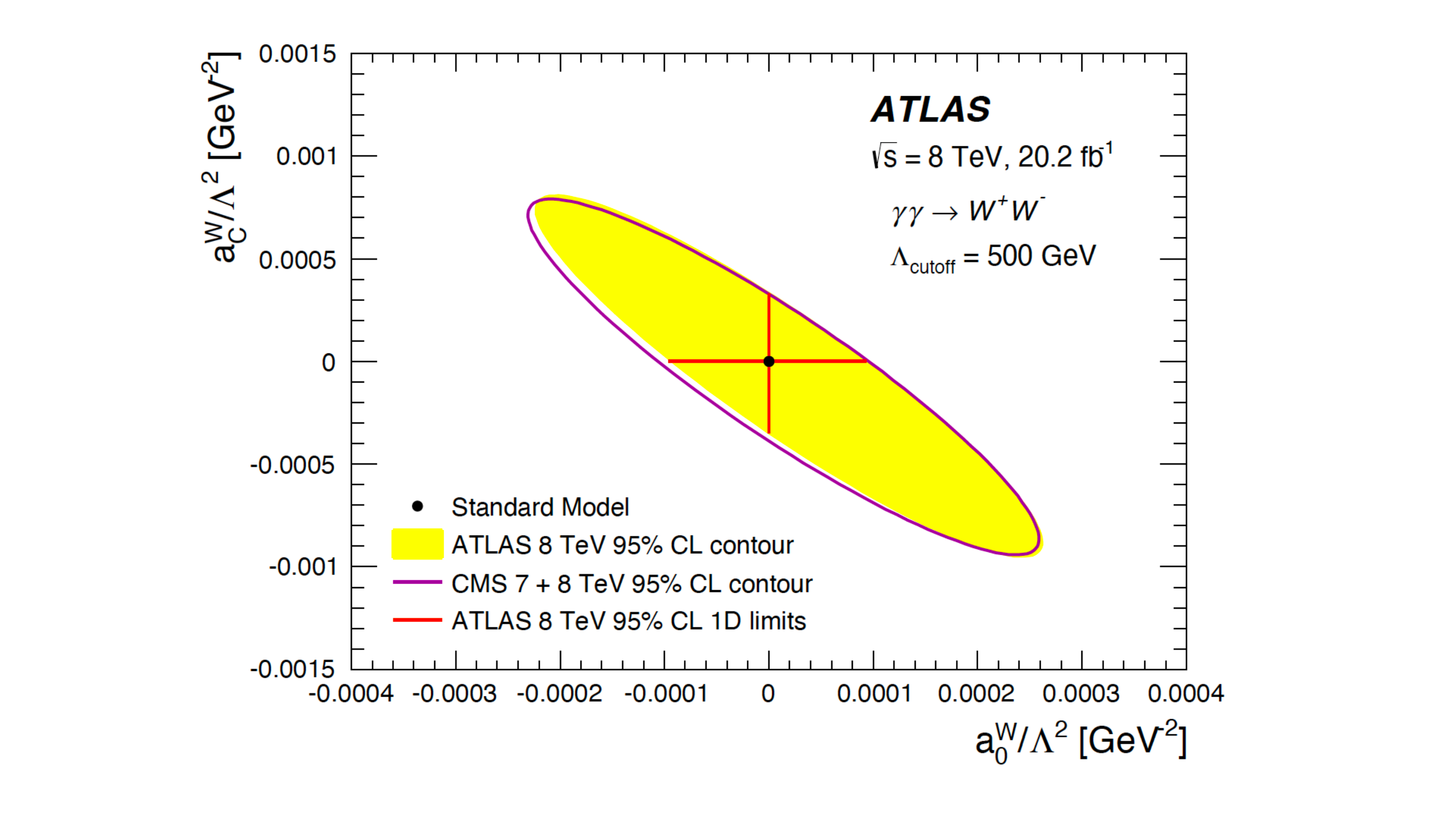}
\vspace*{-1.cm}
  \caption{ 
  {\small The observed  95 \% confidence-level contour and 1D limits   on the couplings
 $\frac{a^{W}_{C}}{\Lambda^{2}}$ and $\frac{a^{W}_{0}}{\Lambda^{2}}$ (results from ATLAS and CMS experiments are shown). }}
  \label{aqgc2}
\end{figure}

The experimental requirements needed to select quartic anomalous gauge coupling $WW$ events are similar as the
ones we mentioned in the previous section, in the domain $p_{\rm T}^{e\mu}>120$ GeV,
where the aQGC contributions
are expected to be important and standard model backgrounds are suppressed \cite{Aaboud:2016dkv}.
The $p_{\rm T}^{e\mu}$ distributions is shown in Fig. \ref{aqgc1}
for data compared to the standard theoretical predictions together with various aQGC scenarios.
We observe that
aQGC predictions enhance the exclusive signal at high $p_{\rm T}^{e\mu}$, while the background is negligible when $p_{\rm T}^{e\mu}>80$ GeV.
The 95 \% Confidence Level limits on the couplings $\frac{a^{W}_{C}}{\Lambda^{2}}$ and $\frac{a^{W}_{0}}{\Lambda^{2}}$  
can then be extracted with a likelihood test using the one
observed data event as a constraint \cite{Aaboud:2016dkv}, see Fig. \ref{aqgc2} where ATLAS and CMS results are shown.  Essentially, no anomalous effect is visible with existing data as foreseen, but the limits themselves are an interesting basis for these kinds of non standard predictions.

\subsection{Exclusive production of pairs of muons at the LHC}

\subsubsection{Proton-proton collisions}

It is simpler to start with proton-proton collisions and then to continue the discussion with ion-ion as it was done in the description of the formalism. The CMS and ATLAS experiments have measured the
exclusive  production of pairs of muons in ultra-peripheral collisions: $pp \rightarrow pp (\gamma \gamma) \rightarrow pp \mu^+ \mu^-$
\cite{Chatrchyan:2011ci,Chatrchyan:2012tv,Aad:2015bwa,Aaboud:2017oiq}.
Below, we present results for collisions at a center of mass energy of $7$ TeV recorded in 2010  by the CMS experiment for an integrated luminosity 
of $40$ pb$^{-1}$ \cite{Chatrchyan:2011ci}.

\begin{figure}[!]   
\centering
\includegraphics[width=0.8\textwidth]{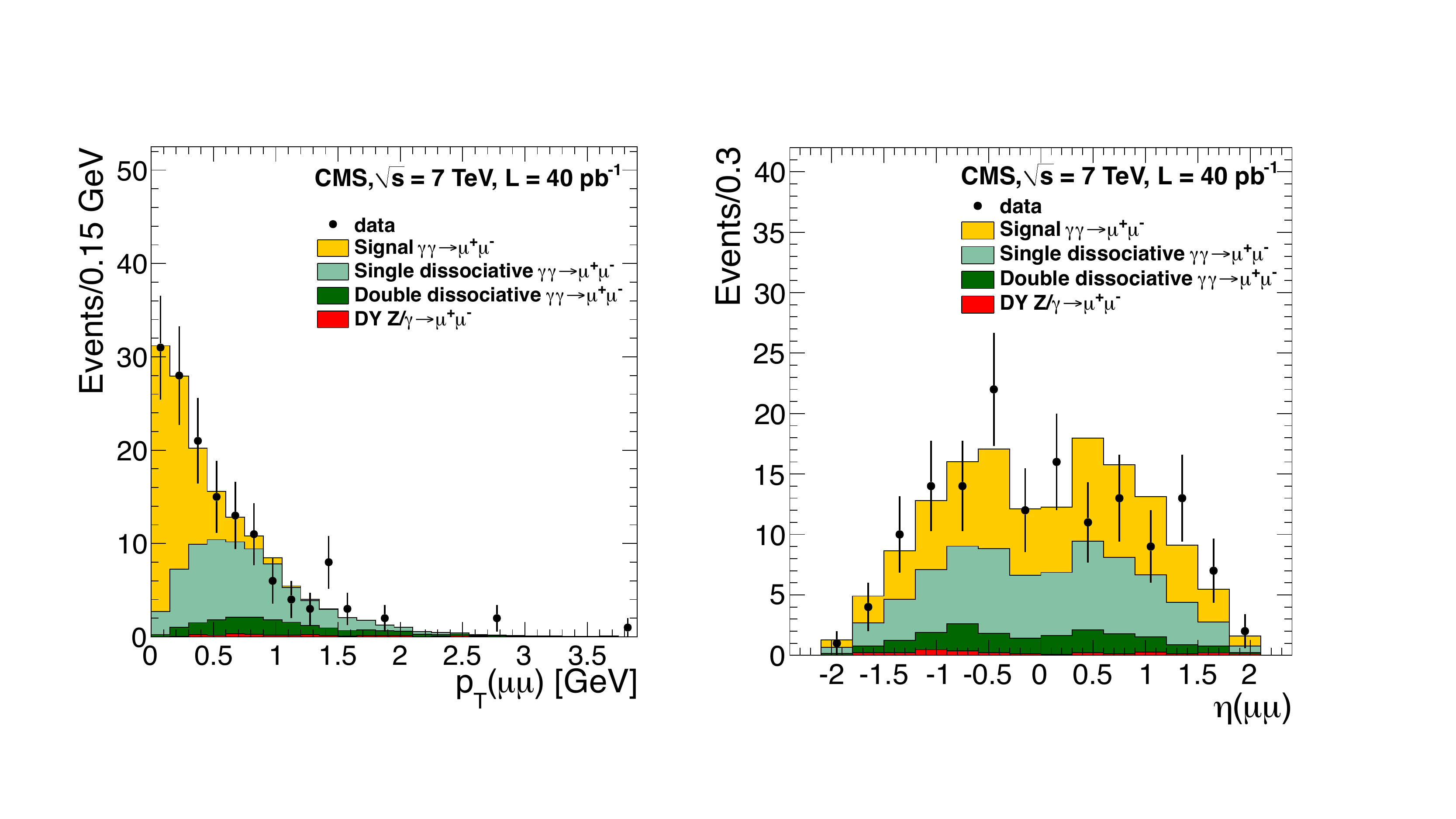}
\vspace*{-1cm}
\caption{\label{dimuons_pp_cms}
{\small
(Left) Distribution of the transverse momentum of the pair of muons for the selected sample (see text). Data are shown as points with statistical error bars. The histograms represent the simulated signal (yellow) {\sc Lpair} \cite{lpair}, single (light green) and double (dark green) proton dissociative backgrounds, and Drell-Yan (red). The yields are determined from a fit using the distributions from simulation.
(Right) Distribution for the pseudo-rapidity of the pair of muons (similar conventions).}
}
\end{figure}

\begin{figure}[!]   
\centering
\includegraphics[width=0.9\textwidth]{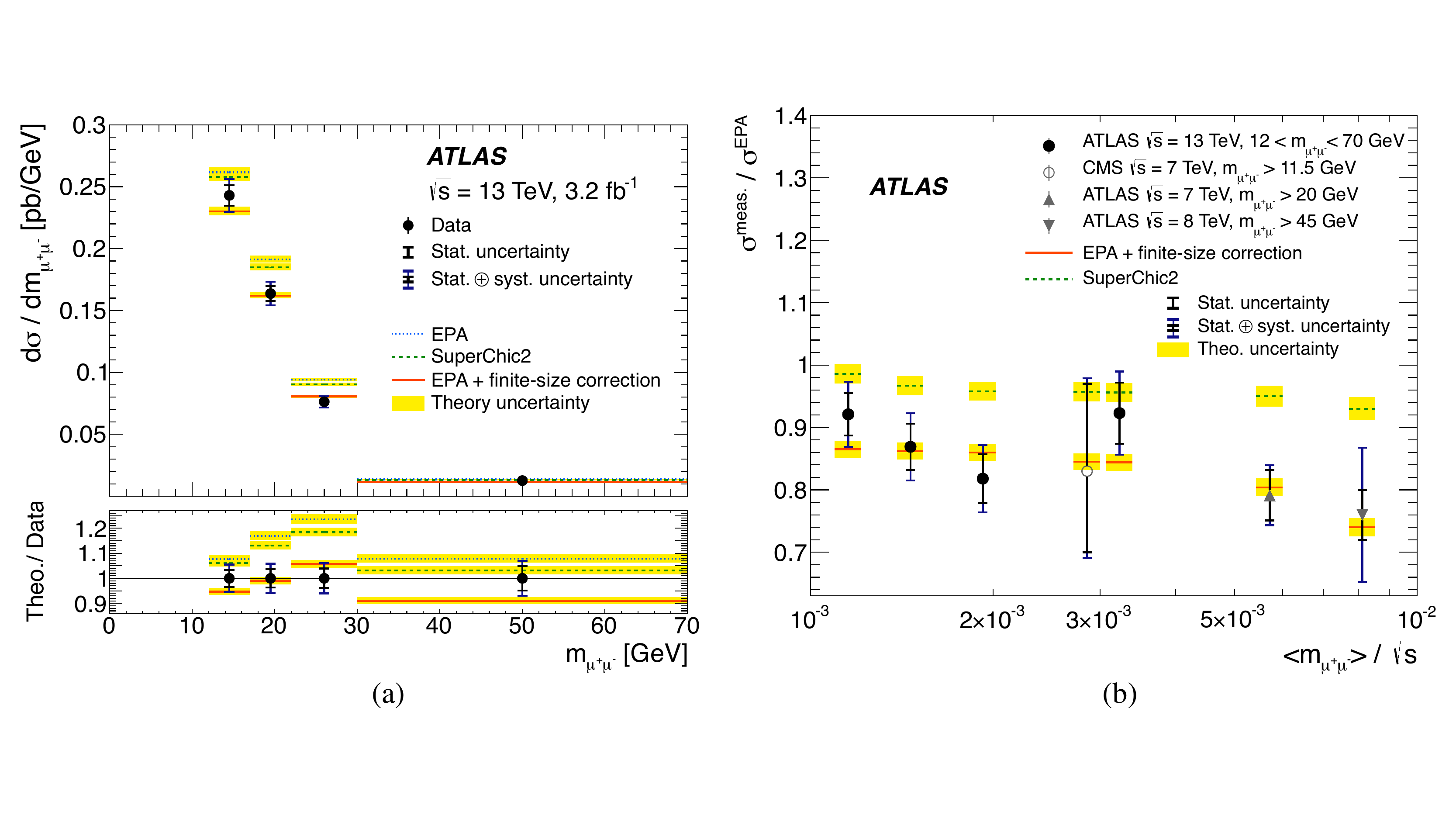}
\vspace*{-1cm}
\caption{\label{dimuonsglobal}
{\small
(a) The exclusive $\gamma \gamma \rightarrow \mu^+ \mu^-$ differential fiducial cross-section measurements as a function of  invariant mass 
of the pairs of muons. (b) Comparison of the ratios of measured and predicted cross-sections to the (bare) EPA calculations as a function of the average  invariant mass of the pairs of muons scaled to the proton-proton centre-of-mass energy used. Data from ATLAS and CMS experiments (markers) are compared to various predictions (lines). 
The inner error bars represent the statistical uncertainties, and the outer bars represent the total uncertainty in each measurement. The yellow bands represent the theoretical uncertainty in the predictions. The bottom panel in (a) shows the ratio of the predictions to the data.}
}
\end{figure}

An important initial task is to record correctly the events of interest. 
They are selected by triggers requiring the presence of two muons with a minimum $p_{\rm T}$ of $3$ GeV 
with some vetoes to avoid a too large fraction of dissociative events.
Then, at the analysis level, by construction,
the  signal is characterized by the presence of two muons, no additional tracks, and no activity above the noise threshold in the calorimeters. 
Of course, the presence of additional interactions in the same bunch crossing would spoil this signature by producing additional tracks and energy deposits in the calorimeters. This is why the selection requires a valid vertex, reconstructed  with exactly two muons and no other associated tracks,
and this vertex is required to be separated by more than $2$ mm from any additional tracks in the event. This value depends  of the 
distribution of additional vertices and thus depends on the running conditions.
Then, in order to enhance the signal efficiency, some kinematic requirements are also imposed:
the transverse momenta
$p_{\rm T}$ of the identified muons is required to be greater than $4$ GeV, with a restriction in pseudo-rapidities as $\vert \eta \vert < 2.1$, 
the invariant mass of the  pair of muons is required to be greater than $11.5$ GeV, the acoplanarity ${\cal A}=1-\Delta \phi/\pi$
is taken to be smaller than $0.1$ (following the same arguments that in the lead-lead case) and, finally, the scalar
difference in the $p_{\rm T}$ of the two muons, $\vert \Delta p_{\rm T} \vert$, is required to be
smaller  $1$ GeV.
The results are presented in Fig. \ref{dimuons_pp_cms}.
The {\sc Lpair} MC \cite{lpair}, including absorptive corrections,  is used to produce the simulated samples shown in Fig. \ref{dimuons_pp_cms}. 
We can observe that the dominating contribution comes from the exclusive production of pairs of muons:
$pp \rightarrow pp (\gamma \gamma) \rightarrow pp \mu^+ \mu^-$, as discussed in the previous section.
However, there is also a non negligible fraction of single dissociative events, namely when one proton is producing 
a low mass excited state, $p^{(*)}$, after the collision: $p p \rightarrow p p^{(*)} \mu^+ \mu^-$. 
For such processes (single proton dissociation), the cross
section calculations depend on the proton structure functions and how the proton is fragmented
into a particular low mass excited state \cite{Chatrchyan:2011ci}. On the principle, this follows the ideas exposed in the section on the formalism, this time with an inelastic photon distribution. We note also  in Fig. \ref{dimuons_pp_cms} that a small contribution comes from  double dissociative events:
$p p \rightarrow p^{(*)} p^{(*)} \mu^+ \mu^-$. All this taken into account, the data are correctly described by the simulations.
In order to illustrate more clearly the effect of the absorptive corrections when compared to a (bare) simulation without such corrections, we present a compilation of
results from the ATLAS and CMS experiments in Fig. \ref{dimuonsglobal}.
In particular, Fig. \ref{dimuonsglobal} (a) illustrates cross section measurements of exclusive production of pairs of muons
using data  in proton-proton collisions collected at a centre-of-mass energy of $13$ TeV during 2015,
for an integrated luminosity of $3.2$ fb$^{-1}$ \cite{Aaboud:2017oiq} . Then, Fig. \ref{dimuonsglobal} (b) is grouping all experimental results obtained at the LHC
\cite{Chatrchyan:2011ci,Chatrchyan:2012tv,Aad:2015bwa,Aaboud:2017oiq}  as a ratio of the measured cross section to the (bare) EPA prediction.
We observe  the effect of order $20$ \% of the absorptive corrections (also called finite-size correction), already mentioned in the formalism part of this report.

\subsubsection{Lead-lead collisions}

We have already mentioned in the introduction that first results  in 
$Au-Au$ collisions and exclusive production of electron-positron pairs have been obtained at RHIC \cite{Adams:2004rz,Afanasiev:2009hy}.
Below, we discuss only the recent high statistics ATLAS analysis. Indeed,
the ATLAS experiment has measured  cross sections for exclusive  production of pairs of muons in ultra-peripheral lead-lead collisions 
at a center of mass energy of $5.02$ TeV per nucleon pair,
for  invariant masses of the pairs of muons larger than $10$ GeV:
$Pb Pb \rightarrow Pb Pb (\gamma \gamma) \rightarrow Pb Pb \mu^+ \mu^-$ \cite{atlaspbpb}. 
These data have been taken in 2015 using an integrated luminosity of $515$ $\mu$b$^{-1}$.

The first essential point in this kind of analysis is to record the needed events during the data taking, what we also call trigger the events. In this sense,
events have been recorded using a trigger specially designed to detect events with a muon and little additional activity in the detector. The primary requirement is for the presence of a region of interest associated with a track in the muon spectrometer, with no lower bound on $p_{\rm T}$  in order to be sensitive to relatively low
transverse momentum muons. 
In order to keep the statistics acceptable on tape,
the events are rejected at this level if the calorimeter system contained $50$ GeV or more transverse energy, integrated over the full calorimeter acceptance. Also, events are rejected if more than one hit was registered in either of the inner rings of two 
MBTS arrays,  scintillator slats positioned between the inner detector and the calorimeter. 
This avoids to keep too many events where the ions dissociate, while the search is for photon-photon interactions for which the ions stay mainly intact.
In addition, in order to ensure the presence of a muon, the event  has to contain at least one track reconstructed  with $p_{\rm T} > 400$ MeV. 

\begin{figure}[!]   
\centering
\includegraphics[width=0.9\textwidth]{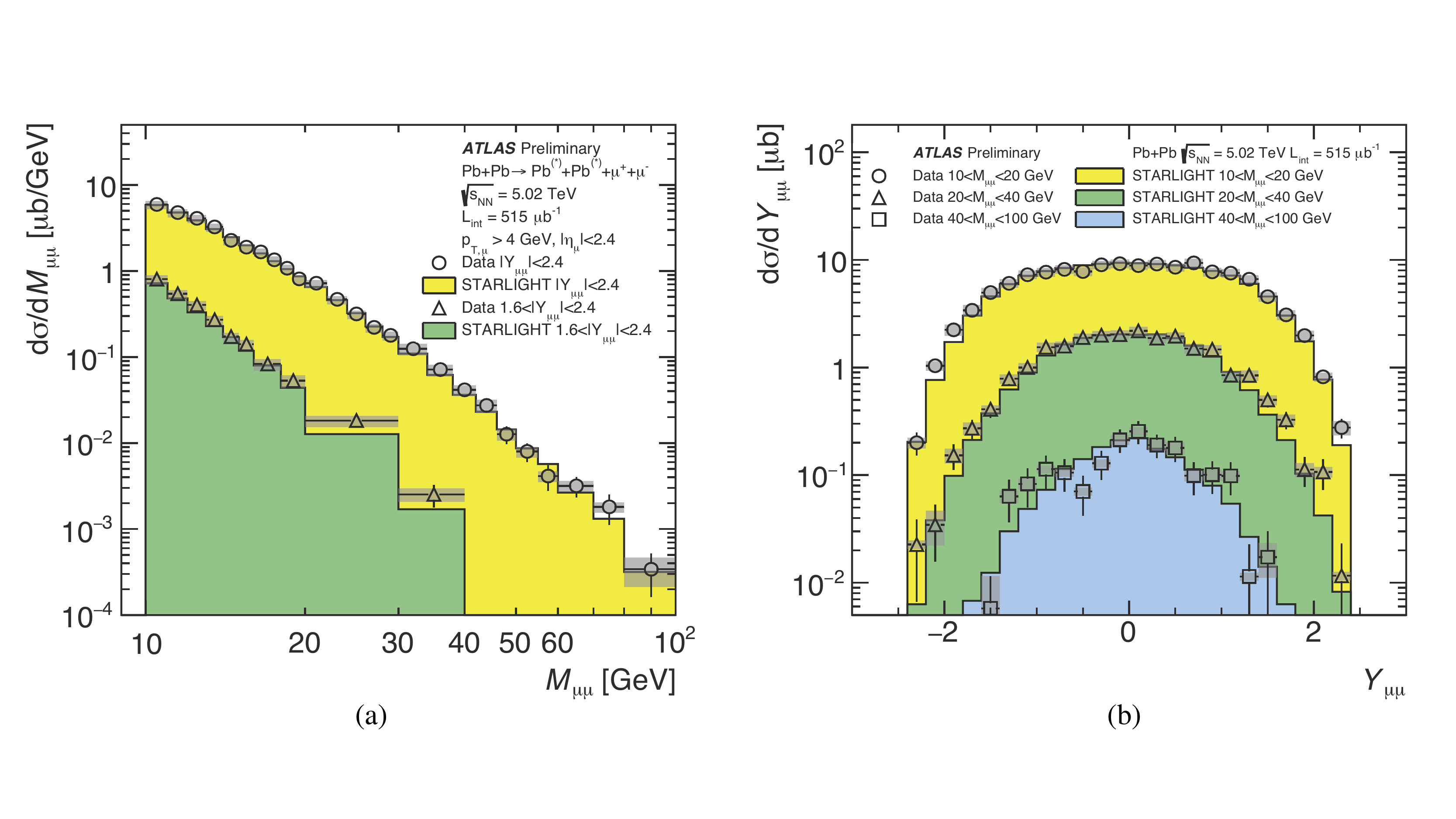}
\vspace*{-1.5cm}
\caption{\label{dimuons_pbpb_atlas}
{\small
The cross-sections for exclusive  production of pairs of muons $Pb Pb \rightarrow Pb Pb (\gamma \gamma) \rightarrow Pb Pb \mu^+ \mu^-$ 
\cite{atlaspbpb} as a function of (a) pair mass and (b) pair rapidity. A possible EM excitation of the outgoing ions is denoted by (*) in the plots
\cite{Auerbach:1983hld}. The data are indicated by the symbols while the MC predictions are shown by solid histograms. Error bars indicate statistical uncertainties while the grey bands indicate the combined systematic uncertainties.}}
\end{figure}

\begin{figure}[!htb]
\centering
\includegraphics[width=0.8\textwidth]{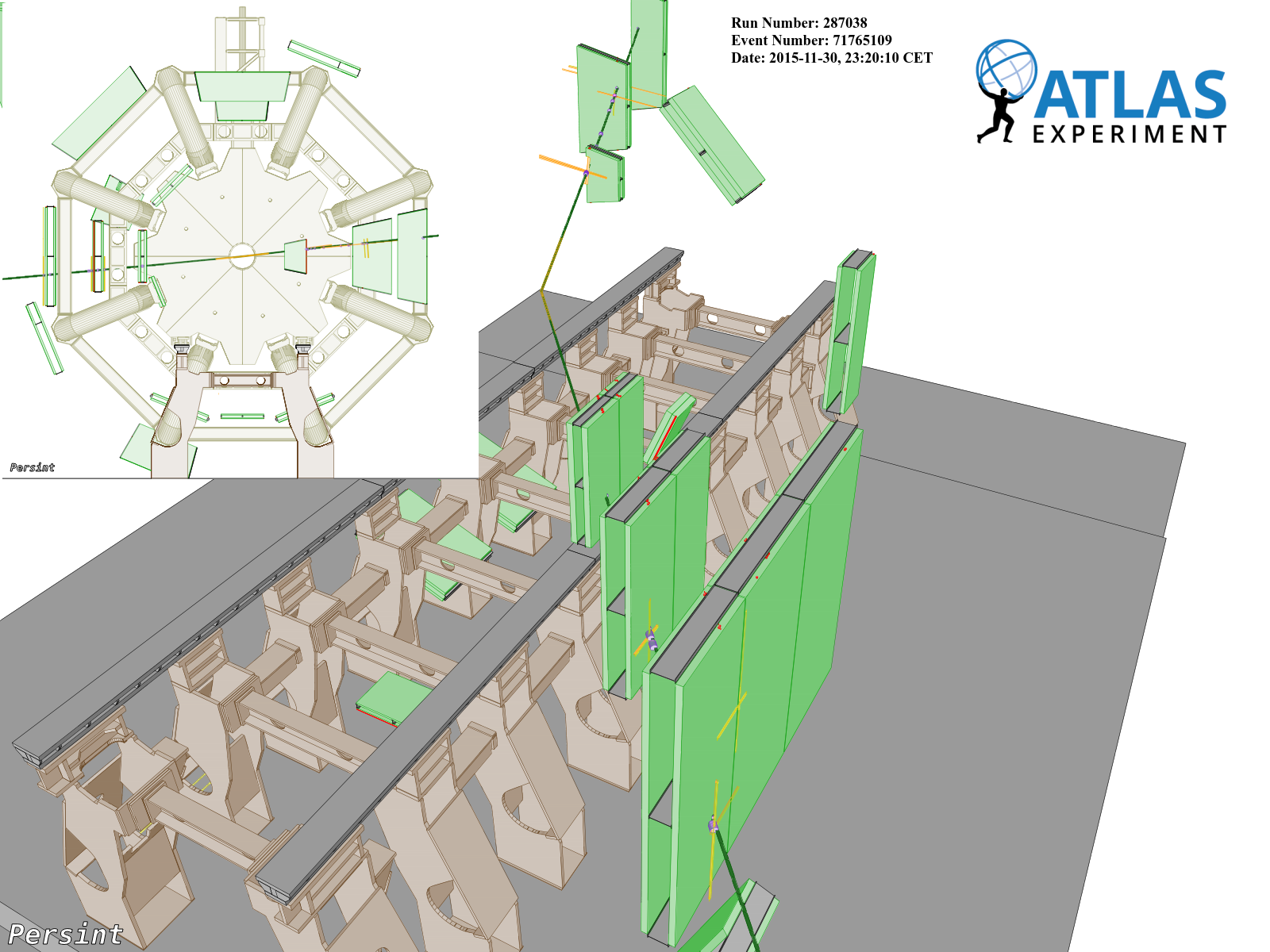}
\caption{\label{event} 
{\small $Pb Pb \rightarrow Pb Pb (\gamma \gamma) \rightarrow Pb Pb \mu^+ \mu^-$ with an invariant mass of the two muons of $173$ GeV \cite{persint}.}}
\end{figure}

This is on this basis, with recorded events, that the analysis is conducted. An event of this kind is pictured in Fig. \ref{event} \cite{persint}.
At this level,
 events are selected using a single-muon trigger, in association with an otherwise low-multiplicity event. The events are then required to have a primary vertex formed solely from two oppositely-charged muons, each having transverse momentum 
$p_{\rm T} > 4$ GeV and pseudo-rapidity smaller that $2.4$. The kinematic variable distributions are then corrected for muon trigger and reconstruction efficiency, and for vertex efficiency.
The nuclear EM form factors ensure that the muons are emitted back-to-back, resulting in a small acoplanarity, defined as 
${\cal A}=1-\Delta \phi/\pi$, where $\Delta \phi$ is the azimutal separation of the two outgoing photons.
The simulation based on {\sc Starlight} \cite{Klein:2016yzr,Auerbach:1983hld} shows that over 99.9\% of pairs have ${\cal A} < 0.008$, while in the data non negligible  tails for the acoplanarity of the pairs of muons  are also observed. This is the reason why the fraction of measured events with ${\cal A} \ge 0.008$ and found to be around 5\%. 
Two assumptions – that the acoplanarity tail is all background, and that it is all signal due to higher-order QED effects – are then tested. Assuming the events with ${\cal A} \ge 0.008$ are primarily background, the data distributions are fit to a functional form comprised of two exponentials. The final result is calculated including the average of the two scenarios and the systematic uncertainties associated with this procedure are defined as half the difference between the two results, such that both extremes are covered.
Finally, at this level, with the selection done and all corrections applied, this is possible to compare the
 results with calculations from {\sc Starlight} \cite{Klein:2016yzr,Auerbach:1983hld}, as a function of the mass for the pairs of muons and pair 
 and as a function of the rapidity of the pairs of muons. This is illustrated  in Fig. \ref{dimuons_pbpb_atlas}. 
 We can observe that a 
 good agreement between the data and the  predictions from the {\sc Starlight} MC is found \cite{atlaspbpb}.
 Here again, in the lead-lead configuration, this gives some credit to the formalism developed in the previous section.
 Using this data set, it was also possible to study precisely the nuclear break-up of the lead ions, producing potentially  neutrons tagged in the forward detectors.
 There are  three topologies to consider: (i) the lead ions are left intact by the interactions, labelled as 0n0n, with no
 neutron produced in either direction, (ii)  one lead ion is dissociated and the other not, labelled as Xn0n, with neutrons in only one direction and
not the other, (iii) the two lead ions are dissociated, labelled as XnXn, with neutrons emitted in both directions.
 In practice, we expect that events
with smaller impact parameter, where the ions are closer together, are more likely to be accompanied by
neutron dissociation in one or both arms and to have photons with higher energies. That is why this is interesting to make the 
 separation between the different topologies as far as this is accessible experimentally \cite{atlaspbpb}. 

\begin{figure}[!htb]
\centering
\includegraphics[width=0.75\textwidth]{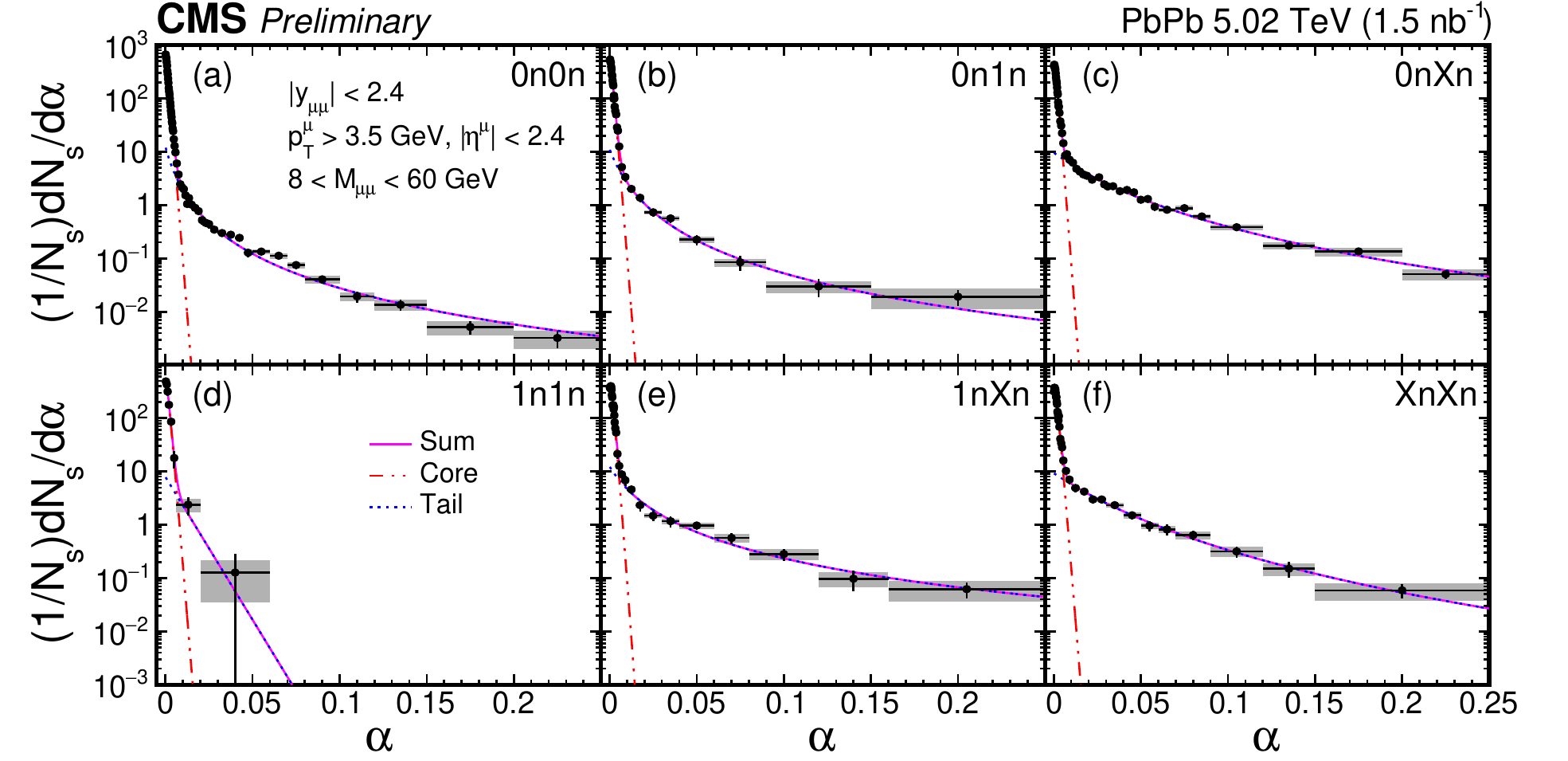}
\caption{\label{acoplanarity} 
{\small Neutron multiplicity dependence of $\alpha$ spectra $\gamma\gamma\rightarrow \mu^+\mu^-$ within the CMS from acceptance for $8 < m_{\mu\mu} < 60$ GeV in PbPb collisions at $\sqrt{s} _\mathrm{NN} = 5.02$ TeV \cite{neutron_dependence}. }}
\end{figure}

\begin{figure}[!htb]
\centering
\includegraphics[width=0.4\textwidth]{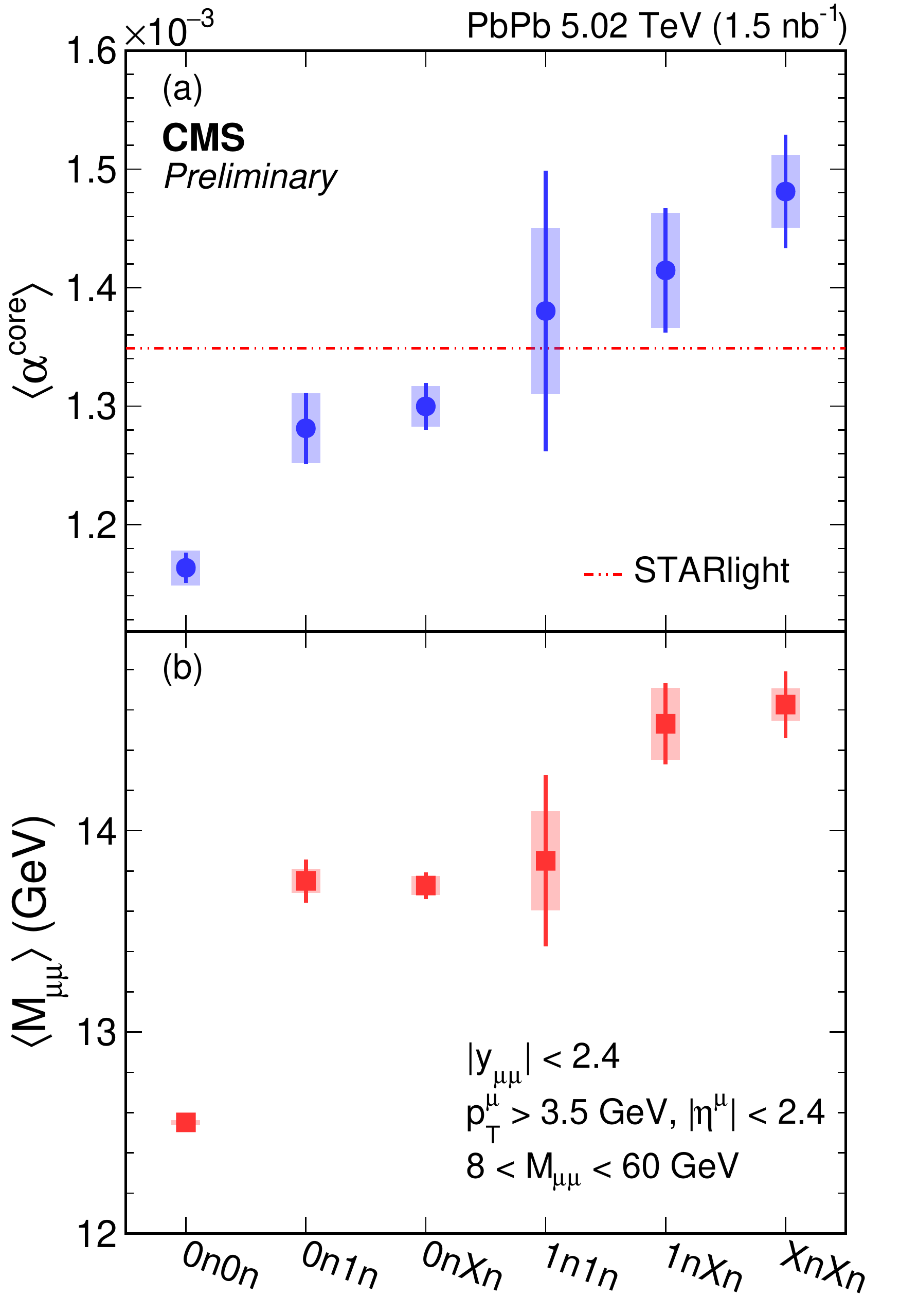}
\caption{\label{neutron_multiplicity} 
{\small Neutron multiplicity dependence on $\langle \alpha^\mathrm{core} \rangle$ and $\langle m_{\mu\mu} \rangle$, within $8 < m_{\mu\mu} < 60$ GeV. The vertical bars represent the statistical uncertainties while the systematic uncertainties are shown in the shaded areas. The red dashed line represents the prediction by {\sc Starlight} MC (no neutron multiplicity dependence) \cite{neutron_dependence}.}}
\end{figure}

In the same spirit, the CMS Collaboration reported a study on forward neutron multiplicity dependence of $\gamma\gamma\rightarrow \mu^+\mu^-$ production in ultra-peripheral PbPb collisions at $\sqrt{s}_\mathrm{NN} = 5.02$ TeV~\cite{neutron_dependence}. The analysis is based on heavy-ion collision data collected in 2018 with an integrated luminosity of 1.5 nb$^{-1}$. These results help better constrain the modeling of photon-induced interactions in ultra-peripheral collisions.
Photon-induced interactions can occur in association with the excitation of the scattered ions via absorption of photons into GDRs or other excited states. The GDRs typically relax by emitting a single neutron, while higher excited states may emit two or more neutrons. A higher neutron multiplicity corresponds to smaller mean average impact parameter $\langle b \rangle$.
For this study, there is no direct online selection on the muon $p_T$. Events with an energy deposit above the calorimeter noise threshold in both forward calorimeters are vetoed. At the analysis level, events must contain exactly two muon candidates with $p_T > 3.5$ GeV each and no additional track in the range $|\eta|<2.4$. The selected events are then classified by neutron multiplicity, which is inferred based on the energy deposited in the ZDCs. The total energy distribution is divided into three neutron multiplicity classes ($0$n, $1$n, and $X$n with  $X \geq 2$) on each side with energy thresholds. The 0n0n class corresponds to no Coulomb break-up of either nucleus and the 1nXn class corresponds to one neutron emitted from one nucleus and at least two neutrons emitted from the other nucleus.
The studied di-muon kinematic range is limited to $8 < m_{\mu\mu} < 60$ GeV and rapidity $|y^{\mu\mu}| < 2.4$. In {\sc Starlight} MC, only muon pairs from the LO $\gamma\gamma$ scattering are generated, and the calculation is performed by integrating over the entire impact parameter space for ultra-peripheral collision events. No differential impact parameter dependence of initial photon $p_T$ is considered in {\sc Starlight} MC.
Figure~\ref{acoplanarity} shows the corrected acoplanarity $\alpha$ distributions of $\mu^+\mu^-$ pairs in lead-lead collisions for different neutron multiplicity classes. Each $\alpha$ spectrum is made up of a narrow core close to zero and a long tail, where the core component mostly originates from the LO $\gamma\gamma\rightarrow \mu^+\mu^-$ scattering while the tail component mostly originates from high-order electromagnetic interactions processes. These high-order processes include, for example, soft photon radiations off the produced lepton and scattering of multiple photons.
The average acoplanarity ($\langle \alpha^\text{core} \rangle$) is determined from fits to the $\alpha$ distribution. The neutron multiplicity dependence of $\langle \alpha^\text{core} \rangle$ is shown in Fig.~\ref{neutron_multiplicity}. Likewise, the mean $\langle m_{\mu\mu} \rangle$ dependence on the neutron multiplicity is shown in Fig.~\ref{neutron_multiplicity}. A clear dependence of either of these variables as a function of the neutron multiplicity is observed. This observation demonstrates the transverse momentum and energy of photons emitted from relativistic ions have impact parameter dependence.

\begin{figure}[!]   
\centering
\hspace*{-1.5cm}
\includegraphics[width=1.1\textwidth]{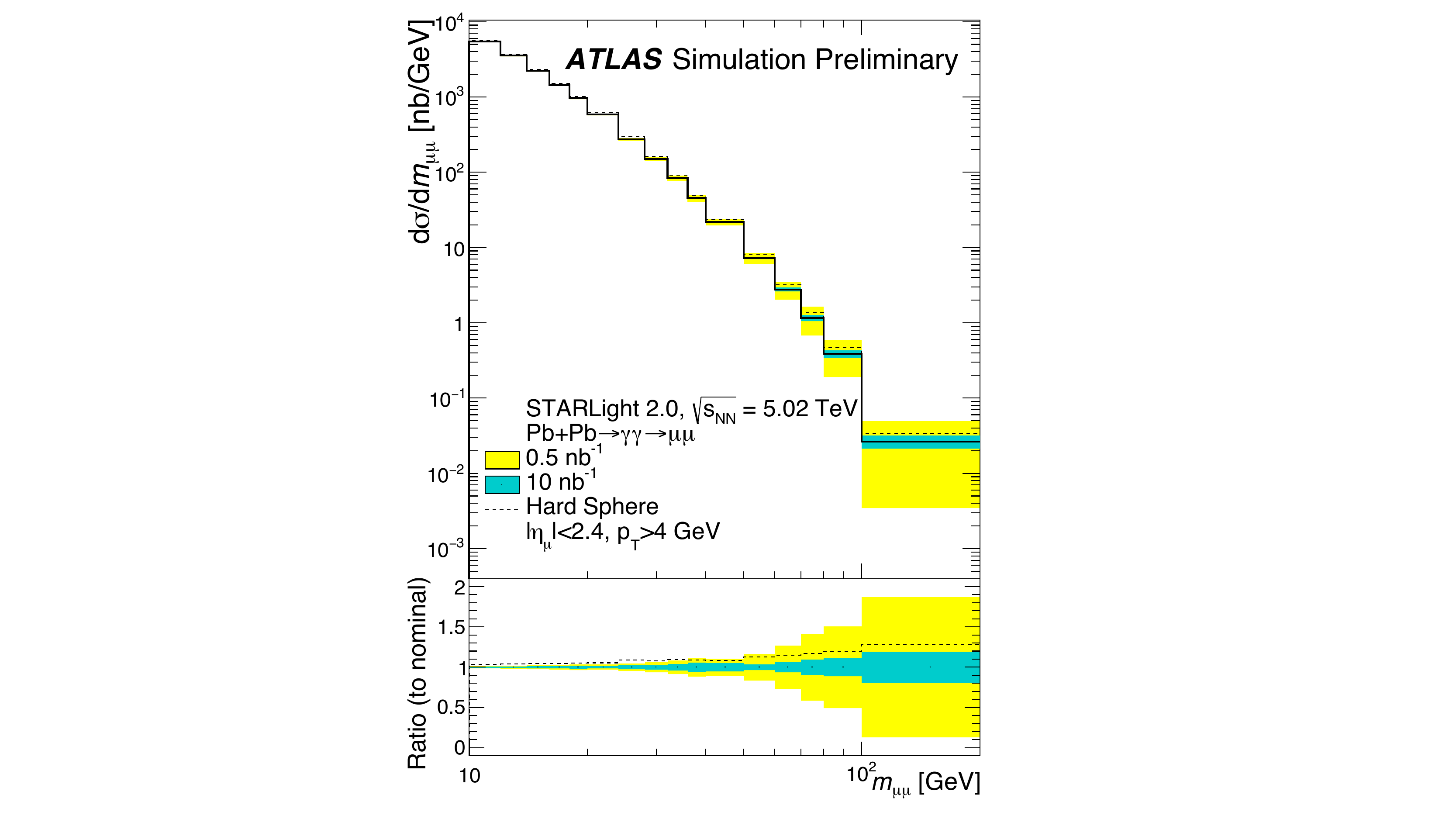}
\caption{\label{yr-pbpb}
{\small
(Upper) Differential cross section for exclusive production of the pairs of muons as a function of the dimuon mass extracted from 
{\sc Starlight}. Two scenarios are considered for the nuclear geometry (form factors) \cite{Citron:2018lsq}: a realistic skin depth of the nucleus (solid line) or a hard sphere (dashed line). (Bottom) Ratio to nominal as a function of the  mass of the pairs of muons, where nominal stands for the realistic skin depth of the nucleus. Shaded bands represent expected statistical uncertainties associated with a number of signal events in each bin for integrated luminosity of $0.5$ nb$^{-1}$, and $10$ nb$^{-1}$.}
}
\end{figure}

Finally, in order to give a perspective on what could be achieved with a larger statistics recorded by the
detectors, a simulation is presented in Fig. \ref{yr-pbpb}. This shows clearly how the uncertainties can be reduced in the large mass domain
and then the potential discrimination power of such measurements for ion form factors (or charge distribution hypothesis).


\vfill
\clearpage
\newpage
\section{Light-by-light scattering at the LHC and beyond}

After having shown that the formalism presented in the first section is able to describe different exclusive processes at LHC energies, we can address a more subtle problem,
with the observation of light-by-light scattering. Indeed, this reaction $Pb Pb \rightarrow Pb Pb (\gamma \gamma) \rightarrow Pb Pb\gamma \gamma$, pictured on the left of Fig. \ref{figslbl} adds one level of complexity to the understanding due to the fact the prediction needs a box calculation. Then we can expect that the rate of production for  exclusive photon-photon final states is thus  weaker than the di-lepton final states by a factor $\alpha_{em}^2$. Moreover, as explained below, in the optical domain, light-by-light scattering in its full glory using laser beams is even more difficult to observe, if not impossible. The purpose of this section is to show in which conditions the observation of light-by-light scattering has been obtained at the LHC and how this result can inspire some ideas for the optical domain.

\subsection{Experimental observations}
\label{lblsec}

A search for the reaction
$Pb Pb \rightarrow Pb Pb (\gamma \gamma) \rightarrow Pb Pb\gamma \gamma$ (at a center of mass energy per nucleon pair of $5.02$ TeV)
has  been performed by the ATLAS and CMS experiments, leading to the observation of the
$\gamma \gamma \rightarrow \gamma \gamma$  process (light-by-light scattering)
\cite{lbl1,lbl2,lbl3,Sirunyan:2018fhl}. Below, we give some precisions for the analysis of the  ATLAS measurement in the field
\cite{lbl1,lbl2,lbl3}, and then we complete this presentation with the CMS results \cite{Sirunyan:2018fhl}
and the recent update by ATLAS \cite{lbl1,lbl2,lbl3}.

Again, prior to any analysis, the first task is to record the events.
For the study described here,
events have been recorded in lead-lead collisions in 2015 ($\sqrt{s_{NN}}=5.02$ TeV),
corresponding to an integrated luminosity of $480$ $\mu$b$^{-1}$, using a dedicated trigger for events with moderate activity in the calorimeter but little additional activity in the entire detector. The interest of the 2015 data taking period is that both ATLAS and CMS experiments have published results on light-by-light scattering based on these data. Thus, they can be put in perspective and compared.
Later in the discussion, we present a new analysis from ATLAS using also data recorded in  lead-lead collisions in 2018. The interest is an increase of the statistics and 
consequently of the observed number of light-by-light scattering candidates. However, the discussion based on the 2015 data contain by itself all the experimental ideas and conclusions.

At a first level of selection, the total transverse energy registered in the calorimeter after noise suppression is required to be between $5$ and $200$ GeV. Then at a next step, events are rejected if more than one hit was found in the inner ring of the MBTS (MBTS veto) or if more than ten hits were found in the pixel detector.
This means that
events  are recorded if they present a moderate activity in the calorimeter (covering the pseudo-rapidity 
range $\vert \eta \vert < 4.9$) with little additional activity in the entire detector. 

\begin{figure}[bh!]   
\centering
\includegraphics[width=0.7\textwidth]{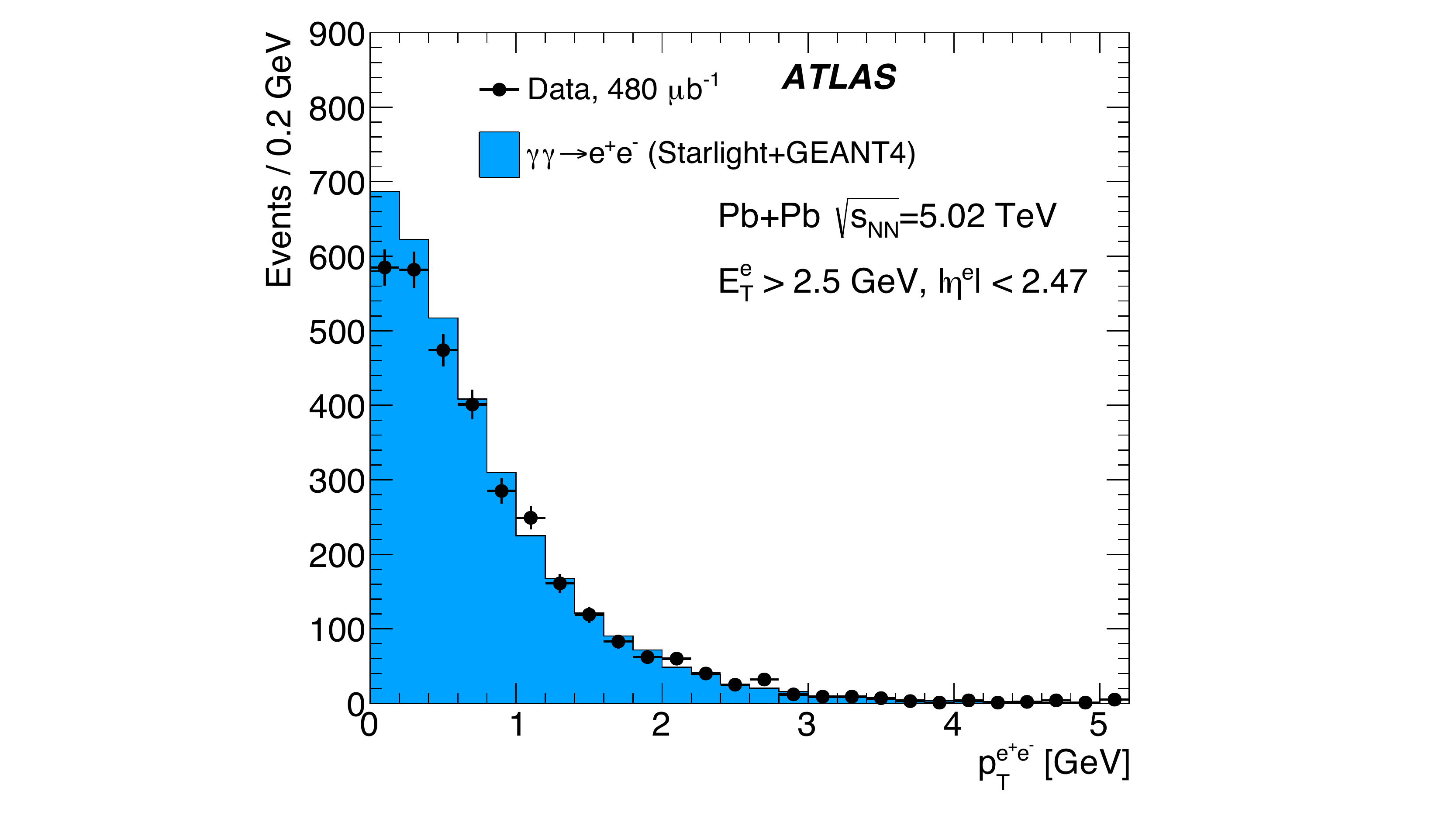}
\vspace*{-0.5cm}
\caption{\label{figlbl1}
{\small
Transverse momentum distribution of the electron-positron pairs for the events corresponding to the reaction
$Pb Pb \rightarrow Pb Pb (\gamma \gamma) \rightarrow Pb Pb e^+ e^-$. Electrons are required to have a transverse energy greater than $2.5$ GeV and pseudo-rapidities
$\vert \eta \vert < 2.47$ with  or $1.37 < \vert \eta \vert < 1.52$ excluded.
Data (points) are compared to simulation (histograms) \cite{lbl1}. The statistical uncertainties on the data are shown as vertical bars.
}
}
\end{figure}

\begin{figure}[!]   
\centering
\includegraphics[width=0.7\textwidth]{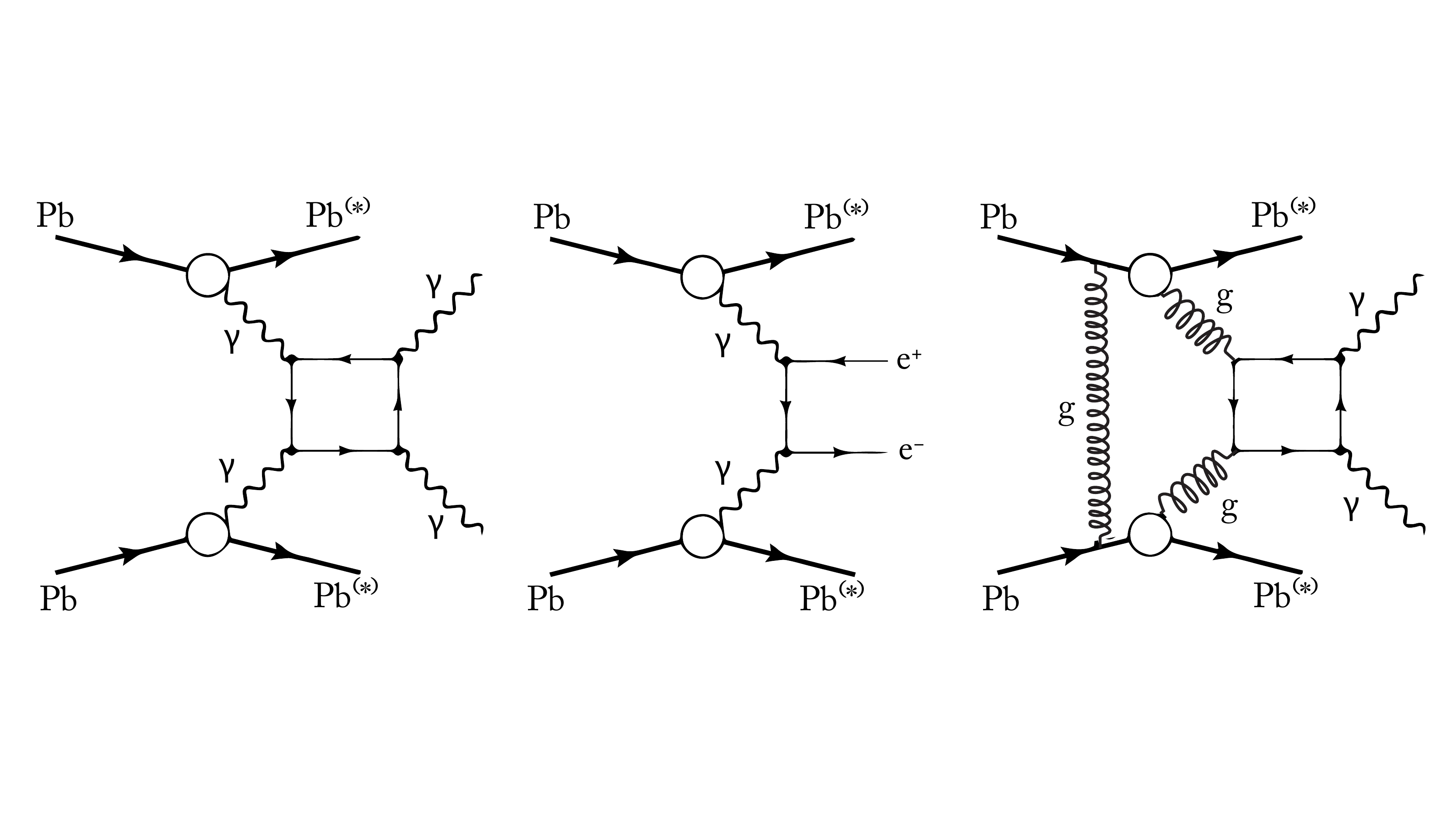}
\vspace*{-0.5cm}
\caption{\label{figslbl}
{\small
Schematic diagrams of light-by-light scattering, QED exclusive production of electron-positron pairs, and CEP $g g \rightarrow \gamma \gamma$ (see text). The $(*)$ superscript indicates a potential electromagnetic excitation of the outgoing ions.
}
}
\end{figure}

Once this is done, the analysis selection can be performed with all the needed corrections.
Then, events are required to have exactly two photons with transverse energy greater than $3$ GeV, pseudo-rapidity $\vert \eta \vert < 1.37$ or $1.52 < \vert \eta \vert < 2.37$, and a photon-photon invariant mass greater than $6$ GeV.
Since the analysis requires the presence of low-energy photons, which are not typically used in ATLAS analyses, detailed studies of photon reconstruction and calibration are also performed. In particular, exclusive  pairs of electron-positron from the reaction $\gamma \gamma \rightarrow e^+ e^-$ are used for various aspects of the analysis, for example to validate the EM calorimeter energy scale and resolution. Fig. \ref{figlbl1} presents the transverse momentum of the exclusive electron-positron pairs. A good agreement between the data and simulation is obtained.
Possible backgrounds can arise mainly from misidentified electrons from the $\gamma \gamma \rightarrow e^+ e^-$  process, as well as from the central exclusive production of two photons from the fusion of two gluons (CEP $g g \rightarrow \gamma \gamma$ ). The signal and the two main backgrounds are pictured in Fig. \ref{figslbl}.

\begin{figure}[!]   
\centering
\includegraphics[width=0.85\textwidth]{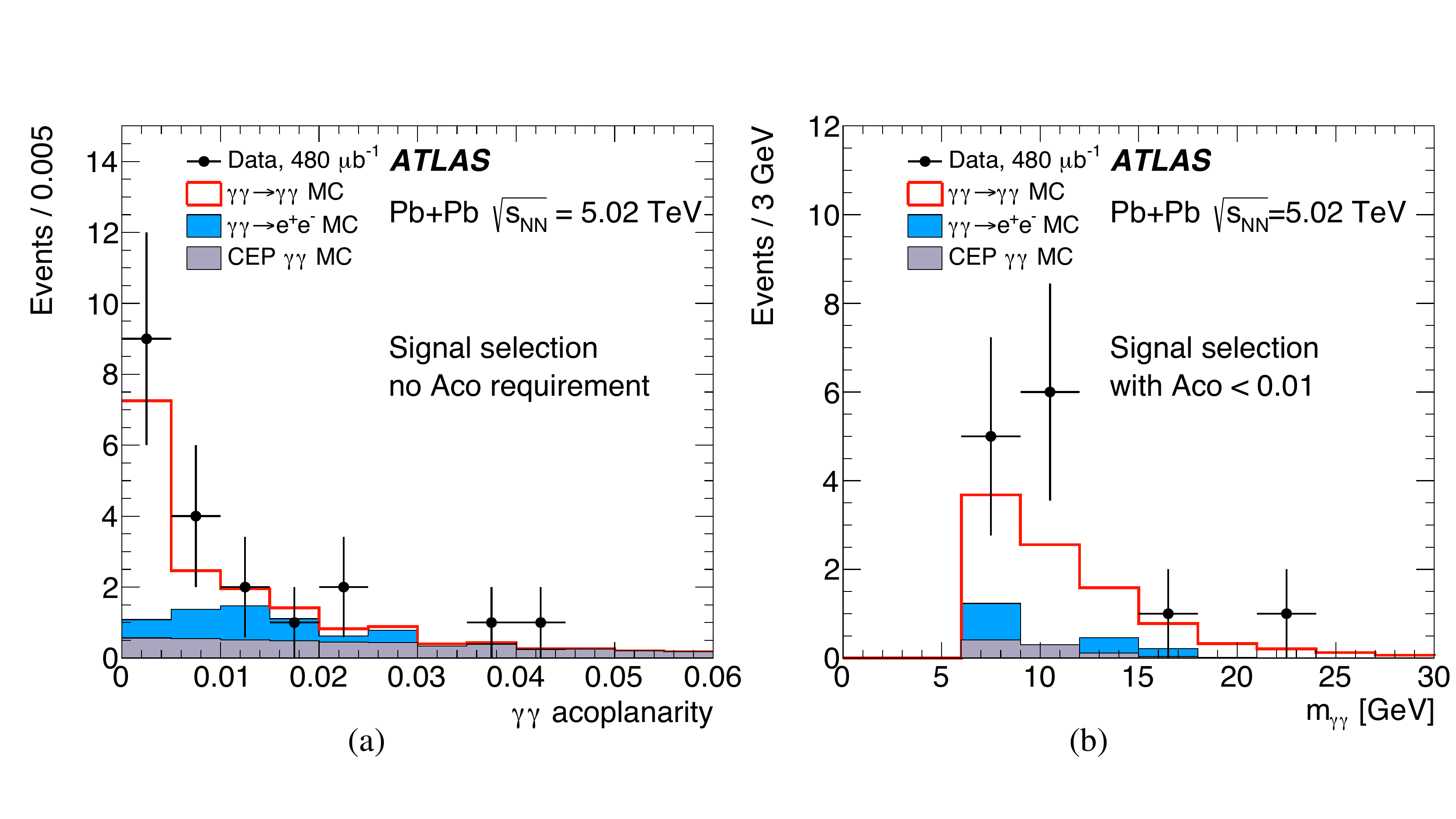}
\vspace*{-0.5cm}
\caption{\label{figlbl2}
{\small
Kinematic distributions for $Pb Pb \rightarrow Pb Pb (\gamma \gamma) \rightarrow Pb Pb\gamma \gamma$ event candidates in the ATLAS experiment: (a)  photon-photon acoplanarity  before applying the acoplanarity (below $0.01$) requirement, (b) photon-photon invariant mass after applying the acoplanarity requirement. Data (points) are compared to MC predictions (histograms) \cite{lbl1}. The statistical uncertainties on the data are shown as vertical bars.
}
}
\end{figure}

In order to reduce the
$\gamma \gamma \rightarrow e^+ e^-$ background, a veto on the presence of any charged-particle tracks is imposed. To reduce other fake-photon backgrounds (for example, cosmic-ray muons), the photon-photon transverse momentum is required to be below $2$ GeV. The photon-photon acoplanarity distribution for events satisfying above mentioned selection criteria is shown in Fig. \ref{figlbl2}. The shape of this distribution is different for the signal and background processes, therefore an additional requirement on photon-photon acoplanarity below $0.01$ is imposed.

The photon-photon invariant mass distribution for events satisfying all selection criteria is shown in Fig. \ref{figlbl2}. In total, $13$ events are observed in data whereas $7.3$ signal events and $2.6$ background events are expected. The statistical significance against the background-only hypothesis is found to be $4.4$ standard deviations.
After background subtraction and detector corrections, 
the cross section for the $Pb Pb \rightarrow Pb Pb (\gamma \gamma) \rightarrow Pb Pb\gamma \gamma$
process is measured in a fiducial phase space defined by the transverse energy of photons above $3$ GeV, the photon absolute pseudo-rapidity below $2.4$, the photon-photon invariant mass greater than $6$ GeV, the transverse photon-photon momentum lower than $2$ GeV and the photon-photon acoplanarity below $0.01$. The measured fiducial cross section is found to be:
$$
\sigma_{{\rm fid}} = 70 \pm 24 {\rm (stat.)} \pm 17 {\rm (syst.)} \ {\rm nb}.
$$
This is in agreement with the predicted values of 
$45 \pm 9$ nb and $49 \pm 10$ nb within uncertainties.

\begin{figure}[!]   
\centering
\includegraphics[width=0.85\textwidth]{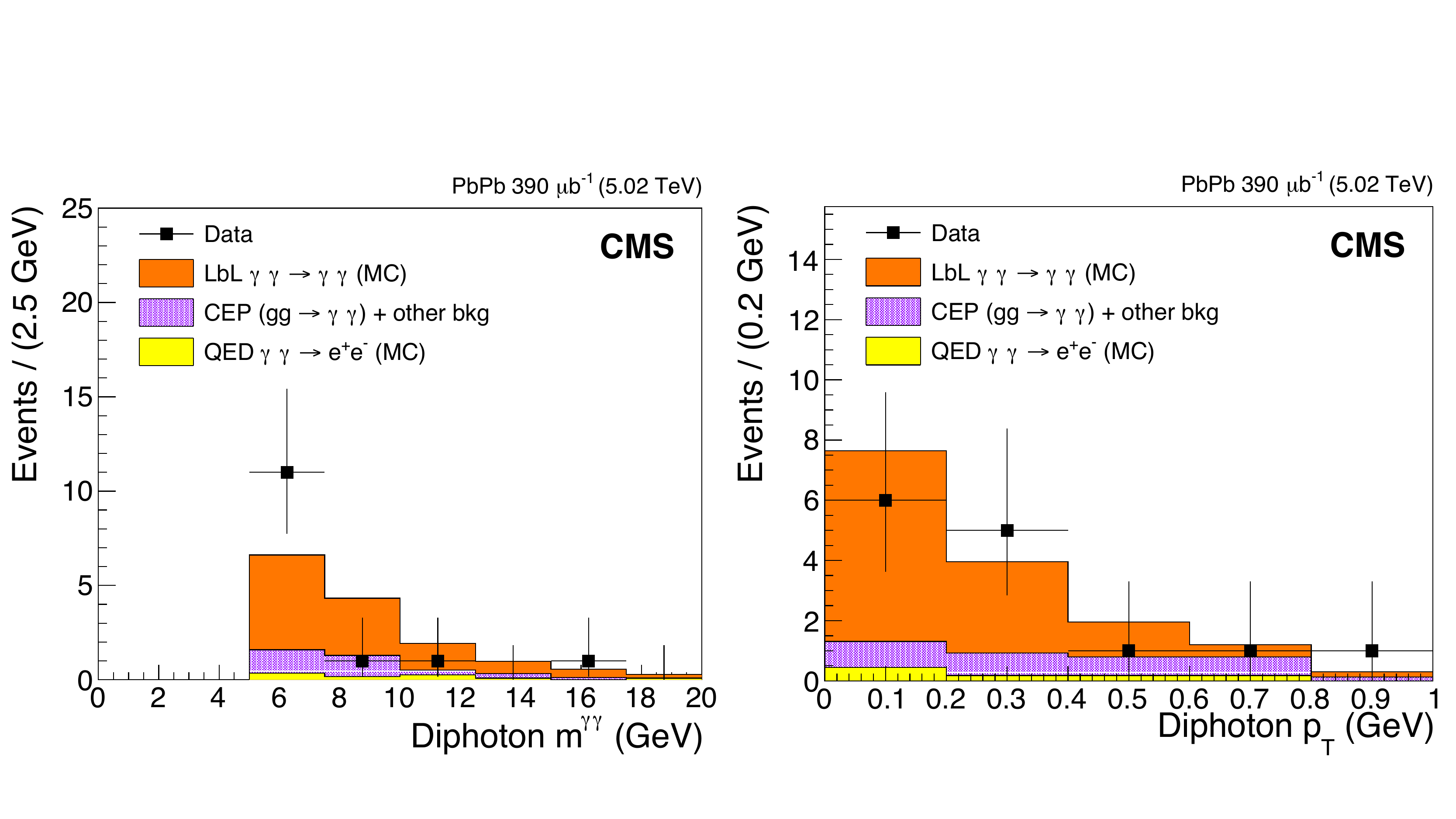}
\vspace*{-0.5cm}
\caption{\label{figlbl4}
{\small
Kinematic distributions for $Pb Pb \rightarrow Pb Pb (\gamma \gamma) \rightarrow Pb Pb\gamma \gamma$ event candidates in the CMS experiment
\cite{Sirunyan:2018fhl}: (Left)  photon-photon invariant mass after applying all selection requirements, (Right) photon-photon transverse momentum. Data (points) are compared to MC predictions (histograms). The statistical uncertainties on the data are shown as vertical bars.
}
}
\end{figure}

A similar analysis has been done by the CMS experiment \cite{Sirunyan:2018fhl}, also using data collected during the same period in 2015 in lead-lead collision
 ($\sqrt{s_{NN}}=5.02$ TeV) and corresponding to an integrated luminosity of $390$ $\mu$b$^{-1}$.
 The fiducial kinematic domain of the measurement is slightly different to the one described above for the ATLAS measurement. In particular the photon-photon invariant mass is requested to be greater than $5$ GeV
 and the transverse energy of photons are requested to be above $2$ GeV. 
 Kinematic distributions are presented in Fig. \ref{figlbl4}.
 Then, the fiducial cross section is found to be:
$$
\sigma_{{\rm fid}} = 120 \pm 46 {\rm (stat.)} \pm 28 {\rm (syst.)} \pm 4 {\rm (theo.)} \ {\rm nb}.
$$
This is consistent with the prediction of $138 \pm 14$ nb.

\begin{figure}[!]   
\centering
\includegraphics[width=0.95\textwidth]{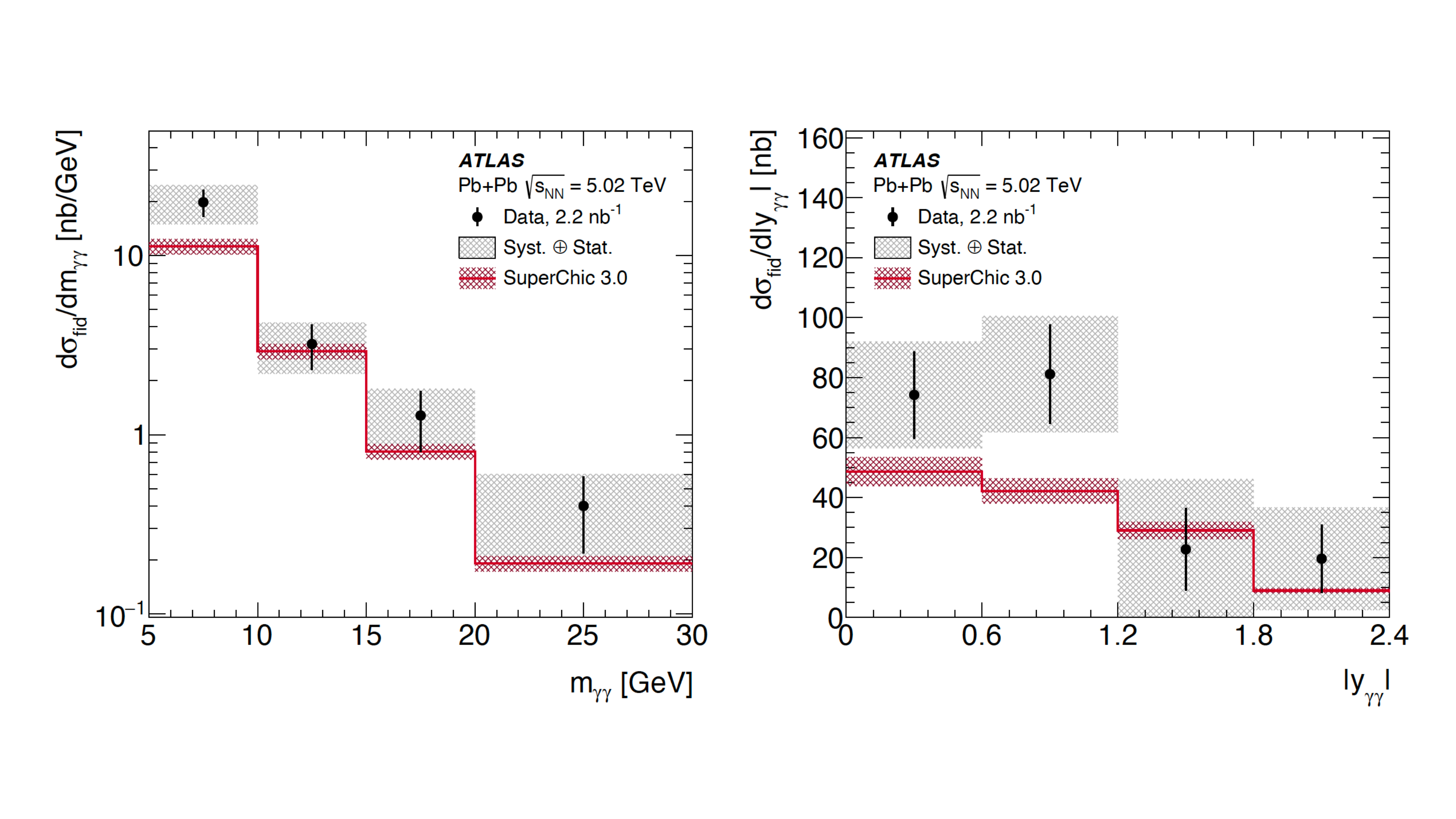}
\vspace*{-1cm}
\caption{\label{figlbldiff}
{\small
Measured differential fiducial cross sections of the reaction $Pb Pb \rightarrow Pb Pb (\gamma \gamma) \rightarrow Pb Pb\gamma \gamma$
at $5.02$ TeV for two observables: the invariant mass  and the absolute rapidity  of the two outgoing photons.
The measured cross-section values are shown
as points with error bars giving the statistical uncertainty and grey bands indicating the size of the total uncertainty \cite{lbl3}.
The results are compared with the MC prediction from the SuperChic v3.0 MC generator (solid line) with bands denoting
the theoretical uncertainty.
}
}
\end{figure}

Recently, by combining the 2015 and 2018 lead-lead collisions data sets (at a center of mass energy per nucleon pair of $5.02$ TeV),
the ATLAS collaboration has been able to produce differential cross sections for light-by-light scattering \cite{lbl1,lbl2,lbl3}. In practice
the differential fiducial cross sections are determined using an iterative Bayesian unfolding method. The idea is to correct 
for bin migrations between particle and
detector-level distributions due to detector resolution effects, and applies reconstruction efficiency as
well as fiducial corrections. The reconstruction efficiency corrects for events inside the fiducial region that
are not reconstructed in the signal region due to detector inefficiencies; the fiducial corrections take into
account events that are reconstructed in the signal region, but originate from outside the fiducial region.
Results are presented in  Fig. \ref{figlbldiff} and compared to MC predictions. A reasonable agreement is obtained.

\subsection{Discussion on the light-by-light scattering}

\subsubsection{LHC versus the optical domain}

Interestingly, we can convert the cross section obtained in lead-lead collisions to the pure light-by-light cross section for the reaction
$\gamma \gamma \rightarrow \gamma \gamma$ at a center of mass energy $\sqrt{\hat s}$ for the initial photon-photon state which is not fixed in this case 
(what is fixed is the ion-ion collision energy). However, it is easy to determine that $\sqrt{\hat s} \sim 20$ GeV. Then, we obtain:
\be
\sigma(\gamma \gamma \rightarrow \gamma \gamma, \sqrt{\hat s} \sim 20 \ {\rm GeV}) \sim 1  \  {\rm pb}.
\label{value1}
\ee
This cross section value can be compared to the predicted value in the optical domain:
\be
\sigma(\gamma \gamma \rightarrow \gamma \gamma, \sqrt{\hat s} \sim 1 \ {\rm eV}) \sim 3 \ 10^{-30}  \  {\rm pb}.
\label{value2}
\ee

\begin{figure}[!]
\centering
  \includegraphics[scale=0.3]{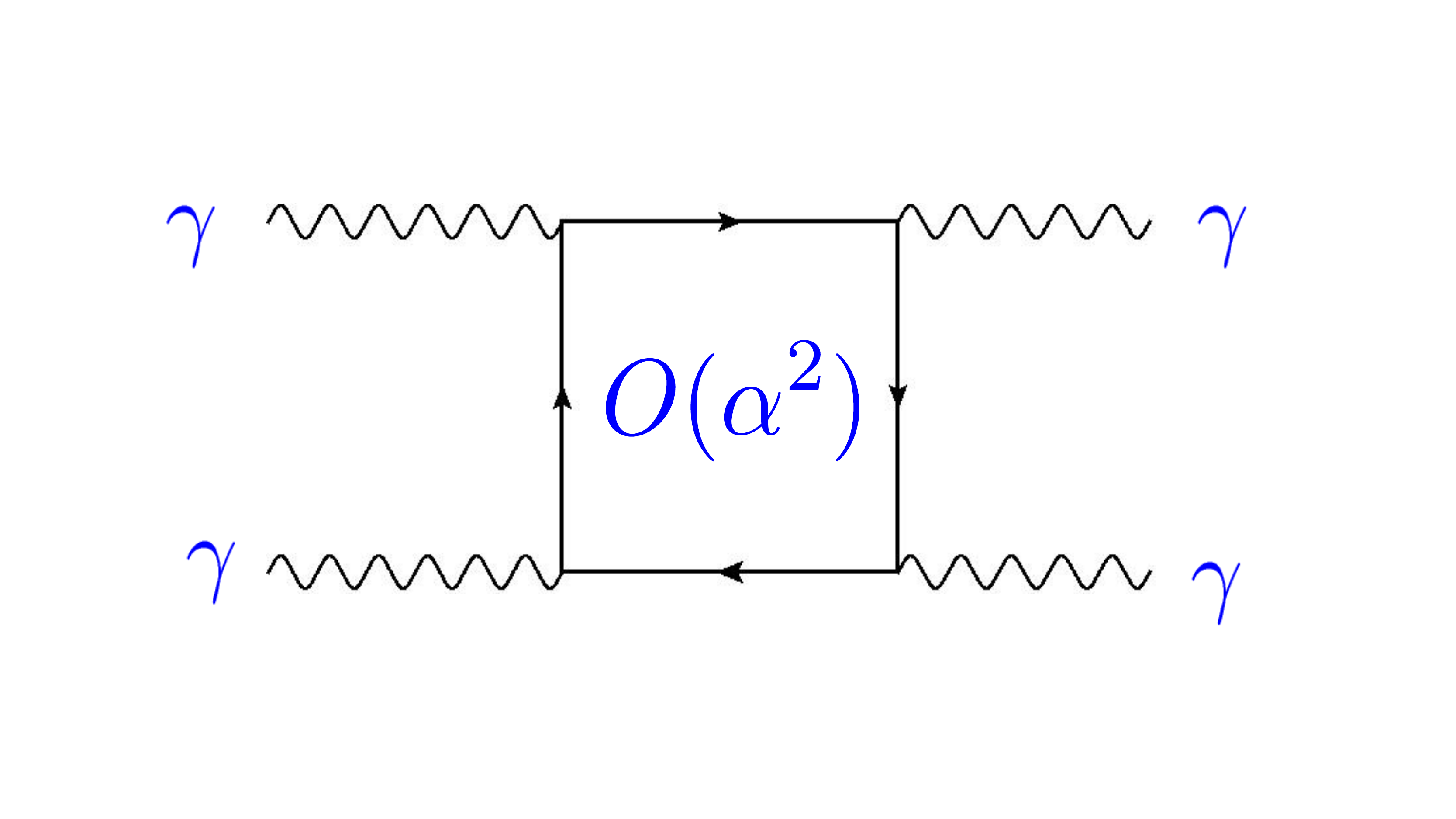}
\vspace*{-1.1cm}
\caption{\label{fig_lbl_basics}
{\small
Generic QED diagram for the light-by-light scattering process. It can be broken up in $6$ different topologies, 
each one of them leading to ${4!}$ contractions in the photon creation-annihilation operators, thus $144$
possible diagrams.}
}
\end{figure}

This is interesting to put in perspective these two cross section values for the reaction $\gamma \gamma \rightarrow \gamma \gamma$.
In Quantum Electrodynamics (QED) the $\gamma \gamma \rightarrow \gamma \gamma$
reaction proceeds at lowest order in the fine-structure constant ($\alpha_{em} \simeq 1/137$)
via virtual  one-loop box diagrams involving fermions, see Fig. \ref{fig_lbl_basics}, which leads to
an amplitude that scales as $\alpha_{em}^2$. And, if the center mass energy of the photon-photon initial state is large enough
this is possible to consider $W^\pm$ bosons in the loop. For this reaction to happen as depicted in Fig. \ref{fig_lbl_basics},
the incident EM fields has to be sufficiently large to create at least an electron-positron pair, which gives: $E>E_c=m_e^2/e \simeq  10^{18}$ V/m and 
$B>B_c \simeq 4 \ 10^9$ G (or in terms of intensity of the field $ I>I_c \simeq 10^{29}$ W/cm$^2$).
For lead-lead  collisions at a center of mass energy per nucleon pair of $5.02$ TeV, the corresponding electric field is about $10^{25}$ V/m and the
previous conditions are fulfilled. The QED calculations (based on the diagram in Fig. \ref{fig_lbl_basics}) are therefore supposed to  
describe the measurements and this is what we have observed in the previous section and this gives the result of Eq. (\ref{value1}).

\subsubsection{Light-by-light scattering in the optical domain}

In the optical domain, typically 
with laser beams,
the conditions are reversed with $E<<E_c$ (and similarly for the magnetic field) and also with fields that vary slowly  over a length
equal  to the reduced electron Compton wavelength. This means that in this physics case, only an effective form of the photon-photon interaction
needs to be considered \cite{theorigins}:
\be
{\mathcal L} = 
-\frac{1}{4} F_{\mu \nu} F^{\mu \nu} + 
\frac{a}{E_c^2} (F_{\mu \nu} F^{\mu \nu})^2
+\frac{b}{E_c^2}({\tilde F}_{\mu \nu} F^{\mu \nu})^2,
\label{eh1}
\ee
where $a$ and $b$ are parameters (of similar units) that can be determined using the limit of the QED calculations in the low energy (optical) domain
(with $4b=7a$). We can formulate
easily the expression (\ref{eh1}) in terms of the EM fields ($\textbf{E}$, $\textbf{B}$):
\be
{\mathcal L} = 
\frac{1}{2} (\textbf{E}^2-\textbf{B}^2)+ 
\frac{2\alpha_{em}^2}{45m_e^4}  \left[
(\textbf{E}^2-\textbf{B}^2)^2 +7(\textbf{E}.\textbf{B})^2
\right].
\label{optical1}
\ee
This expression  leads to the prediction of Eq. (\ref{value2})
once we compute the cross section from the Lagrangian density above:
$$
d\sigma(\gamma \gamma \rightarrow \gamma \gamma, \omega \sim 1 \ {\rm eV}) 
=\frac{1}{512 \pi \omega^2} \vert {\cal A}_{\gamma \gamma \rightarrow \gamma \gamma} \vert^2
d \cos(\phi),
$$
where $\omega$ is the energy of one incoming photon and $\phi$ the scattering angle 
in the center of mass frame. This gives:
\be
\sigma(\gamma \gamma \rightarrow \gamma \gamma, \omega \sim 1 \ {\rm eV}) 
= K \alpha_{em}^2 r_e^2 (\omega/m_e)^6,
\label{value2b}
\ee
where $K$ is a numerical constant ($K=\frac{973}{10125\pi}$) and $r_e=e^2/(4\pi m_e)$ the classical radius of the electron.
Interestingly, we can rewrite Eq. (\ref{value2b}) as:
$$
\sigma(\gamma \gamma \rightarrow \gamma \gamma, \omega \sim 1 \ {\rm eV})  = K \frac{1}{16\pi^2}(\frac{\alpha_{em}}{E_c^2})^2 \omega^6,
$$
with $\omega$ expressed in eV.

Regarding the value obtained in the optical domain (prediction of Eq. (\ref{value2})), we understand that this will be extremely difficult to observe light-by-light scattering in its full glory through collisions of laser beams
\cite{Sarazin:2016zer,King:2012aw,Takahashi:2018uut,felix}.  In principle, the experiment could be done by focusing two laser beams on the same point in space. Then, a third laser beam focused on the same point would stimulate the diffusion in a fourth beam, with the conservation of energy and momentum.
From the expression of the Lagrangian density (\ref{eh1}) or (\ref{optical1}), it is not complicated to compute the value of the electric field for the fourth beam in function of the three others. For any polarisations of the beams, we obtain an expression of the form:
$$
\frac{\partial E_4}{\partial z}+ \frac{\partial E_4}{\partial t} =-i \omega_4 \chi^3 E_1 E_2 \bar{ E_3},
$$
where $\chi^3$ depends linearly of the parameters $a$ and $b$ of Eq. (\ref{eh1}). 
We find an expression of the form:
$$
\chi^3 \simeq \frac{p}{45\pi}\frac{\alpha_{em}}{E_c^2} \simeq 3 \ 10^{-41} p  \  \mathrm{m^2/V^2},
$$
where $p$ is dimensionless form factor of order of unity, which depends on polarisation states of the laser waves.
Then, by measuring the number of emitted photons within the fourth beam, which is proportional to $\vert \chi^3 \vert^2$, we could (in principle) get an evidence of light-by-light scattering in a laser beam experiment. The smallness of $\vert \chi^3 \vert^2$ which is exactly equivalent to the smallness of
$\sigma(\gamma \gamma \rightarrow \gamma \gamma, \omega \sim 1 \ {\rm eV}) $ makes this objective in its full glory difficult to reach.

However, it is possible to find a  subtle way that could be achieved in practice.
Below, we study the interaction of two
counter-propagating EM waves along the $(z)$ axis produced by high intensity laser beams. 
We are quite precise in the development of these classical calculations because we reuse them in a next section (\ref{sectiontoto}) for a non-standard case and obtain a non trivial result.
The two laser beams considered here are supposed to be produced in high finesse cavities.
For simplicity, we expose the calculations for plane waves of different intensities (linearly polarized), defined as:
$$ 
\textbf{E}_0 = E_0 \textbf{e}_x e^{i\omega_0 (t+z)} +c.c.
$$
and 
$$ 
\textbf{E}_1 = (E_{1,x} \textbf{e}_x+E_{1,y} \textbf{e}_y) e^{i\omega (t-z)} +c.c.
$$ 
with $E_0\gg E_{1,x},E_{1,y}$
and different angular frequencies $\omega_0 \ne \omega$. 
Without loss of generality, we can pose: $E_0= \mathrm{constant}$ and $E_{1,x}=E_{1,x}(z)$, $E_{1,y}=E_{1,y}(z)$, where the $z$ dependence for $E_{1,x}$ and $E_{1,y}$ will be a consequence of the bi-quadratic  terms of the EM fields in expression (\ref{eh1}).  
Introducing a transverse profile to the EM fields would not modify the conclusions exposed below.
We obtain $E_{1,x}(z)$ and $E_{1,y}(z)$ by solving the non-linear Maxwell equations derived from the Lagrangian density (\ref{eh1}), namely:
\begin{equation}
\nabla \times \textbf{E}  = -\partial_t{\textbf{B}},
\label{eqb}
\end{equation}
\begin{equation}
\nabla \times \textbf{H} = \partial_t{\textbf{D}},
\label{eqa}
\end{equation}
with the vectors $\textbf{D}$ and $\textbf{H}$ defined as:
\begin{equation}
\textbf{D}=\textbf{E}+ K\left[2\left(\textbf{E}^2-\textbf{B}^2\right)\textbf{E}+7\left(\textbf{E}\cdot\textbf{B}\right)\textbf{B}\right],
\label{eqD}
\end{equation}
\begin{equation}
\textbf{H}=\textbf{B}+ K\left[2\left(\textbf{E}^2-\textbf{B}^2\right)\textbf{B}-7\left(\textbf{E}\cdot\textbf{B}\right)\textbf{E}\right],
\label{eqH}
\end{equation}
where $K=\frac{16\pi}{45}\frac{\alpha_{em}^2}{m_e^4}$. First, the magnetic field 
$\textbf{B}$ is obtained from equation (\ref{eqb}). Then,
$\textbf{E}$ and $\textbf{B}$ are used in equations (\ref{eqD}) and (\ref{eqH}) in order to derive the vector fields $\textbf{D}$ and $\textbf{H}$. Finally, by solving equation (\ref{eqa}), we obtain:

\be
E_{1,x}(z) = E_{1,x}(0) e^{i 16 K \omega |E_0|^2 z},
\ \ \
E_{1,y}(z) = E_{1,y}(0) e^{i 28 K \omega |E_0|^2 z}.
\ee

The interaction of the two counter-propagating EM wave through the creation of virtual electron-positron pairs, leading to light-by-light scattering, generates a birefringence of the vacuum. 
Then, the non-linear coupling of the fields induces an increase of the vacuum refractive index.
In the physics case considered here, the refractive index along the strong electric field is found to be: $n_x = 1+16 K  |E_0|^2$, while the index in the perpendicular direction is: $n_y = 1+28 K  |E_0|^2$.

Experimentally, let us consider that the wave $\textbf{E}_1$ is
initially polarized with an angle $\alpha$ w.r.t. the $(x)$ axis
and then interact with the counter propagating intense wave $\textbf{E}_0$ over a length $L$. After the propagation over $L$:
$$
\textbf{E}_1(L,t) = (A \textbf{e}_x \cos \alpha e^{in_x \omega L}
+A \textbf{e}_y \sin \alpha e^{in_y \omega L}) e^{i\omega t} + c.c.
$$
There will be a field component perpendicular to the initial polarization and the intensity of this mode is proportional to $\sin^2 2\alpha \sin^2\theta$
with 
$$
\theta = \pi (n_y-n_x)L/\lambda.
$$
Then, we can write (up to numerical factors denoted by $k$ and $k'$):
\be
\theta = k \frac{\alpha_{em}^2}{m_e^4} |E_0|^2 L/\lambda = k' \alpha_{em} \frac{|E_0|^2}{E_{cr}^2} L/\lambda.
\label{vacb}
\ee
Numerically, one can consider a field of high intensity of about $I=10^{24}$ W/cm$^2$ focused on a cross section area of $s=10^{-6}$ cm$^2$.
This intensity corresponds to a beam pulse of energy $e=10$ kJ with a duration $\tau=L$, then $I=e/(\tau s)$. 
With $\lambda=0.5$ $\mu$m, we conclude from the previous calculations that: $\theta \sim 10^{-8}$ rad. It seems that these values are non unrealistic.

Let us note that the last form in $\theta \sim \alpha_{em} \frac{|E_0|^2}{E_{cr}^2} L/\lambda$ is very general for any experiment using properties linked to the vacuum birefringence
in intense EM fields. The intense field can be a static magnetic field, a static electric field or the EM field of a counter propagating wave as in the physics case discussed above. The important feature is that the prediction for a typical observable characterizing the birefringence (here $\theta$) will always be of the form of
Eq. (\ref{vacb}).

\subsubsection{Other studies of non-linear QED}
\label{TT}

\begin{figure}[bh!]
\centering
  \includegraphics[scale=0.5]{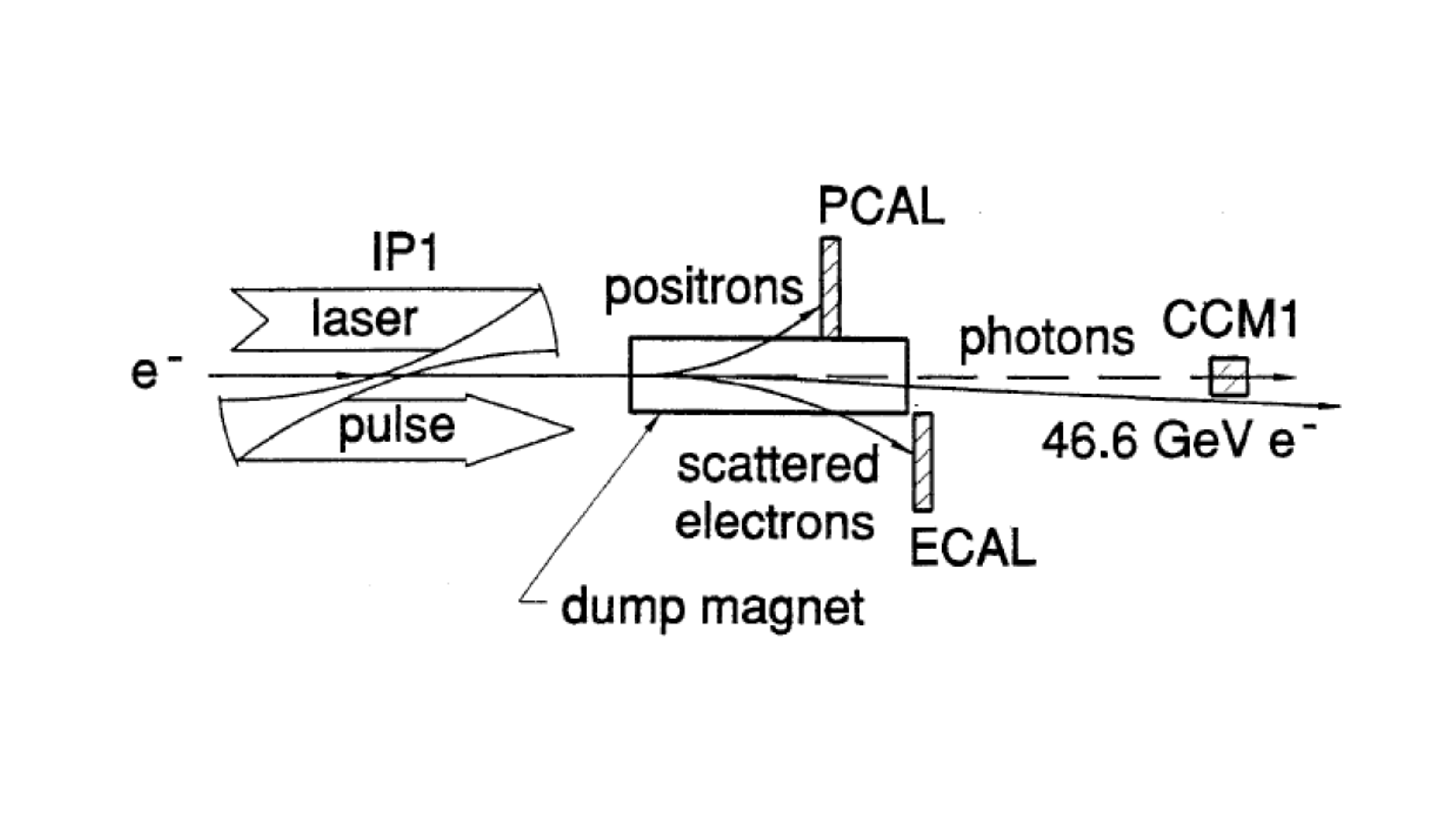}
  \vspace*{-1.6cm}
\caption{\label{slacfig1}
{\small Collision region layout of the SLAC E-144 experiment.
}
}
\end{figure}

The first collaboration to have reported a direct observation of   strong-field effects in QED is the
Experiment-144 (E-144) at the Stanford Linear Accelerator Center (SLAC) \cite{slac}.
In this experiment,
high intensity
laser photon (hereafter labeled as $\Omega_0$) pulses, of power $10^{12}$ W and two possible wavelengths in the optical domain $1054$ and $527$ nm, are produced at a rate available from few other laser
systems. These laser pulses are then brought into collision with high energy electron ($e^-$) bunches. The energy of the electron beam is $46.6$ GeV, a value
available only in the Final Focus Test Beam (FFTB) line of SLAC. The $\Omega_0$ pulses
are directed along a path that is approximately antiparallel to that of the
$e^-$ beam (see Fig. \ref{slacfig1}). This geometry exploits the extremely relativistic nature of the
electron beam. In particular, the Lorentz transformation of the electric field $E$ of the laser pulse in the laboratory frame 
yields, in the $e^-$ rest frame, a field strength on the order of  $10^{16}$ V/cm, an
amplitude that is otherwise unattainable in a laboratory setting.
Indeed, in the $e^-$ rest frame, the electric field of the laser beam is about
$\bar{E} = 2 \gamma_L E_{lab} \simeq 2 \ 10^5 \ E_{lab} \sim 10^{16}$ V/cm
 (where $\gamma_L$ is the Lorentz factor).

\begin{figure}[bh!]
\centering
  \includegraphics[scale=0.4]{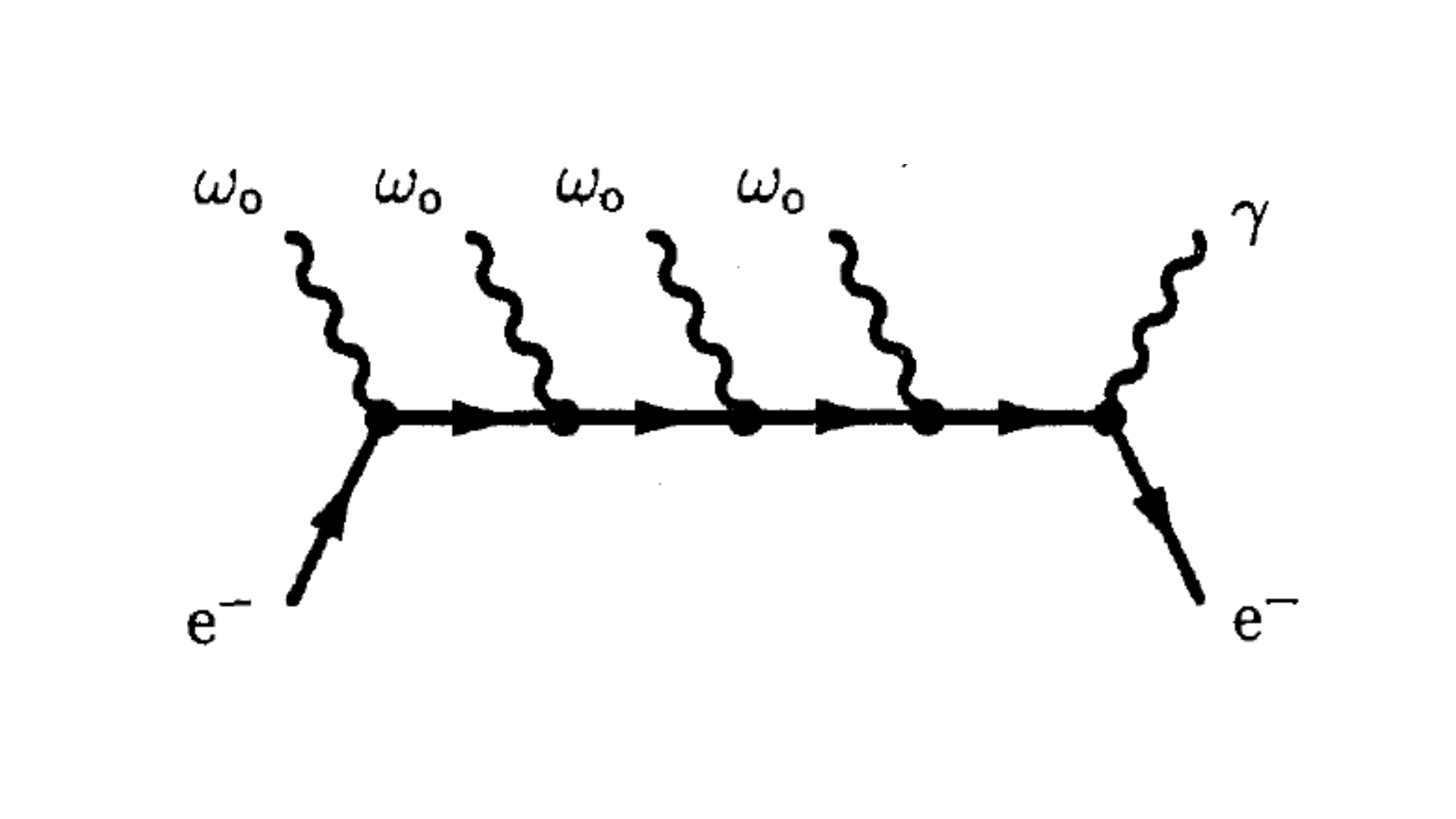}
  \vspace*{-1.5cm}
\caption{\label{slacfig2}
{\small Non-linear Compton scattering process.
}
}
\end{figure}

Then multiple photon effects (strong-field QED) are possible, like non-linear Compton scattering, in which an electron absorbs multiple photons
from the laser field, but radiates only a single (high energy) photon:
$$
e^- + n \omega_0 \rightarrow e^- + \gamma.
$$
In this reaction, $\omega_0$ represents a photon from the strong EM wave, $n \ge 1$ indicates the number of such photons absorbed
and $\gamma$ represents the (high energy) outgoing photon, see Fig. \ref{slacfig2}.
Within the E-144 configuration \cite{slac}, $n$ is shown to be above $3$. Also, the energy of the photon $\gamma$ can be easily computed.
In the rest frame of the electrons, we can write:
$$
{\bar E}_\gamma = \frac{{\bar E}_{\omega_0} m_e }{m_e  + {\bar E}_{\omega_0}(1-\cos(\theta^*))}.
$$
Where ${\bar E}_{\omega_0}= 2 \gamma E_{\omega_0}$ is the energy of the laser photons (in the rest frame of the electrons).
This gives the value $E_\gamma$ in the laboratory frame:
$$
E_\gamma = \frac{4 \gamma_L^2 m_e }{m_e  + 4\gamma_L E_{\omega_0}} E_{\omega_0}.
$$
Then $E_\gamma \simeq 30$ GeV \cite{slac}, which proves the statement above on the the energy of the outgoing photon.

An important dimensionless quantity can be defined as:
\be
\eta = \frac{e E_{rms}}{ \omega_0 m_e }.
\label{etaslac}
\ee
Here, $E_{rms}$ is the electric field value of the laser beam (in the laboratory frame), namely: 
$$
I_{laser} (\rm{W/cm^2})=  \epsilon_0 \ E_{rms}^2.
$$
With an intensity of the laser beam that can vary from $3 \ 10^{-16}$ W/cm$^2$ to $47 \ 10^{-16}$ W/cm$^2$
at E-144, depending on the focal area.
In the weak field regime of QED, $\eta<<1$ while in the strong-field regime it is expected that $\eta \sim 1$. 
With the parameters of  E-144 \cite{slac},  $\eta$ can be varied from $0.1$ to $0.4$ (see below).
We can relate $\eta$ to another dimensionless quantity that incorporates the critical field of QED:
\be
{\cal Y} = \frac{{\bar E}_{\omega_0}}{E_{cr}} = \frac{2 \gamma_L \omega_0}{m_e} \eta.
\label{formulaT}
\ee
Where we recall that ${\bar E}_{\omega_0}= 2 \gamma E_{\omega_0}$ is the energy of the laser photons (seen in the rest frame of the electrons).
This parameter can reach $0.2-0.3$ at E-144.

\begin{figure}[bh!]
\centering
   \hspace*{-6cm}
  \includegraphics[scale=0.9]{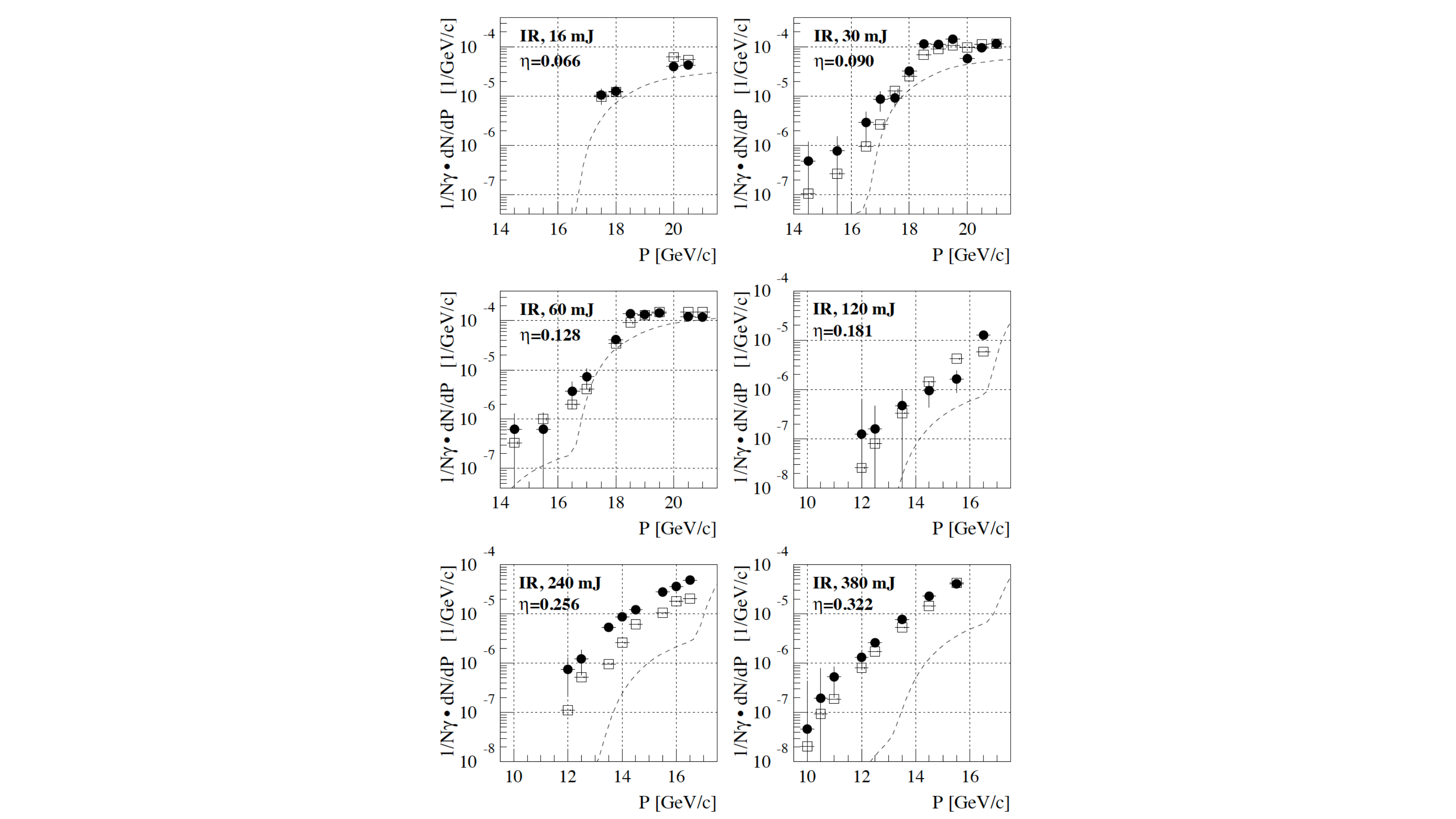}
  \vspace*{-0.3cm}
\caption{\label{slacfig2b}
{\small 
The yield of non-linearly scattered electrons
$\frac{1}{N_\gamma}\frac{dN}{dP}$ vs. momentum $P$ for six different circularly
polarized IR laser energies. The data are the solid circles with
vertical error bars corresponding to the statistical and reconstruction
errors added in quadrature \cite{slac}. The open boxes are the
simulation with error estimates indicated by the horizontal
and vertical lines. The effect of systematic uncertainty in the
laser intensity is not shown. The dashed line is the simulation
of plural (linear) Compton scattering.
}
}
\end{figure}

Let us describe the experimental observation and why this is consistent with the detection of non-linear Compton scattering and thus with the appearance of the strong-regime of QED. First, results for non-linear Compton scattering are best done in terms of rates rather than cross sections, as the later are not well defined for initial states involving multiple (unknown) laser photons.
We can formulate in a very concise form the rates for the reaction of non-linear Compton scattering with $n$ laser photon absorbed scattering (Fig. \ref{slacfig2}) as:
$$
P_n \simeq \eta^{2(n-1)} C.
$$
Where $C$ is the case for $1$ photon absorbed, thus linear Compton scattering. 
This relation is correct in the limit $\eta<<1$. 
The total scattering rate is therefore:
$$
P = \sum_n P_n.
$$
Also valid in the domain $\eta \sim 1$ upon the condition that the sum is not truncated.
Data have been recorded for circularly infrared IR ($\lambda=1053$ nm) and green ($\lambda=527$ nm) laser beams.
Then, non-linear Compton scattering can be studied by measuring the scattered electrons, as well as observing the forward high energy photons
(see Fig. \ref{slacfig2}).
Fig. \ref{slacfig2b} shows the differential yield $\frac{1}{N_\gamma}\frac{dN}{dP}$
 for electrons scattered from the IR 
laser beam at six different laser intensities
(similar results have been obtained for green laser beam).
The observed yield (Fig. \ref{slacfig2b}) is shown as a function of momentum
by the solid circles. The horizontal error bars give
the width of the corresponding momentum bin, and the
vertical bars include systematic errors in the reconstruction.
The simulation, including both non-linear Compton
scattering, and plural Compton scattering (several linear Compton scattering processes one after the other) is
shown by the open boxes. For each event, the simulation
incorporates the measured laser and electron beam parameters.
Importantly, a simulation that ignores non-linear Compton scattering
is shown by the dashed curve.
The difference between this prediction and the data is the experimental proof of the observation
of non-linear Compton scattering.

\begin{figure}[bh!]
\centering
  \includegraphics[scale=0.2]{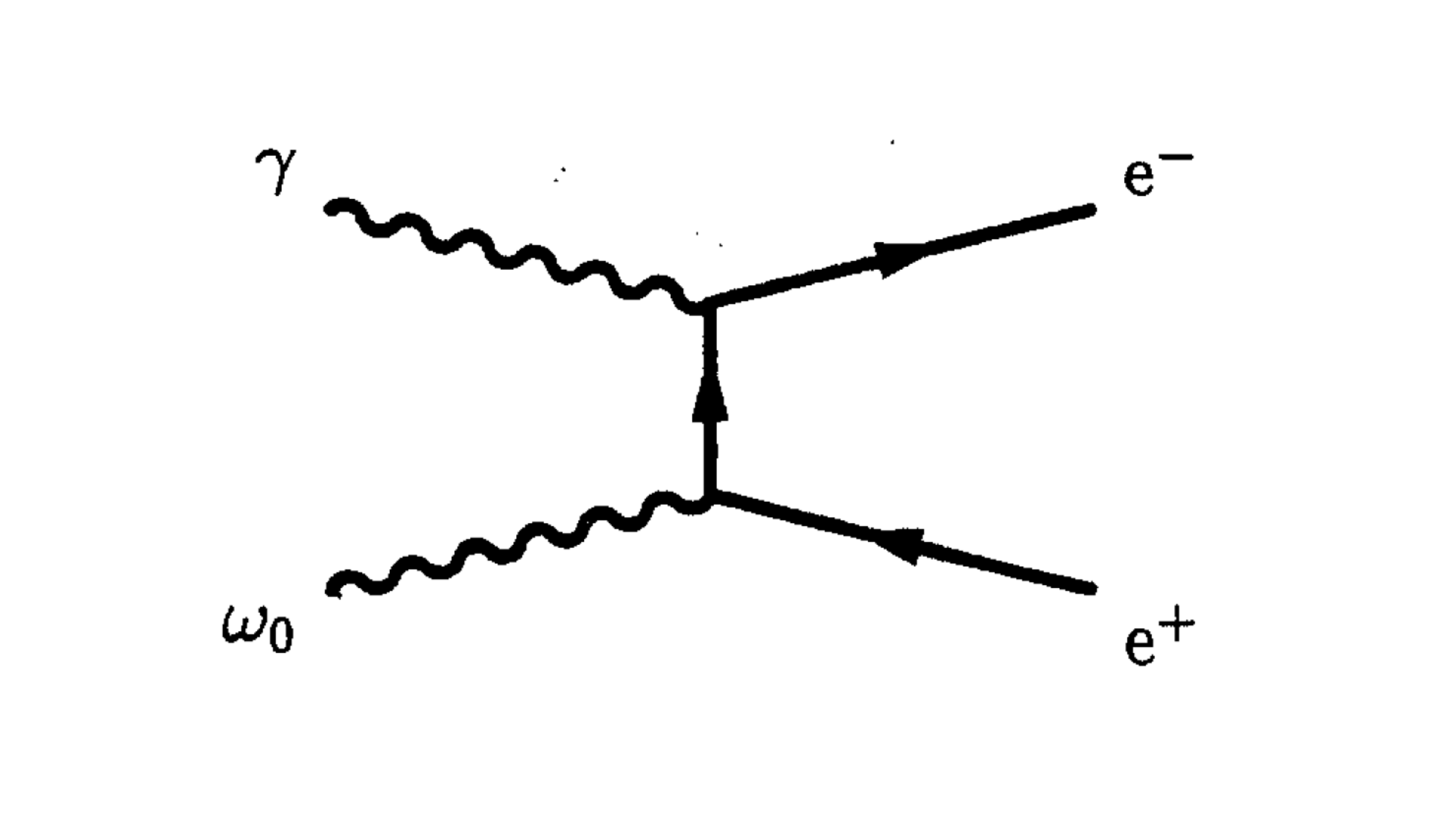}
  \includegraphics[scale=0.2]{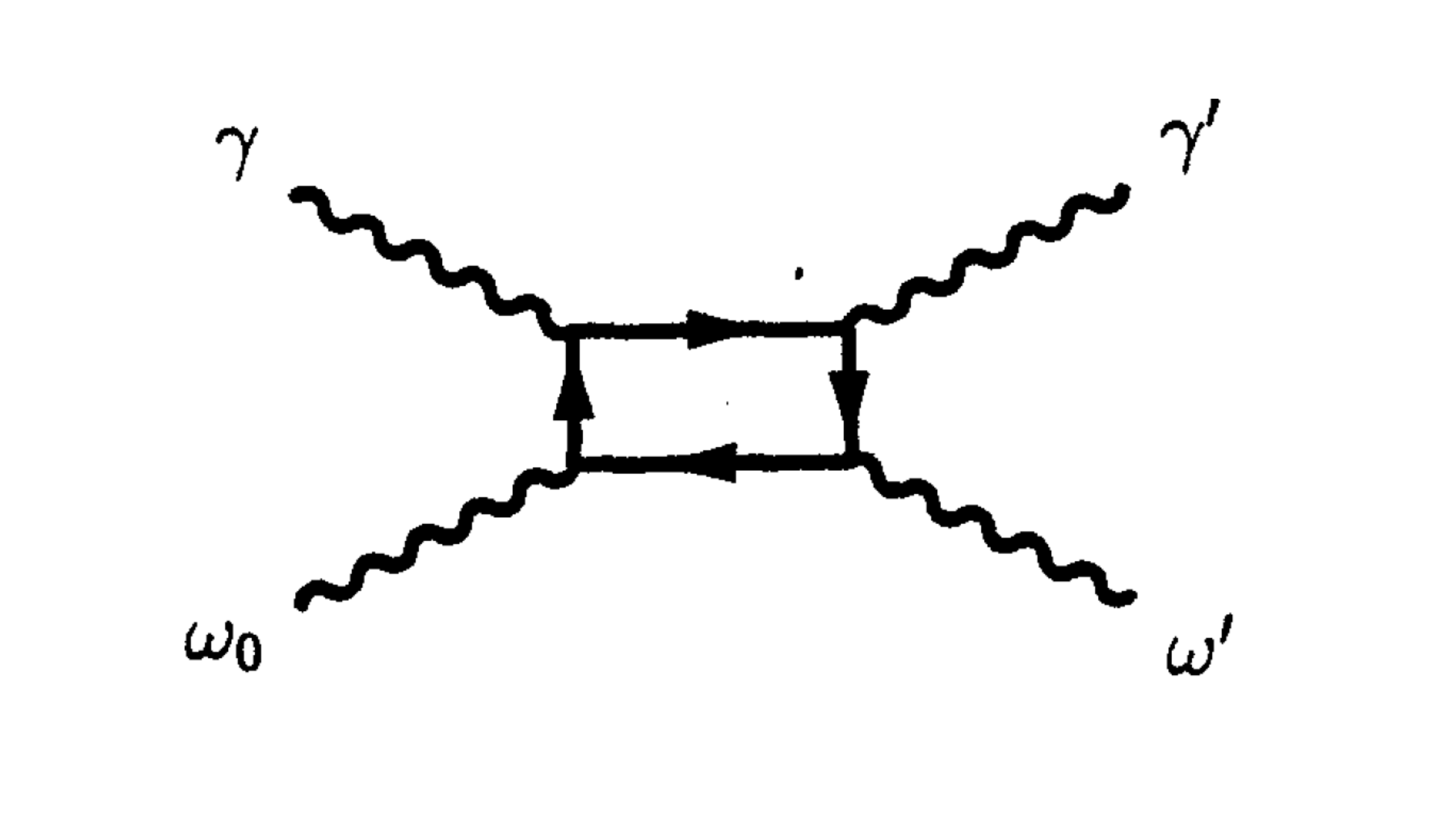}
  \vspace*{-0.5cm}
\caption{\label{slacfig2bb}
{\small Left: Inelastic photon-photon scattering process. Right: Elastic photon-photon scattering (light-by-light) process.
}
}
\end{figure}

\begin{figure}[bh!]
\centering
  \includegraphics[scale=0.5]{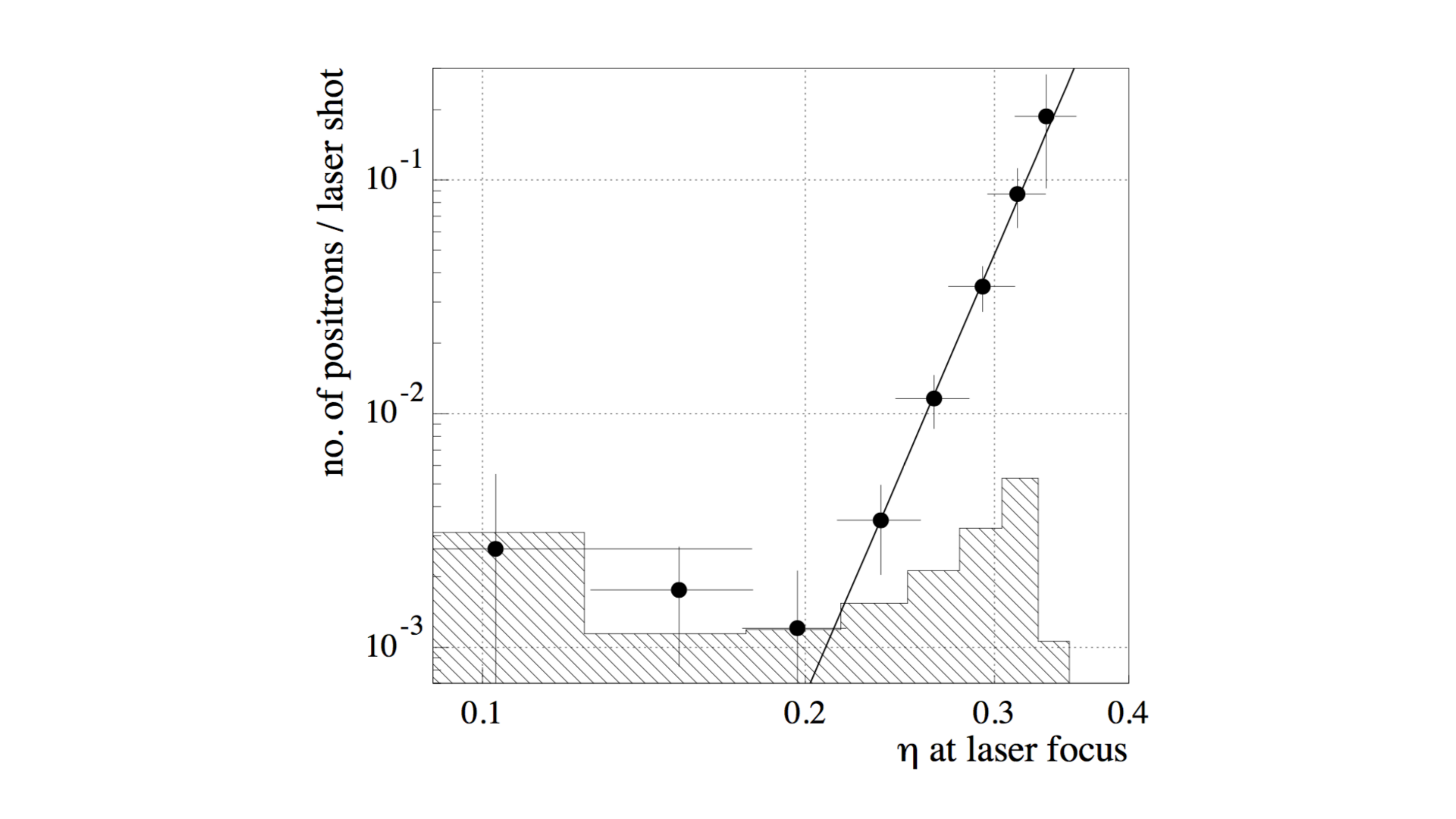}
  \vspace*{-0.5cm}
\caption{\label{slacfig4}
{\small Dependence of the positron rate production per laser shot
on the laser field-strength dimensionless quantity $\eta$ (see text).
The shaded distribution is the $95$ \% confidence limit 
on the residual background from showers of lost beam
particles after subtracting the laser-off positron rate \cite{slac}.
}
}
\end{figure}

\begin{figure}[bh!]
\centering
  \includegraphics[scale=0.5]{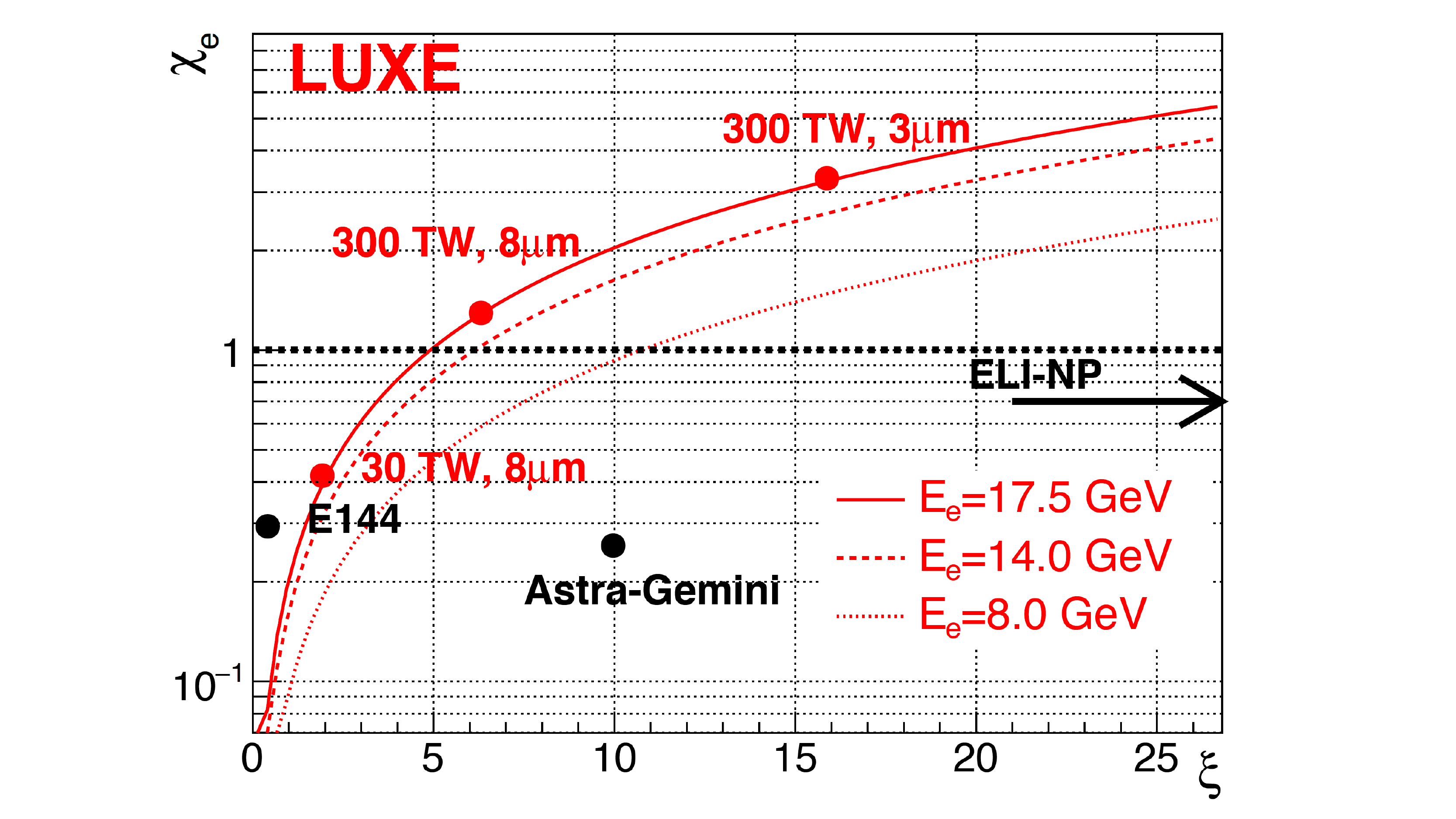}
    \vspace*{-0.1cm}
\caption{\label{luxe}
{\small 
The $(\chi_e,\xi) \equiv ({\cal Y},\eta))$ parameter space accessible for LUXE (red points) and E-144 (black point). The three red lines show the parameters accessible
to LUXE for three electron beam energies. Indicated are the parameters corresponding to the foreseen laser configurations of
LUXE: a $30$ TW laser focused to a FWHM of $8$ mm, or a $300$ TW laser focused to either a FWHM of $8$ mm or $3$ mm
\cite{luxe}.
}
}
\end{figure}

Once the non-linear Compton scattering reaction has occurred, the high energy photon can propagate through the laser field and thus can interact to produce an electron-positron pair (see Fig. \ref{slacfig2bb}-left) as
$\gamma + n\omega_0 \rightarrow e^+ e^-$ or light-by-light scattering (see Fig. \ref{slacfig2bb}-right).
The reaction $\gamma + n\omega_0 \rightarrow e^+ e^-$ is the one that is studied at E-144 in order to study non-linear effects in QED,
see Fig. \ref{slacfig4}. The key observation is to identify positrons produced after each laser pulse. Fig. \ref{slacfig4}
shows the yield $R_{e^+}$ of positrons per laser shot as a function of $\eta$. The line is a power law fit to the data and gives
$R_{e^+} \propto \eta^{2n}$ with $n=5.1 \pm 0.2 \pm 0.8$. Thus, the observed positron production rate is highly non-linear.
This is in good agreement with the theoretical rate of multi-photon reactions involving $n$ laser photons in $\eta^{2n}$.
and with the kinematical requirement that five photons are needed to produce a pair (electron-positron) near threshold.
The detailed simulation indicates that, on average, $1.5$ photons are absorbed from the laser field and $4.7$ in the reaction 
$\gamma + n\omega_0 \rightarrow e^+ e^-$, thus in agreement with the predicted value.
These results are therefore a first observation of inelastic photon-photon scattering with real photons.
Interestingly, a new experiment is being build at DESY with the idea to use the high-quality and
high-energy electron beam of the European XFEL and a powerful laser beam \cite{luxe}.
Indeed, as in the E-144 experiment,
high-energy electrons, accelerated by the European XFEL
linear accelerator, and high-energy photons, produced via Bremsstrahlung of those beam electrons, colliding with a laser
beam will give access to the strong regime of QED. A potential reach is shown in Fig. \ref{luxe}.
In the LUXE experiment, values up to $\eta=16$ and ${\cal Y} = 3$ are expected., thus clearly in the strong regime of QED.

\vfill
\clearpage
\newpage

\subsection{Axion like particles (ALPs) in light-by-light scattering}

\subsubsection{ALPs of masses of a few GeV}

\begin{figure}[!bh]
\centering
  \includegraphics[scale=0.5]{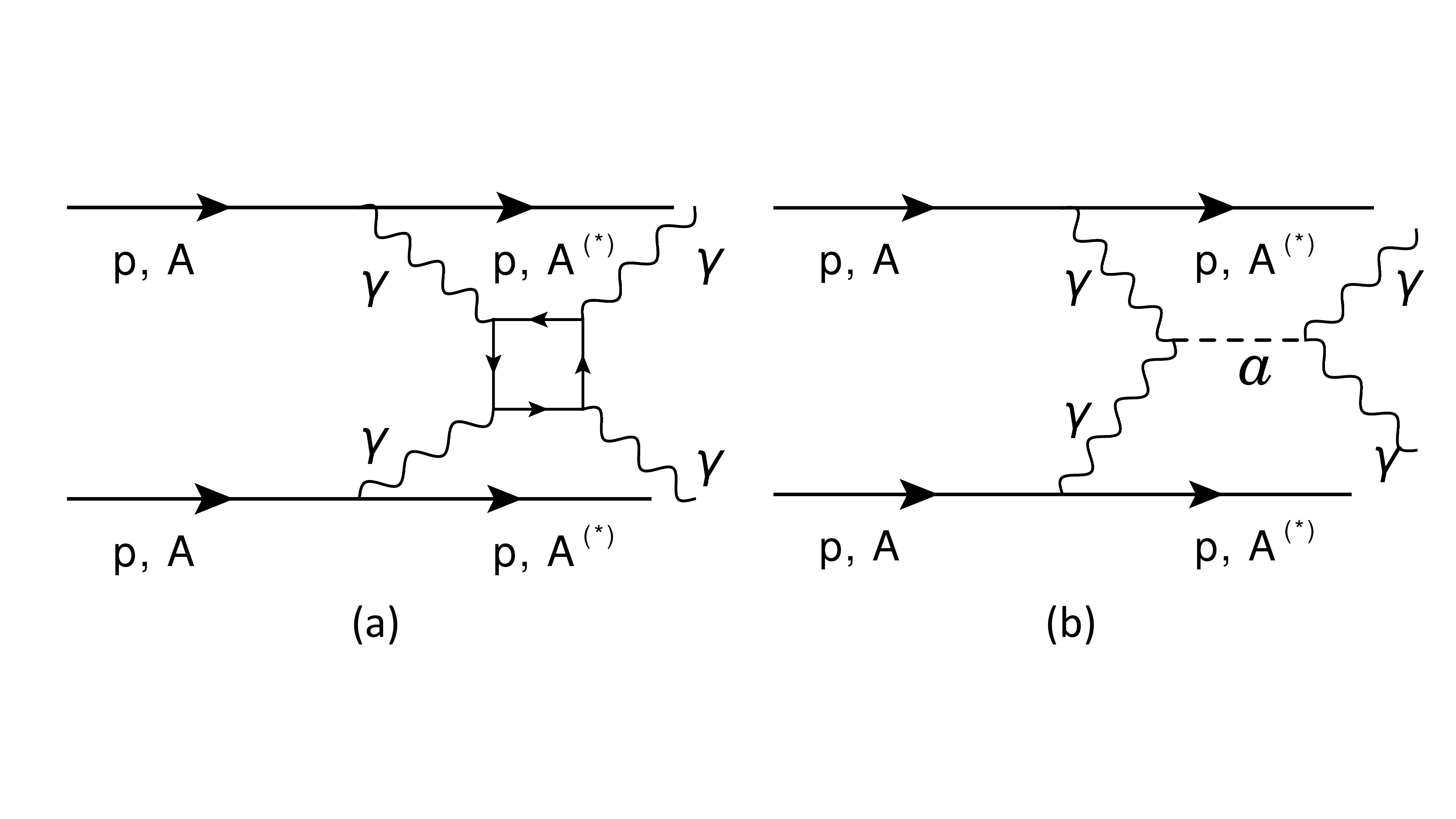}
\vspace*{-1.7cm}
\caption{\label{alpfig}
{\small
(a) Schematic diagram of light-by-light scattering in  proton-proton, proton-ion or ion-ion collisions. The box diagram includes quarks, charged leptons and $W$ boson contributions.
(b) Schematic diagram for central exclusive production of an axion-like (ALP) particle. Hereafter, we discuss only the $s$-channel exchange (as pictured). The field $a(.)$ could also be exchanged in the $t$-channel but would not contribute to the analysis presented in the text.}
}
\end{figure}

It is well known that central exclusive production of photon pairs in lead-lead collisions at the LHC,
namely $Pb Pb \rightarrow Pb Pb (\gamma \gamma) \rightarrow Pb Pb\gamma \gamma$,
has an interesting discovery potential for axion-like particles (ALP)
\cite{Baldenegro:2019whq,Bauer:2017ris,Jaeckel:2015jla,Baldenegro:2018hng,inan2,Knapen:2016moh}. Let us remind that
ALPs, for  masses accessible through this reaction at the LHC, are pseudo-scalars weakly coupled to Standard Model (SM) fields that 
appear in theories with spontaneously broken global, approximate, symmetries as pseudo Nambu-Goldstone bosons. For instance, ALPs appear in supersymmetric extensions of the SM or string theories \cite{Witten:1984dg,Conlon:2006tq,Svrcek:2006yi,Arvanitaki:2009fg}.
At the LHC,
ALPs  (below labeled as the field $a(.)$) of masses above a few GeV  could be produced as intermediate states in the reaction:
$\gamma \gamma \rightarrow a \rightarrow \gamma \gamma$ (see Fig. \ref{alpfig}),
thus modifying the photon-photon invariant mass distribution. 
The purpose of an experimental analysis would then be to detect these
potential tiny deviations in these invariant mass spectrum, knowing that the standard (light-by-light scattering) spectrum has itself some
non negligible irreductible uncertainties.

As in Ref. \cite{Baldenegro:2019whq},  we parameterize the photons-ALP interaction by a Lagrangian
density of the form:
\be
{\mathcal L}_{a \gamma\gamma} 
= \frac{4}{f} a \ \textbf{E} \cdot \textbf{B},
\label{alp}
\ee
where $a$ is the massive scalar ALP field (of mass $m_{a}$),
($\textbf{E}$, $\textbf{B}$) the electric and magnetic components of the EM field, 
 and $1/f$ the coupling of the interaction (the unit of $f$ is the energy).
This means in particular that the equation of motion of the field $a$ reads:
\be
(\partial_\mu \partial^\mu +m^2)a = -\frac{4}{f} \ \textbf{E} \cdot \textbf{B}.
\label{alp2}
\ee

\begin{figure}[!]
\centering
\includegraphics[scale=0.38]{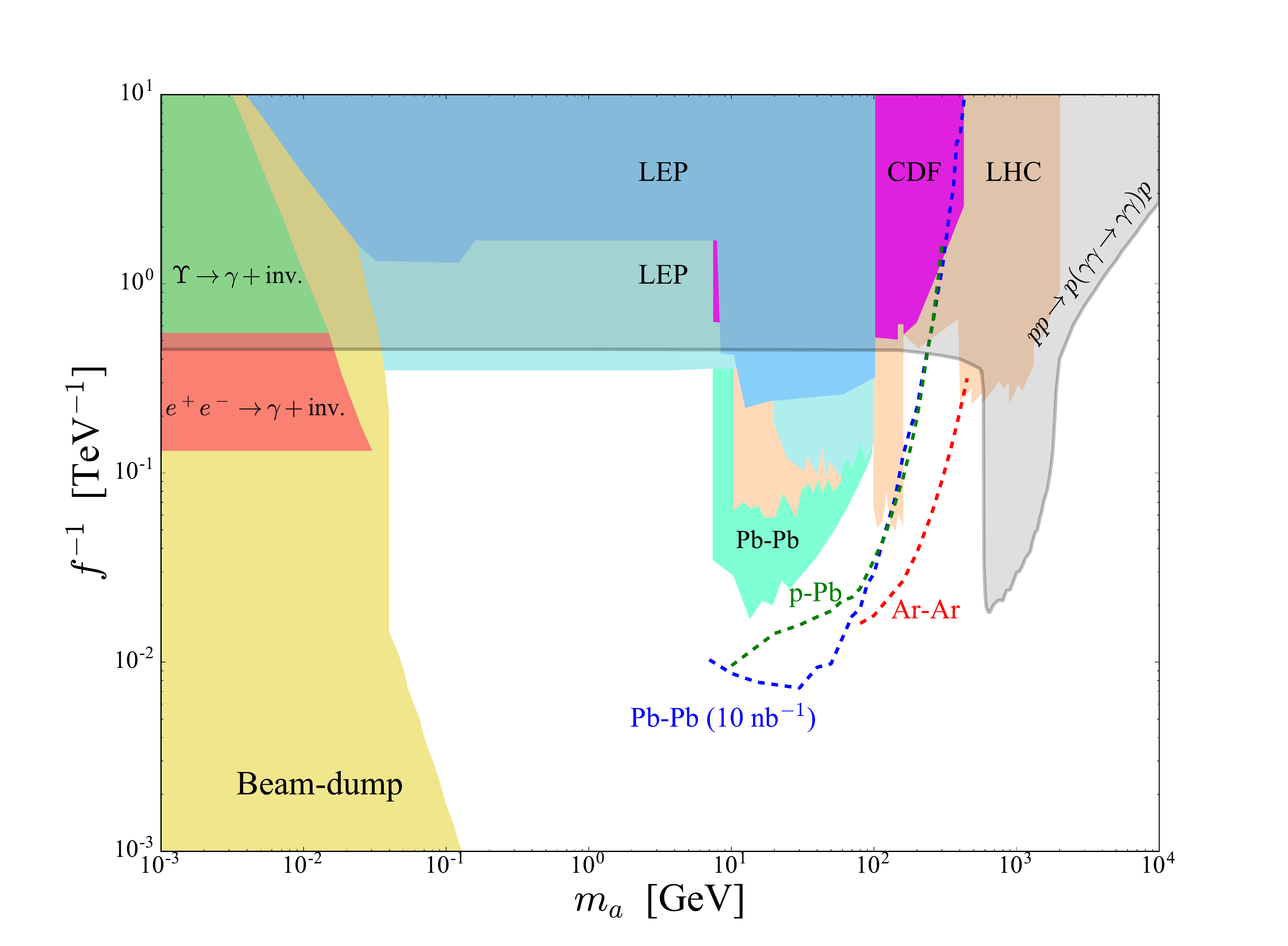}
\caption{
  {\small
  Exclusion limits on the ALP-photon coupling as a function of the ALP mass derived from several particle physics and astro-particle physics experiments (see Ref. \cite{Bauer:2017ris,Jaeckel:2015jla}).
  Projections for proton-proton collisions with the proton tagging technique are drawn from Ref. \cite{Baldenegro:2018hng,inan2}.
   The projections derived in Ref. \cite{Baldenegro:2019whq} for
$pPb$ collisions (at $8.16$ TeV per nucleon pair and for a luminosity of $5$~pb$^{-1}$),
$PbPb$ collisions 
(at $5.02$ TeV per nucleon pair and for a luminosity of $10$~nb$^{-1}$) 
and
 $ArAr$ collisions 
(at $7$ TeV per nucleon pair and for a luminosity of $3$~pb$^{-1}$)  are  shown as dotted lines, under the assumption that
   $\mathcal{B}(a \rightarrow\gamma\gamma)=1$ \cite{Baldenegro:2019whq}. 
}}
  \label{superTOTO}
\end{figure}

\begin{figure}[!]
\centering
\hspace*{-1.3cm}
\includegraphics[scale=0.6]{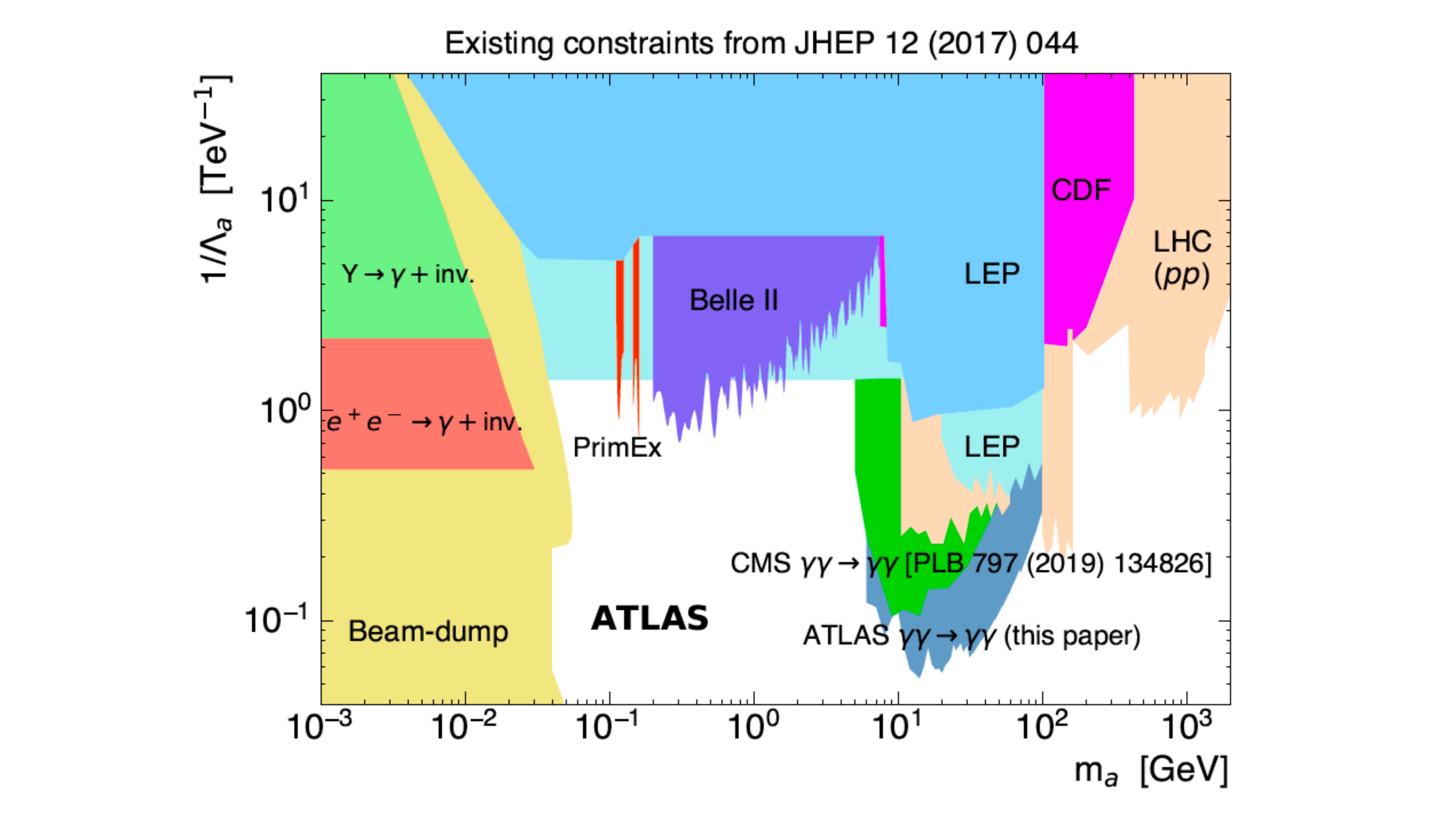}
\vspace*{-0.8cm}
\caption{
  {\small
    Exclusion limits on the ALP-photon coupling as a function of the ALP mass  
    as presented in Fig. \ref{superTOTO} including the most recent ATLAS 
    \cite{lbl1,lbl2,lbl3} and Belle results \cite{belle}.
}}
  \label{superTOTO2}
\end{figure}

In the high energy limit, 
for the narrow width approximation, in which we suppose that
$\sigma_{\gamma \gamma \rightarrow a \rightarrow \gamma \gamma}$ to be non-zero only when
the invariant mass of the two photons is equal to $m_{a}$ plus/minus the
width of the resonance, this cross section can be shown to be of the form:
$
\sigma_{\gamma \gamma \rightarrow a \rightarrow \gamma \gamma}
\propto \frac{1}{f^2} {\cal B}_{a \rightarrow \gamma \gamma}
$,
where ${\cal B}_{a \rightarrow \gamma \gamma}$ is the branching ratio of the
ALP into photons.
In the extraction of the limit for the coupling in $1/f$ (as a function of
$m_{a}$), we pose ${\cal B}_{a \rightarrow \gamma \gamma}=1$.
Indeed, we consider that ALPs predominantly couple to photons and have a negligible coupling strength 
to a photon and a $Z^0$ boson.
Then,
the invariant mass distribution can be  used as the discriminating variable, with bin widths
comparable to the expected resolution of a narrow
resonant signal and exclusion limits can be derived for the coupling $1/f$ 
 \cite{Baldenegro:2019whq}.  Results and presented in 
 Fig. \ref{superTOTO} and \ref{superTOTO2}.

\subsubsection{ALPs of masses of a few eV }
\label{sectiontoto}

We understand that in lead-lead collisions at the LHC, ALPs of masses of a few GeV could be accessible 
in light-by-light scattering
because the photons energies are also of a few GeV.
The principle can be extended at smaller energies using  the laser beam experiment described above using 
counter-propagating EM waves.
In this context, we can study how the results obtained above are  modified 
in the presence of ALPs, also considered as a pseudo-scalar field $a(.)$ 
of mass $m_a$  with a coupling to the EM fields of the form:
${\mathcal L}_{int}=- {\eta} \ a \textbf{E} \cdot \textbf{B}$,
where we have redefined $\eta \equiv 4/f$.
The coupling of the fields $a$, $\textbf{E}$ and $\textbf{B}$ is then encoded in two equations:
\begin{equation}
(\square +m_a^2)a = -\eta  \ \textbf{E} \cdot \textbf{B},
\label{phi1}
\end{equation}
\begin{equation}
\nabla \times \textbf{H} = \partial_t{\textbf{D}} + {\eta} \ [\textbf{E} \times 
\nabla a - \textbf{B} \frac{\partial a}{\partial t}].
\label{phi2}
\end{equation}

In Eq. (\ref{phi1}), we see that the right-hand side of the equation contains terms in $e^{\pm i(\omega_0+\omega)t}$ and terms in $e^{\pm i(\omega_0-\omega)t}$, which correspond to possible time dependence for the ALP field. 
Then, the equation of motion for $a$ (\ref{phi1}) gives:
\begin{equation}
(\square +m_a^2)a = 2\eta  e^{-i(\omega_0-\omega)t-i(\omega_0+\omega)z} \bar{E_0}E_{1,y}+
2\eta  e^{i(\omega_0+\omega)t+i(\omega_0-\omega)z} {E_0}\bar{E}_{1,y} + c.c.
\label{phi1b}
\end{equation}
Following this, we can write:
$$
a = a_0 e^{-i(\omega_0-\omega)t-i(\omega_0+\omega)z}
+a_0' e^{i(\omega_0+\omega)t+i(\omega_0-\omega)z}
 +c.c.
$$
with:
$$
a_0 = \frac{2\eta \bar{E_0}E_{1,y}}{4\omega_0 \omega +m_a^2}
\ \ \ \ \
a_0' = \frac{2\eta {E_0}\bar{E}_{1,y}}{-4\omega_0 \omega +m_a^2}
$$
Here, it is possible to inject this expression into equation (\ref{phi2})
and  derive the modified refractive index due to the presence of the ALP field $a$.
 We find the index along the $(x)$ axis is left unchanged while 
 $n_y$ is modified as:
\begin{equation}
n_y = 1+28 K  |E_0|^2+ \frac{ 4 \eta^2 m_a^2 |E_0|^2 }
{ m_a^4- (4\omega_0 \omega)^2}.
\label{nyalp}
\end{equation}
In the experimental conditions considered above, $\lambda=0.5$~~$\mu$m~$\sim 2.5$~eV, we obtain $(4\omega_0 \omega)^\frac{1}{2} \sim 5$ eV. Then, there is a resonant effect for ALPs of this mass, that will dominate the non-linear contribution in $K  |E_0|^2$.
For deep red of  $\lambda=0.7$~~$\mu$m~$\sim 1.75$~eV,
the resonance will be for $m_a=3.5$ eV.  This means 
that this experimental configuration is  interesting in order to obtain a  sensitivity (through a resonant effect) to ALP of masses of the order of eV, when there is the possibility to scan several values of laser wavelengths.
This strategy follows what has been developed for the LHC era: the search for ALPs as a resonant deviation in the light-by-light scattering.
In order to quantify this sensitivity, we need to compare the second term in equation (\ref{nyalp}), namely 
$\frac{ 4 \eta^2 m_a^2 |E_0|^2 }
{ m_a^4- (4\omega_0 \omega)^2}$
with 
the light-by-light term:
$28 K  |E_0|^2$. Thus, we need to compare:
 $\sqrt{K} \sim \frac{10^{-5} \mathrm{GeV}^{-1}}{\mathrm{eV}}$
 to 
 $\frac{\eta}{m_a} \cal{A}$, where $\cal{A}$ is the amplification factor due to the resonant effect: the better is the finesse of the cavities, the largest is $\cal{A}$. 

 The existing and potential limits in the search for ALPs of masses of the eV or much below
 \cite{Asztalos:2011bm,Wagner:2010mi,Anastassopoulos:2017ftl,DellaValle:2015xxa,Ehret:2010mh} are of the order
 $\frac{\eta}{m_a} \sim \frac{10^{-10} \mathrm{GeV}^{-1}}{\mathrm{eV}}$ for ALPs masses of order of the eV. Therefore, we need an amplification factor $\cal{A}$ of order $10^5$ so that one may obtain competitive sensitivity using this technique, which is an experimental challenge. An amplification factor of $10^3$ to $10^4$  is already feasible with high quality cavities where the laser beams propagate.
Also, using astrophysical probes, some prospects have been extended for ALP in the eV range  down to photon-ALP coupling of $\frac{\eta}{m_a} \sim \frac{10^{-12} \mathrm{GeV}^{-1}}{\mathrm{eV}}$ \cite{Ringwald:2012hr}. However, the analysis using light-by-light scattering stays a complementary approach, that could already provide first results in a near future
\cite{Sarazin:2016zer}.

\vfill
\newpage

\subsection{Other exotic intermediate states in light-by-light scattering}
 
In Fig. \ref{alpfig} (b), the exchanged field $a(.)$ have been considered to be a pseudo-scalar ALP (of spin $0$). However, nothing prevents to consider the exchange
of spin $2$ states with the condition that such fields would be acceptable in theory. Such fields can be considered: they are called the states of Kaluza-Klein (KK)
graviton, hereafter labeled as $g_{KK}$, have spin 2 and appear in large extra dimension (LED) extensions of the standard model
\cite{Witten:1984dg,KK1,KK}.
They can be exchanged in the $s$-channel (as in Fig. \ref{alpfig} (b)) or in the $t$-channel. In this context, only proton-proton collisions 
at the LHC make sense as the masses of $g_{KK}$
are expected to be of a few TeV (see below), with the consequence that the light-by-light $M_{\gamma \gamma}$ spectrum could be modified only at very large masses
typically above $500$ GeV. In this section, we use the previous experimental results to obtain some constraints on the graviton. This is not a trivial translation and that is the reason why we give some explanations on the ideas that bring to the results at the end.

All consistent string theories
necessarily includes extra dimensions. In order for such theories with extra dimensions
 not  to be in contradiction with the measurements  done in the (standard) four dimensions space-time, it is needed
to hide the existence of the extra dimensions. That's why this is usually assumed that these hypothetical extra dimensions are
 compactified. Then,  length scales
corresponding to the size of the extra dimensions can be defined and potentially detectable. If the size of the
extra dimensions is small, $R \sim 1/M_{Pl}$ where $M_{Pl}$ is the (standard) Planck mass, then 
the consequences of the extra dimensions are out of reach. Therefore, by making the size of the extra dimensions
very small, one can effectively hide these dimensions. 
But we can imagine that there exists large extra dimensions but still compatible with observations.
Let us briefly recall the basic ideas.
We assume that there are $n$ extra dimensions or radii of scale $R$. The principle is to match the coupling of the effective theory (here the SM)
to the fundamental parameters of the hypothetical fundamental theory of higher dimensions. 
First,
by comparing the Einstein-Hilbert action
in $4+n$ dimensions with its four dimensional expression, it can be shown that:
\be
M_{pl}^2 \sim R^n M_S^{n+2}.
\label{pl2}
\ee
Where $M_S$ is the fundamental Planck energy scale of the higher dimensional theory.
Indeed, the action in $4+n$ dimensions reads:
$$
S_{4+n} = -{M_S^{n+2}}V_n \int d^4x \sqrt{g^{(4)}} R^{(4)}.
$$
Here, $V_n \simeq (2\pi R)^n$ represents the volume of the extra dimensions space
and the integral is taken on the standard four dimensions space-time.
This relation has to be matched with the action in four dimensions:
$$
S_{4} = -M_{Pl}^2 \int d^4x \sqrt{g^{(4)}} R^{(4)}.
$$
Which concludes.
Then, if we assume in addition that the SM fields also live in the extra dimensions, another relation can be obtained.
Indeed the SM action that we take only for gauge fields for simplicity can be written as:
$$
S^{sm}_{4+n} = -\frac{1}{4g_*^2} V_n \int d^4x  \sqrt{g^{(4)}} F_{\mu \nu} F^{\mu \nu}
$$
Where the indices $(\mu,\nu)$ run over $0$ to $3$ (standard space-time).
This implies, after matching the gauge coupling with the four dimensions physics case:
 $R \sim 1/M_{Pl}$. In this case, any effect of extra dimensions is out of reach.
However, we can also assume that the SM fields are localized in a brane, here in the four dimensions (standard) space-time.
In this case, only gravity propagates in the higher dimensions. We can also consider a reasonable value for $M_S \sim 1$ TeV.
In this view, the largeness of the four dimensional Planck scale $M_{pl}$
could be attributed to the existence of extra dimensions of volume $R^n$.
If $R$ is large enough to make $M_S$ on the order of the electroweak symmetry breaking scale ($\sim 1$ TeV), the hierarchy problem
(gravity is exceedingly weak compared to other forces)
would also be naturally solved. For example, with $M_S \sim 1$ TeV, we obtain from Eq. (\ref{pl2}) a set of possible couples ($n, R$): $n=1$ for $R \sim 10^{13}$ cm, which is  ruled out, 
$n=2$ for $R \sim 1$ mm, also ruled out by dedicated experiments, or $n \ge 3$ for $R < 1$ nm (but still $R>>1/M_{Pl}$), which lets the opportunity to detect graviton signal 
at colliders like the LHC. 

The next problem is how to couple graviton to SM fields (like photons).
It can be shown that the graviton field corresponds to the fluctuation of metric around the standard Minkowski $\eta_{\mu\nu}$
\cite{KK1,KK,Fichet:2014uka,inan}. Then, the action that describes the coupling is of the form:
\be
S_{int} \simeq  \frac{1}{M_S^{n/2+1}} \int d^4 x T^{\mu\nu} h_{\mu\nu}(x_\mu).
\label{eh2}
\ee
Here $h_{\mu\nu}$ represents the graviton field, $T^{\mu\nu}$ is the energy-momentum tensor and the 
integral is taken on the four dimensions (standard) space-time.
The coordinates $x_\mu$ are also labelling the  four dimensions space-time.
Let us recall that we have assumed the SM fields are living on this four dimensions space-time.
Also, $T^{\mu\nu}$ is simply a re-writting of the standard model action as:
\be
\sqrt{g} T^{\mu\nu} = \frac{\delta S^{sm}}{\delta g_{\mu \nu}}.
\label{eh3}
\ee
Here $g_{\mu \nu}$ is the entire metric tensor including the graviton fluctuations:
\be
g_{\mu \nu} = \eta_{\mu \nu} + \frac{1}{M_S^{n/2+1}}  h_{\mu \nu} (x_\mu).
\label{eh4}
\ee
In particular Eq. (\ref{eh2}), with Eq. (\ref{eh3}) and Eq. (\ref{eh4}), takes a simple form in some models \cite{KK1,KK,Fichet:2014uka,inan}:
$$
{\cal L}_{int} = \frac{1}{f_2} h^{\mu\nu} \left( (-F_{\mu\rho} F^\rho_\nu) + \eta_{\mu\nu} F_{\rho\lambda}F^{\rho\lambda}/4 \right).
$$
With  $1/f_2$ a coupling factor in units of 1/energy$^4$.
From this expression, it is possible to extract the contribution of the denumerable set of $g_{KK}$
to light-by-light scattering. We obtain:
\be
\mathcal{L}_{4\gamma}= 
= -\frac{\kappa^2}{64 {\tilde k}^4} F_{\mu\nu}F^{\mu\nu}F_{\rho\sigma}F^{\rho\sigma}
+\frac{\kappa^2}{16 {\tilde k}^4} F_{\mu\nu}F^{\nu\rho}F_{\rho\lambda}F^{\lambda\mu},
\label{zetas2}
\ee
where $\kappa$ is the coupling strength of the order of unity in the form given above and ${\tilde k}$ is related to
the extra dimensional curvature and gives the mass of the first mode for the $g_{KK}$ ($m_{KK}$).
It can be shown that ${\tilde k} \simeq m_{KK}/3.8$ \cite{KK1,KK,Fichet:2014uka,inan}.
Interestingly, the interaction described by the Lagrangian density (\ref{zetas2}) can be transformed into a modification for
the differential cross section of the light-by-light scattering as:
\be
  \frac{d\sigma}{d\Omega}
  =\frac{a}{16 \pi^2\,s}(s^2+t^2+st)^2
   \frac{\kappa^4}{{\tilde k}^8},
  \label{xsec2}
 \ee
where $s$, $t$ are the usual Mandelstam variables and $a$ is a constant $a=1.56 \ 10^{-2}$.
The key point here is that the contribution of Eq. (\ref{xsec2}) modifies the light-by-light cross section
only for large invariant masses of the outgoing photon-photon pair $M_{\gamma\gamma}$, in a kinematic domain where the
standard theory prediction is negligible. For example, in lead-lead collisions at $5.02$ TeV (center of mass energy of the collision per nucleon pair) described above
(section \ref{lblsec}),
the contribution of Eq. (\ref{xsec2}) is non zero for $M_{\gamma\gamma} > 30$ GeV. 
From the experimental results presented in section \ref{lblsec}, we can use Eq. (\ref{xsec2}) in order to extract 
a 95\% CL limit for the upper value of $m_{KK}$. We find: $m_{KK} < 5.1$ TeV at $95$ \% CL. This result is competitive with previous results using other 
methods \cite{Aaboud:2017yyg}.
It is  interesting to put this result in perspective with cosmological constraints.
For simplicity, we assume that both the SM and the dark matter (DM) fields are localized in the same
four-dimensional brane (the standard space-time), and by definiteness we consider real scalar DM.
In cosmology, we need to consider the radion (scalar field) also.
Besides the interaction through KK gravitons, we also take into
account that the SM and DM fields can interact with the radion field. 
Then, it is possible to obtain the correlation between the effective Planck scale and the mass of the first KK mode
in a given cosmological scenario \cite{Bernal:2020fvw}.

We can extend this derivation by the observation that  Eq. (\ref{zetas2}) contains the most general dimension 8 operators that 
are needed for four-photons interaction. Thus, we can write the same relation but with another view:
\be
\mathcal{L}_{4\gamma}= 
\zeta_1 F_{\mu\nu}F^{\mu\nu}F_{\rho\sigma}F^{\rho\sigma}
+\zeta_2 F_{\mu\nu}F^{\nu\rho}F_{\rho\lambda}F^{\lambda\mu}.
\label{zetas}
\ee
where $\zeta_1$ and $\zeta_2$ are parameters of units 1/GeV$^4$. Then, this Eq (\ref{zetas}) the most general effective
Lagrangian density that describes light-by-light scattering and any possible theory would lead to a definite expression of the
parameters $\zeta_1$ and $\zeta_2$, in the same way as it has been done above the the KK gravitons.
Here also, the contribution of Eq (\ref{zetas}) to light-by-light scattering would manifest at large $M_{\gamma\gamma}$ and we can thus
consider this form (Eq (\ref{zetas})) and use the results of section \ref{lblsec} in order to extract upper limits on the general parameters
$\zeta_1$ and $\zeta_2$.

\section{A few perspectives and prospects}

\begin{figure}[!bp]
\centering
\includegraphics[width=0.7\textwidth]{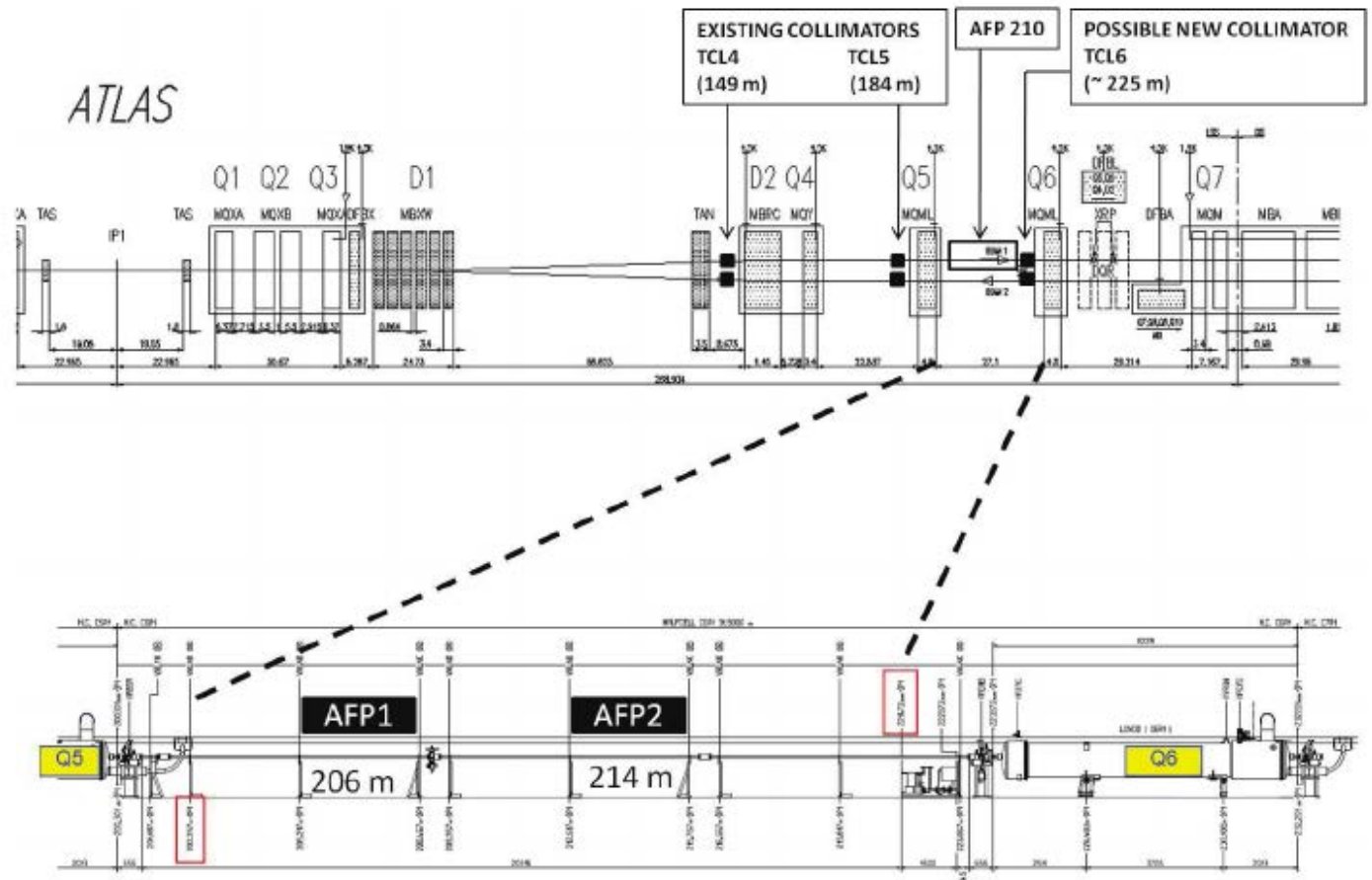}
\caption{\label{fig:afp_schematic} 
{\small Schematic diagram of ATLAS Forward Proton (right arm only). Figure extracted from Ref.~\cite{afp}.}}
\end{figure}

\begin{figure}[!bp]
\centering
\includegraphics[width=0.8\textwidth]{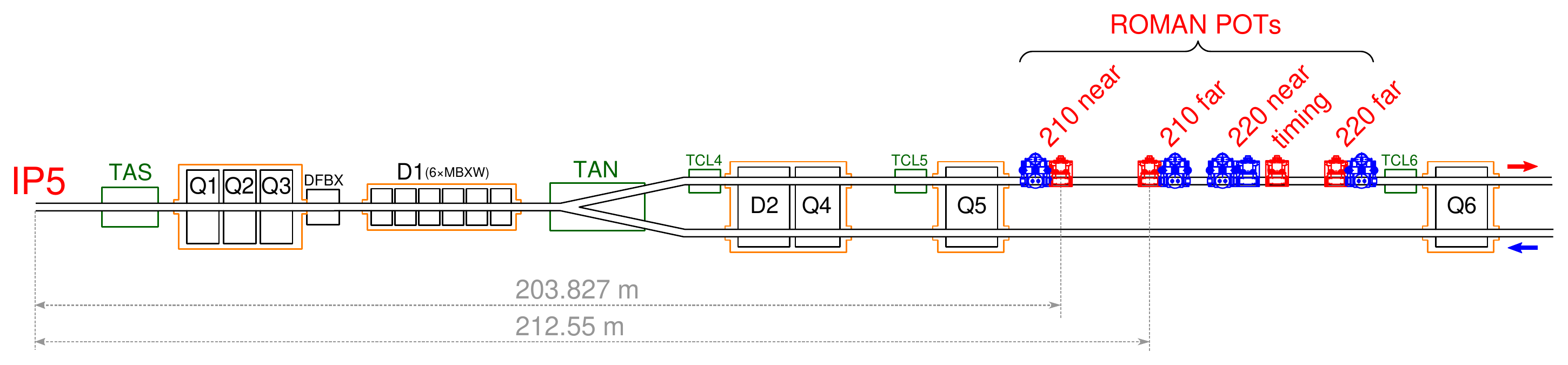}
\caption{\label{fig:pps_schematic} 
{\small Schematic diagram of the CMS-TOTEM PPS (right arm only). Figure extracted from Ref.~\cite{cmspps}.}}
\end{figure}

In proton-proton collisions, 
there is a clear experimental possibility in order to increase the precision of the above measurements.
Indeed the detection of the outgoing scattered proton that survives the quasi-real photon exchange
is already a realistic technology both in ATLAS and CMS experiments. 
Then, the proton that survives this interaction, owing to its small momentum loss, can be separated from other protons in the beam envelope by using the CERN LHC magnets around the interaction point of CMS or ATLAS. The protons, once separated from the beam, can be directly reconstructed with near-beam tracking detectors hosted in Roman pot (RP) detectors. The latter are located at 200 m (210 m) from the ATLAS IP (CMS IP). The deflection of the proton depends on the magnitude of the energy loss suffered, and also on the emission angle at the IP. The chosen RP locations are selected due to the available space, easy site-access, and because the proton kinematics allow so.  The proton spectrometer of ATLAS is called ATLAS Forward Proton (AFP) \cite{afp}, whereas the CMS and TOTEM proton spectrometer is called Precision Proton Spectrometer (PPS) \cite{cmspps}. Schematic diagrams of the ATLAS and CMS RP systems are shown in Figs. \ref{fig:afp_schematic} and \ref{fig:pps_schematic} respectively.

\begin{figure}[!bp]
\centering
\includegraphics[width=.4\textwidth]{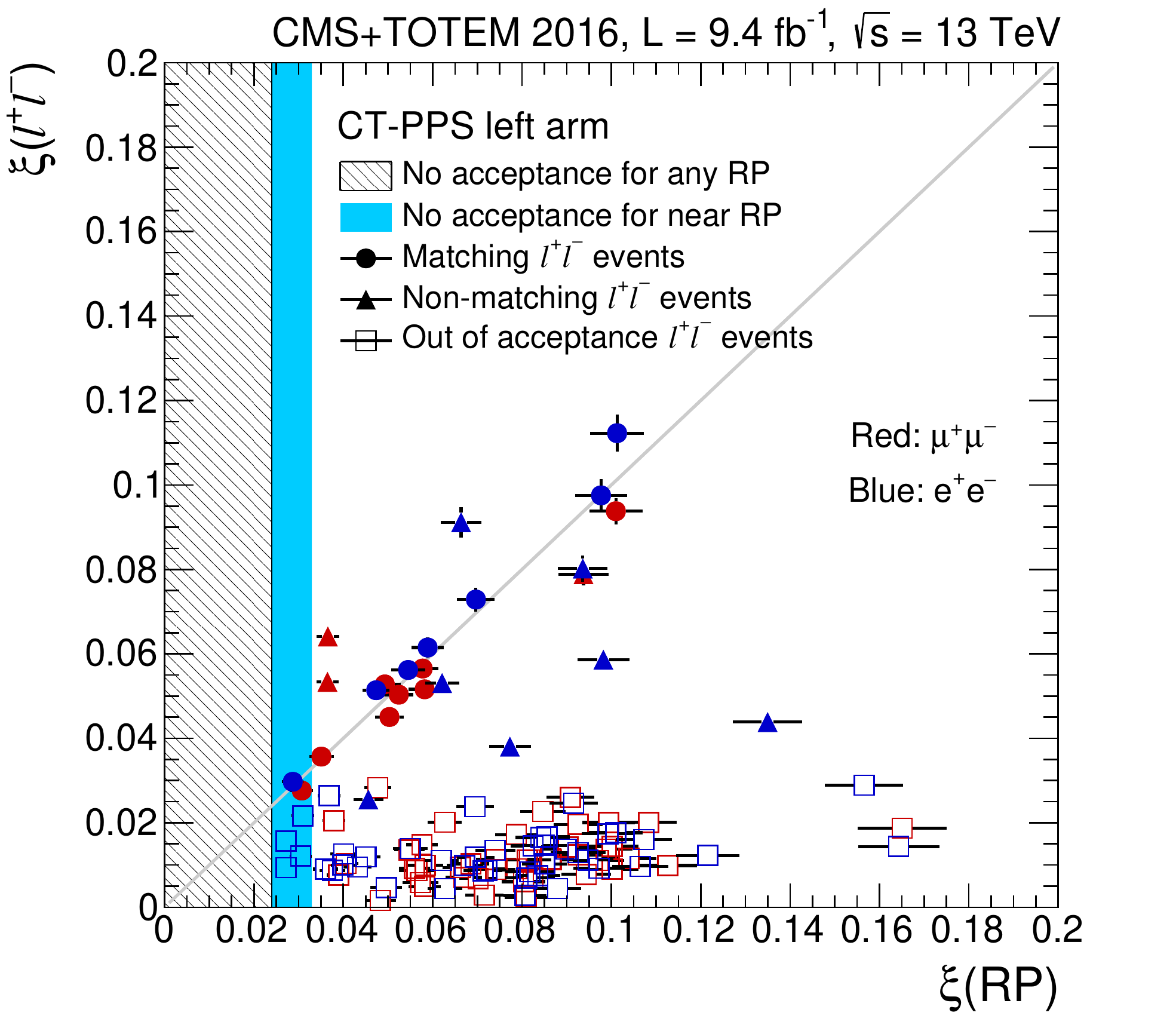}
\includegraphics[width=.4\textwidth]{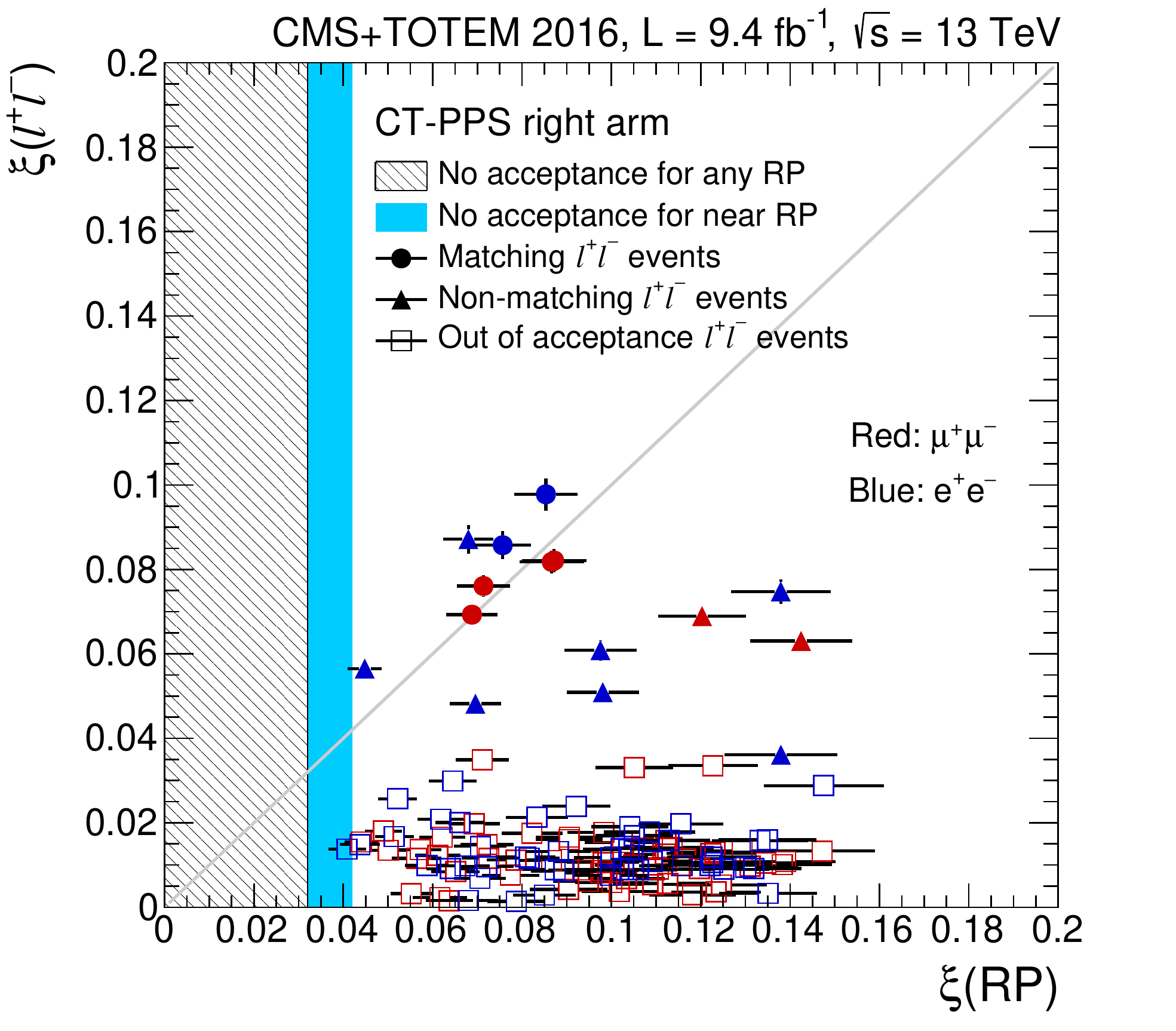}
\caption{\label{fig:xi_correlations_PPS} 
{\small Correlations of the proton fractional momentum loss reconstructed with the di-lepton system  
and with the RP detectors (in sector 45 and 56), which correspond to a proton scattered towards positive and negative pseudo-rapidities, respectively. 
Central exclusive di-lepton candidates satisfy nearly correspond.}}
\end{figure}

\begin{figure}[!thbp]
\centering
\includegraphics[width=0.8\textwidth]{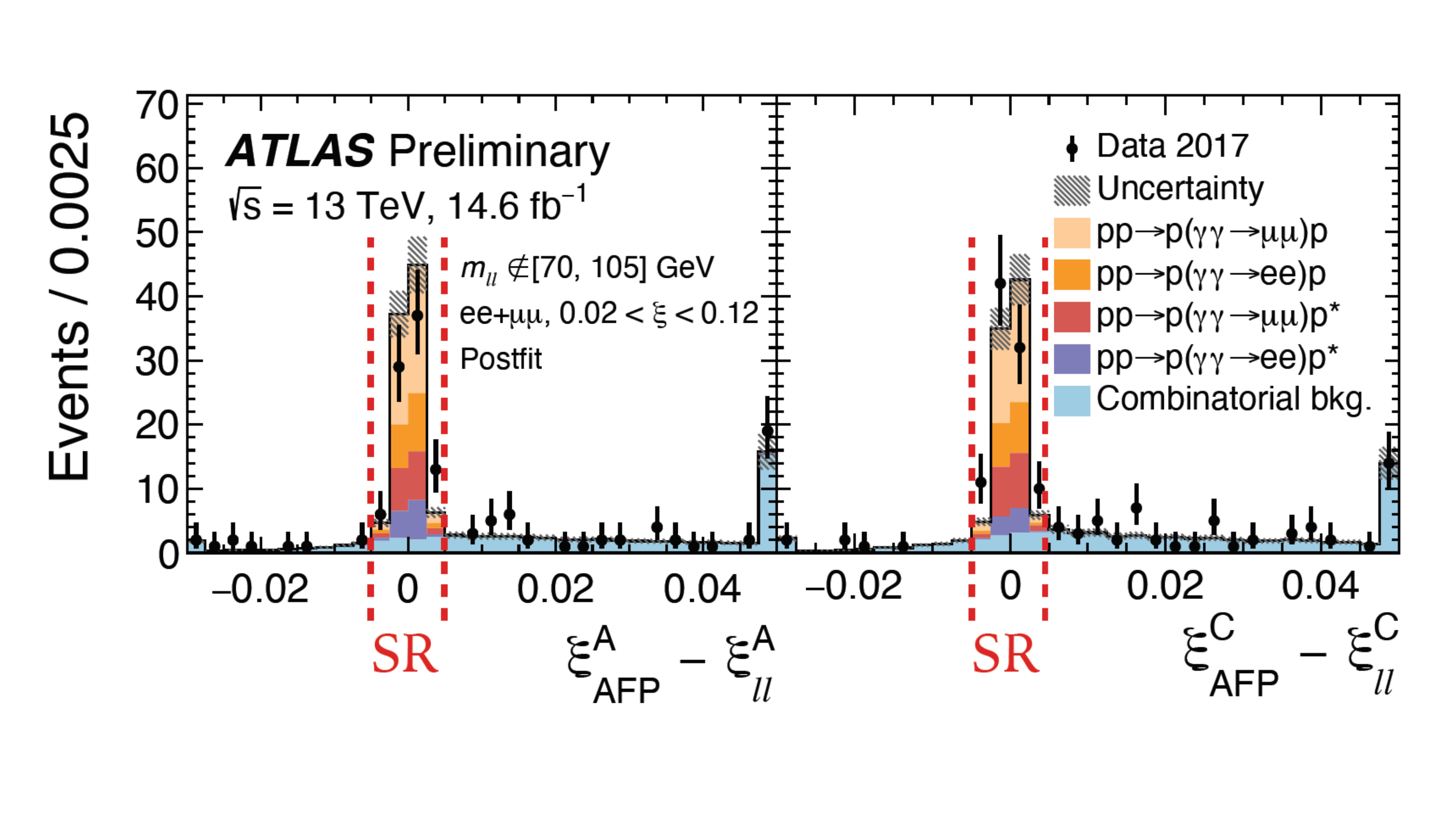}
\vspace*{-1cm}
\caption{\label{afp20} 
{\small 
Distributions of the difference of the momentum loss obtained using the centrally measured lepton pairs or protons in the RP detectors
(left) side A and (right) side C
($p^*$
denotes a dissociating proton).}}
\end{figure}

In particular, the CMS and TOTEM Collaborations have presented the physics results of (semi)-exclusive production of lepton pairs in proton-proton collisions at $13$ TeV, where at least one of the protons were tagged with the PPS system \cite{cmspps}. For an integrated luminosity of 9.4 fb$^{-1}$, a total of 12 $\mu^+\mu^-$ and 8 $e^+e^-$ pairs with at least one proton with matching kinematics are found. The dilepton pair has invariant mass of more than $110$ GeV, and the estimated number of background events is 1.49 $\pm$ 0.07 (stat) $\pm$ 0.53 (syst) and 2.36 $\pm$ 0.09 (stat) $\pm$ 0.47 (syst) for the di-muon and di-electron channels, respectively. The fractional momentum loss of the proton $\xi(\mathrm{RP})$ has values between 2\% and 10\% for candidate events. The kinematic correlation of candidate events between the central di-lepton system and the forward proton(s) is shown in Fig.~\ref{fig:xi_correlations_PPS}. The study proves the feasibility of operating a set of RP detectors during standard luminosity runs at the CERN LHC, which can be used to study photon-exchange interactions with very low production rates. Similar results have been obtained by the ATLAS experiment using the same technology \cite{afp}. Using proton-proton collisions recorded in 2017 with a center of mass energy of $13$ TeV, corresponding to an integrated luminosity of $14.6$ fb$^{-1}$, cross sections in the fiducial detector acceptance
are measured to be: $\sigma_{ee} = 11.0 \pm 2,6 \pm 1.2 \pm 0.3$ fb and $\sigma_{\mu \mu} = 7.2 \pm 1.6 \pm 0.9 \pm 0.2$ fb in the di-electron and di-muon channels respectively.
Fig. \ref{afp20}  shows the difference between the momentum loss measured from the di-lepton pairs or with the RP detectors. This figure is equivalent to
Fig. \ref{fig:xi_correlations_PPS} (for CMS). 

\begin{figure}[tbp]
\centering
\includegraphics[width=0.5\textwidth]{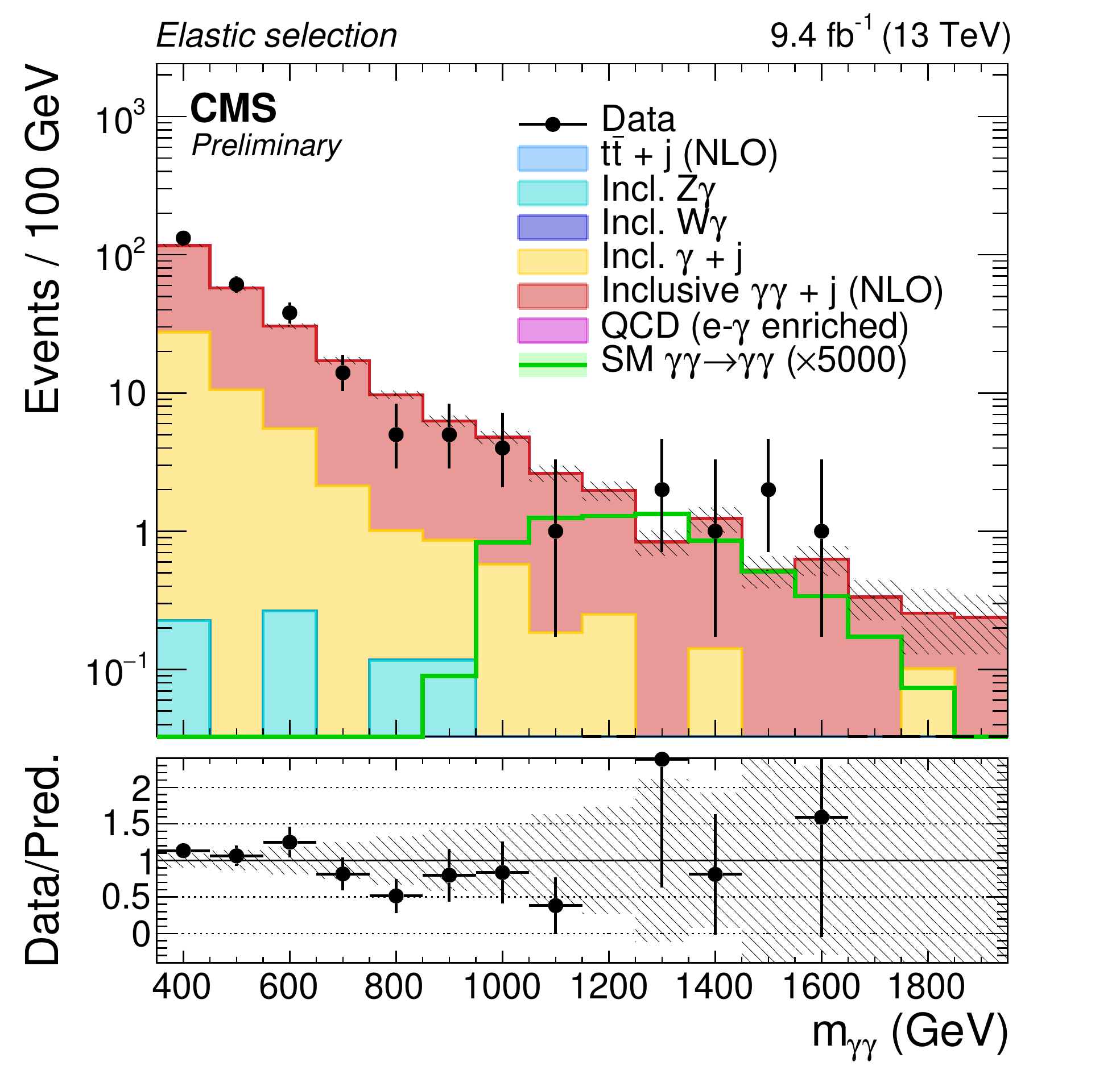}
\caption{\label{fig:diphoton_mass} 
{\small Invariant mass distribution of photon pairs for events satisfying the selection criteria up to $1-|\Delta\phi_{\gamma\gamma}/\pi|<0.005$, as described in the text. The filled histograms represent the MC predictions, and the circle markers represent the data. Figure extracted from Ref.~\cite{PPS_diphoton}.}}
\end{figure}

\begin{figure}[tbp]
\centering
\includegraphics[width=0.4\textwidth]{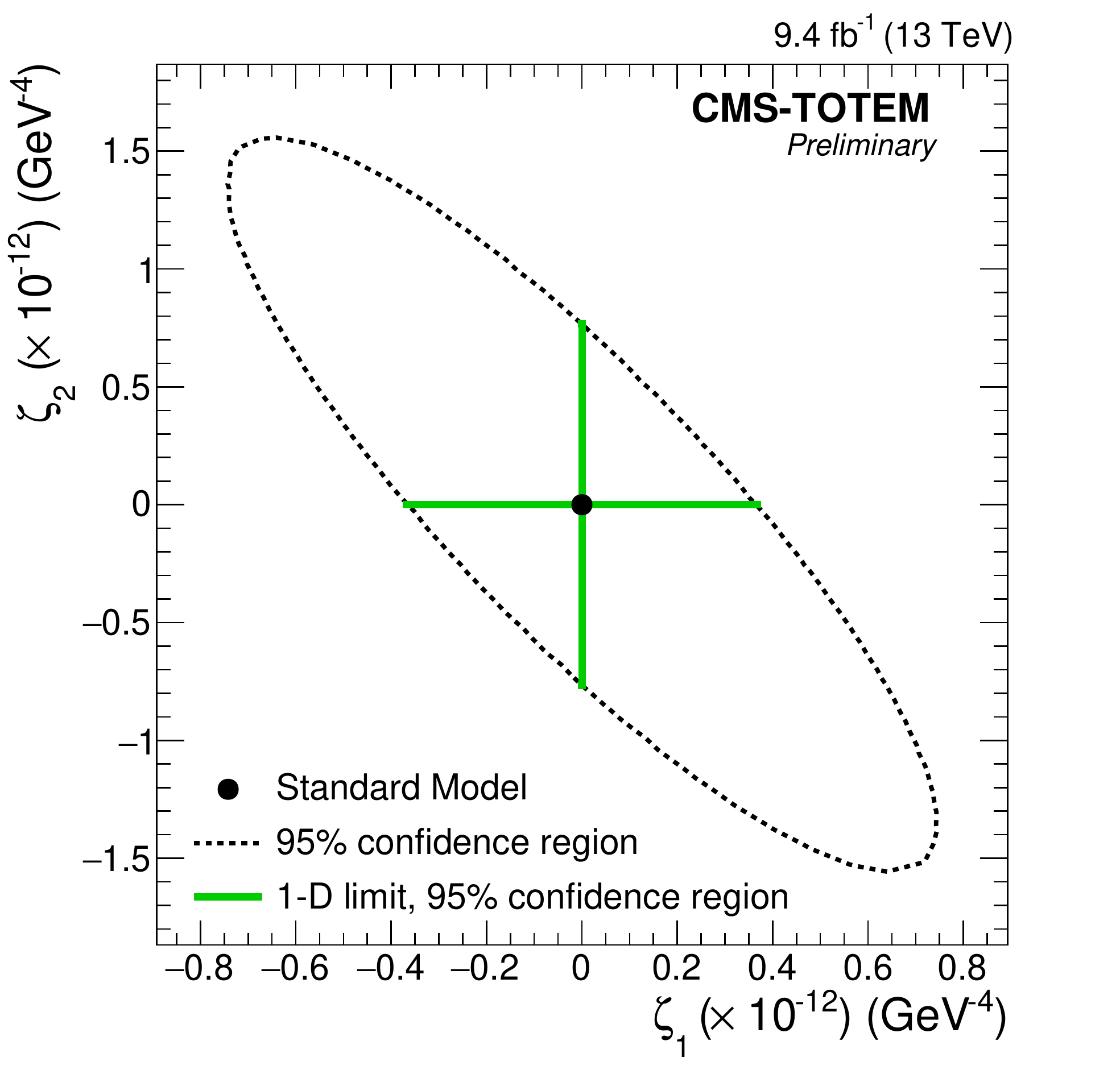}
\caption{\label{fig:couplings_4photon_PPS} 
{\small Two-dimensional limits on anomalous four-photon coupling derived from the search of high-mass light-by-light scattering. The limits are shown in terms of the dimension-8 coupling parameters $\zeta_{1,2}$. Figure extracted from Ref.~\cite{PPS_diphoton}.}}
\end{figure}

In addition, recently, the CMS and TOTEM Collaborations have reported a preliminary search for exclusive two-photon production via photon exchange in proton-proton collisions using intact protons~\cite{PPS_diphoton}. The data correspond to an integrated luminosity of 9.4 fb$^{-1}$ collected in 2016 at 13 TeV, when the LHC operated at standard instantaneous luminosity.
A preselection of events requires each photon to have $p_T > 75$ GeV (where the trigger is efficient),
pseudorapidity $|\eta|<2.5$ with a veto on the CMS electromagnetic calorimeter barrel-endcap transition region, and a diphoton mass of $m_{\gamma\gamma} > 350$ GeV. The diphoton acoplanarity is defined as $a \equiv 1-|\Delta\phi_{\gamma\gamma}/\pi|$, where $\Delta\phi_{\gamma\gamma}$ is the azimuthal separation of the two photons. This is one of the discriminating variables used in the analysis, with exclusive diphoton candidates satisfying $a < 0.005$ (back-to-back). The invariant mass distribution, up to this stage of event selection, is shown in Fig.~\ref{fig:diphoton_mass}.
The kinematics of an opposite-arm, two-proton system is converted into missing mass and rapidity
of the central system kinematics through $m_{pp} = \sqrt{s\xi^+\xi^-}$, and $y_{pp} = (1/2) \log (\xi^+/\xi^-)$. In the case of exclusive diphoton production, both systems are correlated through $m_{pp} = m_{\gamma\gamma}$ and $y_{pp} = y_{\gamma\gamma} $. In this search, a 2$\sigma$ uncertainty window is used in matching the mass and rapidity between the central two photons and the two-proton systems within their uncertainties.
Only two events remain with an expected background prediction of $2.11$ $^{+0.96}_{-0.66}$ (stat) when no matching criteria is applied. Of these events, none of them contains a pair of forward proton tracks. In the 2$\sigma$ and 3$\sigma$ matching windows, the background prediction is respectively 0.23 $^{+0.08}_{-0.04}$ (stat) and 0.43 $^{+0.14}_{-0.08}$ (stat) events. No diphoton candidates with exclusive kinematic features are observed in either of the windows.
A 95\% confidence level observed upper limit of 3 fb is quoted on the light-by-light cross section within the fiducial region. This result permits the derivation of upper limits on anomalous four-photon coupling parameters, which are found to be $|\zeta_1|< 3.7 \times 10^{-13}$ GeV$^{-4}$ and $|\zeta_2|< 7.7 \times 10^{-13}$ GeV$^{-4}$ at 95\% CL. A similar procedure is used to derive the two-dimensional limits on the $\zeta_{1,2}$ parameters. The resulting two-dimensional 95\% confidence region is shown in Fig.~\ref{fig:couplings_4photon_PPS}.

The results by ATLAS and CMS demonstrate the feasibility of operating near-beam RP detectors in standard luminosity runs at the LHC for further photon-photon physics studies.
A perspective for these techniques is then to obtain measurements of the light-by-light scattering using proton tagging, as it has been started with the study mentioned above. We have shown in this review that
with photon-photon invariant mass above $1$ TeV would allow to study physics at the TeV scale beyond the SM.
In particular, we have explained how large extra dimension theories could be examined in this context. The limits we have extracted in this report could be improved by a factor $10$ using these new techniques and a luminosity of $200$ fb$^{-1}$. This would be  a breakthrough in this domain.
Let us also recall that in the context light-by-light scattering, we are at  the edge of observing the process in its full glory through collisions of laser beams
\cite{Sarazin:2016zer,King:2012aw,Takahashi:2018uut,felix}.

\vfill
\clearpage

In a longer future, huge improvements in the study of exotic states can be expected, for example with the
 Future Circular Collider (FCC). This machine is designed  such that it could start as an $e^+ e^-$ collider (FCC-ee) 
for a center of mass energy ranging from $90$ GeV to $365$ GeV. In the second
stage, the FCC would be turned into a $100$ TeV proton-proton machine (FCC-hh) with high-field
magnets of up to $16$ T, also suitable for heavy-ion collisions. With the addition of an energy
recovery electron linac (ERL) of $60$ GeV, also electron-proton interactions could be explored providing additional
input to achieve the ultimate precision of the FCC-hh. 
Using similar strategies as exposed in the previous sections, ALPs studies can be performed leading to improved 
 constraints \cite{pbb}.

 

\providecommand{\href}[2]{#2}\begingroup\raggedright

\end{document}